\documentclass[12pt]{article}
\usepackage{amssymb,amsmath,latexsym}
\usepackage{epsfig}
\usepackage{fancybox}
\usepackage{verbatim}
\usepackage{multirow}
\usepackage{ dsfont }

\usepackage{color,graphicx}
\usepackage{todonotes}
\usepackage{comment}
\usepackage{mathtools}
\usepackage{ulem}
\usepackage{float}

\usepackage{tikz}
\usepackage{tikz-cd}
\usepackage{todonotes}
\usepackage{multicol}

\usepackage{xcolor}
\definecolor{MyBlue}{cmyk}{1,0.13,0,0.63}
\definecolor{MyGreen}{cmyk}{0.91,0,0.88,0.52}
\newcommand{\mylinkcolor}{MyBlue}
\newcommand{\mycitecolor}{MyGreen}
\newcommand{\myurlcolor}{black}

\usepackage{hyperref}
\hypersetup{%
  bookmarksnumbered=true,bookmarksopen=false,%
  plainpages=false,% necessary to prevent duplicate page identifiers
  linktocpage=true,%
  colorlinks=true,breaklinks=true,%
  linkcolor=\mylinkcolor,citecolor=\mycitecolor,urlcolor=\myurlcolor,%
  pdfpagelayout=OneColumn,%
  pageanchor=true,%
}

\title{Random M\"obius dynamics on the unit disc
\\
and perturbation theory for Lyapunov exponents
}

\author{Florian Dorsch, Hermann Schulz-Baldes
\\
\\
{\small Department Mathematik, Friedrich-Alexander-Universit{\"a}t Erlangen-N{\"u}rnberg, Germany}
}

\date{ }

\newtheorem{theo}{Theorem}
\newtheorem{defini}[theo]{Definition}
\newtheorem{proposi}[theo]{Proposition}
\newtheorem{lemma}[theo]{Lemma}
\newtheorem{coro}[theo]{Corollary}
\newtheorem{rem}[theo]{Remark}

\newcommand{\ZZ}{{\mathbb Z}}

\newcommand{\RR}{{\mathbb R}}

\newcommand{\CM}{{\mathbb C}}
\newcommand{\NM}{{\mathbb N}}
\newcommand{\RM}{{\mathbb R}}
\newcommand{\SM}{{\mathbb S}}

\newcommand{\ZM}{{\mathbb Z}}

\newcommand{\PM}{{\mathbb P}}
\newcommand{\DM}{{\mathbb D}}

\newcommand{\EM}{{\mathbb E}}

\newcommand{\PP}{{\mathbb{P}}}

\newcommand{\EE}{{\bf E}}

\newcommand{\Dd}{{\cal D}}

\newcommand{\Ss}{{\cal S}}
\newcommand{\Oo}{{\cal O}}
\newcommand{\Tr}{\mbox{\rm Tr}}

\newcommand{\Nn}{{\cal N}}

\newcommand{\Ll}{{\cal L}}

\def\esssup{\mathop{\rm ess\,sup}}

\newcommand{\diag}{{\mbox{\rm diag}}}

\newcommand{\baret}{\hat{t}}
\newcommand{\barev}{\hat{v}}
\newcommand{\Ccont}{C}  % \mathsf{C}
\newcommand{\AT}{a}  
\newcommand{\CT}{c}  
\newcommand{\BT}{b}  

\begin{document}

\maketitle

%%%%%%%%%%%%%%%%%%%%%%%%%%%%%%%
\begin{abstract}
Randomly drawn $2\times 2$ matrices induce a random dynamics on the Riemann sphere via the M\"obius transformation. Considering a situation where this dynamics is restricted to the unit disc and given by a random rotation perturbed by further random terms depending on two competing small parameters, the invariant (Furstenberg) measure of the random dynamical system is determined. The results have applications to the perturbation theory of Lyapunov exponents which are of relevance for one-dimensional discrete random Schr\"odinger operators.

% 37H15  	Random dynamical systems aspects of multiplicative ergodic theory, Lyapunov exponents 
% 37C40  	Smooth ergodic theory, invariant measures for smooth dynamical systems
% 37M25  	Computational methods for ergodic theory (approximation of invariant measures, computation of Lyapunov exponents, entropy, etc.)
\end{abstract}
%%%%%%%%%%%%%%%%%%%%%%%%%%%%%%%

%%%%%%%%%%%%%%%%%%%%%%%%%%%%%%%
\section{Set-up, intuition and main results}

Perturbation theory for Lyapunov exponents associated to products of random matrices is of relevance for spectral analysis, random dynamical systems and quantum dynamics in random media, as well as numerous other physics related questions. Due to the tight connection of Lyapunov exponents and invariant measures on the projective space via the Furstenberg formula, one is naturally led to study these invariant measures in a perturbative regime. If there is a unique invariant measure, it is referred to as the {\it Furstenberg measure} and this is known to be the case under a variety of sufficient conditions \cite{BL,BQ,DKW}. When dealing with real $2\times 2$ matrices, the real projective space is a one-dimensional circle and the analysis of the Furstenberg measures becomes particularly trackable. The first rigorous contribution going back to Pastur and Figotin \cite{PF} considers a situation stemming from the one-dimensional Anderson model in which the random  $2\times 2$ matrices are given by a rotation with non-vanishing rotation angle perturbed by a small random term. In this situation, the invariant measure is given by the Lebesgue measure in a weak sense (testing only low frequencies and only up to error terms and provided the frequency is smaller than the rotation angle without perturbation). This is based on the so-called {\it oscillatory phase argument}, and the outcome is often also referred to as the {\it random phase property}. This technique can be pushed to deal with large deviations \cite{JSS} and a perturbative analysis of the variance in the central limit theorem \cite{SSS}. When the rotation angle is trivial so that the real random $2\times 2$ matrices are given by a random perturbation of the identity matrix, one deals with a so-called anomaly \cite{KW} and then the random phase property does not hold and alternative methods using second order ordinary differential equations on the unit circle have been developed to deal with this case \cite{BK,SB,DS}. Small random perturbations of a Jordan block are of relevance for band edges of random Jacobi matrices and can be dealt with by analyzing a singular differential equation \cite{DG,SS}. The perturbative analysis of larger random matrices is quite involved and exhibits a rich variety of dynamical behaviors resulting from the competition between hyperbolic and elliptic parts of the dynamics, see \cite{SB1,SV,DSB0} for analytical and \cite{RS} for numerical results.

\vspace{.2cm}

This work considers random $2\times 2$ matrices with \textit{complex} entries, again in a perturbative regime. The suitable projective space is then of real dimension $2$. Under the stereographic projection it can be identified with the Riemann sphere, on which the dynamics is then given by the M\"obius transformation. Sufficient conditions for the uniqueness of the invariant measure are given by Proposition~4.7 in \cite{BQ}.  For a perturbative analysis, it is then also natural to consider random matrices depending smoothly  on two small real parameters $\epsilon$ and $\delta$ (see assumption (i) below), rather than just one. A typical situation of this kind is the single-site Anderson model, in which the parameter $\epsilon$ measures the size of the randomness and the other parameter $\delta$ is the complex part of the energy, see Section~\ref{sec-Anderson}. In this latter situation, a non-negative value of $\delta$,  moreover, assures that the unit disc is an invariant subset of the Riemann sphere and hence carries the invariant measure. This is another important element of the set-up considered here (see assumption (v) below). The dynamics at $\epsilon=\delta=0$ is given by a rotation, while a non-vanishing $\epsilon$ adds randomness and $\delta$ introduces some hyperbolicity to the dynamics.  Hence the two-parameter random family is supported near a bifurcation point from an elliptic matrix (rotation as normal form) to a hyperbolic one. The competition between these two parameters leads to striking differences between the random dynamics and an interesting crossover regime which is analyzed in detail in this paper by combining oscillatory phase arguments and the techniques based on differential equations. A novel element is an oscillatory phase argument to {\it second} order in $\epsilon$, namely to functions on the unit disc that only depend on the modulus of their argument. This allows to derive a differential equation that approximately describes the radial distribution of the invariant measure based on the relative size of $\delta$ and $\epsilon^2$. The weightier the parameter $\delta$ is, the stronger the dynamics is pressed to the center of the unit disc by the hyperbolicity. The differential equation is then dealt with by techniques developed in \cite{SS} in order to obtain an approximate radial density of the invariant measure. This then also allows to compute the leading orders of the Lyapunov exponent in $\epsilon$ and $\delta$ around $\epsilon=\delta=0$, see Section~\ref{Lyapunov}. In the remainder of this introduction, the random dynamics will first be described in a qualitative and intuitive manner, and then the main rigorous results are stated and illustrated by numerical results.

%%%%%%%%%%%%%%%%%%%%%%%%%%%
\subsection{M\"obius action}
\label{sec-Moeb}

Let us begin by recalling the framework. The standard action of a matrix $T\in\CM^{2\times 2}$ with $\det(T)=1$ on the cover $\mathbb{S}_{\mathbb{C}}^1=\left\{x\in\mathbb{C}^2: \|x\|=1\right\}$ of the complex projective space $\mathbb{C}\mathsf{P}^1\cong\mathbb{S}^{1}_{\mathbb{C}}/\textnormal{U}(1)$ is given~by
$$
T\star x
\;=\;
Tx\|Tx\|^{-1}
\,.
$$
The stereographic projection $\pi:\mathbb{S}_{\mathbb{C}}^1\rightarrow\overline{\mathbb{C}}=\CM\cup\{\infty\}\cong\mathbb{C}\mathsf{P}^1$, which is given by
\begin{equation}
\label{eq-Stereo}
\pi(x)
\;=\;
\begin{cases} ab^{-1} \,, \quad &b\neq 0\,,\\\infty\,,\quad &b=0\,,\end{cases}
\qquad
x\,=\,\begin{pmatrix}a\\b\end{pmatrix}\;,
\end{equation}
satisfies $\pi(T\star x)=T\cdot \pi(x)$. Here, $T\cdot$ denotes the M\"obius action given by
\begin{equation}
\label{eq-Moebius}
T\cdot z
\;=\; \frac{az+b}{cz+d}
\;,
\qquad
T
\;=\;
\begin{pmatrix}
a & b \\ c & d
\end{pmatrix}
\;,
\end{equation}
when $z\not\in\{-dc^{-1},\infty\}$, for which one sets $T\cdot (-dc^{-1})=\infty$ and $T\cdot \infty=ac^{-1}$. 

\vspace{.2cm}

Of importance will be two subsets of $\mbox{\rm SL}(2,\CM)=\left\{T\in\mathbb{C}^{2\times 2}\,:\, \det(T)=1\right\}$, namely the Lorentz subgroup
$$
\mbox{\rm SU}(1,1)
\;=\;
\left\{T\in\mathbb{C}^{2\times 2}: T^*JT=J\,,\,\det(T)=1\right\}\;,
\qquad
J
\;=\;\begin{pmatrix} 1 & 0 \\ 0 & -1 \end{pmatrix}
\;,
$$
and the semigroup of sub-Lorentzian matrices
$$
\textnormal{SU}_{\leq}(1,1)
\;=\;
\left\{T\in\mathbb{C}^{2\times 2}: T^*JT\leq J\,,\,\det(T)=1\right\}
\;.
$$

The M\"obius action with $T\in \mbox{\rm SU}(1,1)$ leaves the unit circle invariant, \textit{i.e.}, it obeys $T\cdot \SM^1\subset \SM^1$, where $\SM^1=\{z\in\CM:|z|=1\}$. Furthermore, the action with $T\in \mbox{\rm SU}_{\leq}(1,1)$ leaves the open unit disc invariant, \textit{i.e.}, it satisfies $T\cdot \DM\subset \DM$, where $\DM=\{z\in\CM:|z|<1\}$. The latter fact follows from the inequality
\begin{align}\label{invariance-of-disc}
\left(1-|T\cdot z|^2\right)\left|\begin{pmatrix}0\\1\end{pmatrix}^*T\begin{pmatrix}z\\1\end{pmatrix}\right|^2
\;=\;
-\,\begin{pmatrix}z\\1\end{pmatrix}^*T^*JT\begin{pmatrix}z\\1\end{pmatrix}
\;\geq\;
-\,\begin{pmatrix}z\\1\end{pmatrix}^*J\begin{pmatrix}z\\1\end{pmatrix}
\;=\;1-|z|^2
\;,
\end{align}
holding true for all $z\in\overline{\mathbb{D}}$ and all $T\in\textnormal{SU}_{\leq}(1,1)$.  Note that for $T\in\textnormal{SU}(1,1)$, inequality~\eqref{invariance-of-disc} becomes an equality, hence implying $T\cdot \SM^1\subset \SM^1$.

%Note that $T\cdot z\neq \infty$ holds for all $z\in\overline{\mathbb{D}}$.

%%%%%%%%%%%%%%%%%%%%%%%%%%%
\subsection{Random dynamical system}
\label{sec-RandAct}

If now a sequence $(T_n)_{n\geq 1}$ of  complex $2\times 2$ matrices with unit determinant is drawn independently and identically distributed from a family $(T_\sigma)_{\sigma\in\Sigma}$ according to a law $\PM$ on a probability space $\Sigma$, one obtains a random dynamical system on  $\overline{\CM}$ by setting
\begin{equation}
\label{eq-RDS}
z_n\;=\;T_n\cdot z_{n-1}
\;,
\end{equation}
where $z_0\in\overline{\CM}$ is some initial starting point. As $z_n=\pi(x_n)$, one also has an associated dynamics on $\mathbb{S}^1_{\mathbb{C}}$ satisfying $x_n=T_n\star x_{n-1}$. The average w.r.t. $\mathbb{P}$ is denoted by $\mathbb{E}$. If such a random sequence $(T_n)_{n\geq 1}$ satisfies $\mathbb{E}\max\{\log\|T_1\|,0\}<\infty$, the associated Lyapunov exponent is defined by
\begin{align}\label{Lyapunov-exponent-definition}
\gamma
\;=\;
\lim_{N\to\infty}
\;\frac{1}{N}\,\log\big(\|T_{N}\cdots T_1\|\big)
\end{align}
with convergence either almost surely or in expectation \cite{BL}. The second objects of interest here are the invariant probability measures $\mu$ on $\overline{\CM}$ defined by
\begin{align}\label{invariant-probability-measure-definition}
\int_{\overline{\mathbb{C}}}\mu(\textnormal{d}z)\,f(z)\;=\;
\EM\,\int_{\overline{\mathbb{C}}}\mu(\textnormal{d}z)\,f(T_\sigma \cdot z)
\;,
\qquad
f\in\Ccont(\overline{\CM})
\;.
\end{align}
If the support of $T_\sigma$ is \textit{irreducible}, namely all $z\in\overline{\mathbb{C}}$ obey $\textnormal{supp}(T_\sigma)\cdot z\not\subset\{z\}$, then any such invariant probability measure $\mu$ satisfies
\begin{align*}
\gamma\;=\;\mathbb{E}\,\int_{\overline{\mathbb{C}}}\mu(\textnormal{d}z)\,\log\big(\|T_\sigma  \overline{\pi^{-1}(z)}\|\big)
\;,
\end{align*}
where $\overline{\pi^{-1}(z)}$ denotes an arbitrary element of $\pi^{-1}(z)$ (see~\cite{BQ}, Theorem 4.28). This implies
\begin{align}\label{Furstenberg-formula}
\gamma
\;=\;
\EM\,\int_{\mathbb{S}^1_{\mathbb{C}}} {\nu}(\textnormal{d}x)\,\log\big(\|T_\sigma  x\|\big)
\end{align}
for any corresponding probability measure $\nu$ on $\SM^1_\CM$ satisfying $\pi_*(\nu)=\mu$.
If $\textnormal{supp}(T_\sigma)$ is even \textit{strongly irreducible}, namely all finite $F\subset\overline{\mathbb{C}}$ obey $\textnormal{supp}(T_\sigma)\cdot F\not\subset F$, and if the semigroup generated by $\textnormal{supp}(T_\sigma)$ is not relatively compact, then the invariant probability measure $\mu$ on $\overline{\mathbb{C}}$ satisfying~\eqref{invariant-probability-measure-definition} is unique and called the {\it Furstenberg measure} (see~\cite{BQ}, Proposition 4.7).

%%%%%%%%%%%%%%%%%%%%%%%%%%%
\subsection{Two-parameter family and list of assumptions}
\label{sec-2para}

As already stated above, examples such as the ones described in Section~\ref{sec-Anderson} lead us to consider a family $(T^{\epsilon,\delta}_\sigma)_{\sigma\in\Sigma}$ satisfying the following assumptions:

\begin{itemize}

\item[{\rm (i)}] There is a neighborhood of $(0,0)\in\RM^2$ on which $(\epsilon,\delta)\mapsto T^{\epsilon,\delta}_\sigma$ is twice continuously differentiable for all $\sigma\in\Sigma$.

\item[{\rm (ii)}] At $\epsilon=\delta=0$, one has $T^{0,0}_\sigma=\diag(e^{\imath \eta_\sigma},e^{-\imath \eta_\sigma})$ for some $\eta_\sigma\in[0,2\pi)$. Here, $\imath=\sqrt{-1}$.

\item[{\rm (iii)}]  For $\delta=0$, $T^{\epsilon,0}_\sigma\in\mbox{\rm SU}(1,1)$.

\item[{\rm (iv)}] For $\epsilon=0$, $T^{0,\imath \delta}_\sigma\in\mbox{\rm SU}(1,1)$. 

\item[{\rm (v)}] For $\delta\geq 0$, $T^{\epsilon,\delta}_{\sigma}\in\textnormal{SU}_{\leq}(1,1)$.

\end{itemize}

For each pair $(\epsilon,\delta)$ of fixed parameters, one now obtains by \eqref{eq-RDS} a random dynamical system, the orbits of which will be denoted by $(z_n )_{n\geq 0}$ and $(x_n )_{n\geq 0}$. Thus their dependence on $\epsilon$ and $\delta$ is suppressed. Nevertheless, the dependence of other quantities such as the Lyapunov exponent $\gamma^{\epsilon,\delta}$ and the invariant measure $\mu^{\epsilon,\delta}$  will be kept as it is precisely the main object of the paper to study their dependence on these parameters.  The monotonicity assumption (v) is motivated by applications to random Schr\"odinger operators, see Section~\ref{sec-Anderson}. For $\delta\geq 0$ and an initial condition $z_0\in\DM$, assumption (v) combined with the comments in Section~\ref{sec-Moeb} assure that the random dynamics stays in the unit disc $\DM$. Hence one is indeed in the situation described in the title and the abstract. Section~\ref{sec-Drop(v)} briefly discusses what happens if (v) is dropped. The assumptions (i)-(iv) allow to expand the matrices $T^{\epsilon,\delta}_\sigma$ in a Taylor expansion
\begin{equation}
\label{first-basis-change}
T^{\epsilon,\delta}_\sigma
\;=\;
R_{\eta_{\sigma}}\,
\exp\left[\epsilon {P}_{\sigma}+\epsilon^2{P}^{\prime}_{\sigma}+\imath\delta {Q}_{\sigma}\;+\;\mathcal{O}(\epsilon^3,\epsilon\delta,\delta^2)\right]\,,
\end{equation}
where $R_{\eta_{\sigma}}=\diag(e^{\imath \eta_\sigma},e^{-\imath \eta_\sigma})$ and ${P}_{\sigma}$, ${P}_{\sigma}^{\prime}$, ${Q}_{\sigma}$ and the terms of higher order are random variables with values in the Lie algebra 
$$
\textnormal{su}(1,1)\;=\;
\left\{A\in\mathbb{C}^{2\times 2}: A^*J=-JA\,,\,\Tr(A)=0\right\}
\;
$$
{of $\textnormal{SU}(1,1)$.} In equation~\eqref{first-basis-change} and in a similar manner many times below, we use the notation $\mathcal{O}(\epsilon^3,\epsilon\delta,\delta^2)=\mathcal{O}(\epsilon^3)+\mathcal{O}(\epsilon\delta)+\mathcal{O}(\delta^2)$. These coefficient matrices are assumed to satisfy the following properties, which imply, in particular, that the support of $T^{\epsilon,\delta}_{\sigma}$ is compact:
\begin{itemize}
\item [{\rm (vi)}] The random matrices $P_{\sigma}$, $P^{\prime}_{\sigma}$ and $Q_{\sigma}$ are uniformly bounded in norm.
\item [{\rm} (vii)] The error term $\mathcal{O}(\epsilon^3,\epsilon\delta,\delta^2)$ in \eqref{first-basis-change} is bounded by $C(\epsilon^3+\epsilon\delta+\delta^2)$ for a uniform constant~$C$.
\end{itemize}
Now, if either $\epsilon$ or $\delta$ is non-zero, the following additional assumptions are generic:
\begin{itemize}
\item[{\rm (viii)}]  The support of $T^{\epsilon,\delta}_{\sigma}$ is \textit{strongly irreducible},~\textit{i.e.}, all finite $F\subset\overline{\mathbb{C}}$ obey $\textnormal{supp}(T^{\epsilon,\delta}_{\sigma})\cdot F\not\subset F$.
\item[{\rm (ix)}] {The semigroup generated by $\textnormal{supp}(T^{\epsilon,\delta}_{\sigma})$ is not relatively compact.}
\end{itemize}
Under the assumptions (viii) and (ix), the $(T^{\epsilon,\delta}_{\sigma}\cdot)$-invariant probability measure $\mu^{\epsilon,\delta}$ is unique~\cite{BQ}.

\vspace{.2cm}

%%%%%%%%%%%%%%%%%%%%%%%%%%%
\subsection{Qualitative and intuitive description of the dynamics}
\label{sec-Qualitative}

For $\epsilon=\delta=0$, the M\"obius dynamics is merely given by the multiplication by the random phase $e^{2\imath \eta_\sigma}$. Hence, the points $0$ and $\infty$ are fixed points of the action $T^{0,0}_{\sigma}\cdot$ and its non-trivial orbits lie on circles $r\,\mathbb{S}^1$, $r>0$. Neither (viii) nor (ix) hold in this situation. On each such circle $r\,\mathbb{S}^1$, a $(T^{0,0}_{\sigma}\cdot)$-invariant probability measure is given by the normalized spherical measure, which is unique on $r\mathbb{S}^1$ if the support of $\eta_\sigma$ contains an irrational multiple of $\pi$ (if the support of $\eta_\sigma$ is a finite subset of ${\pi}\mathbb{Q}$, then there are many $(T^{0,0}_{\sigma}\cdot)$-invariant probability measures on $r\mathbb{S}^1$). The Lyapunov exponent vanishes in this case.

%%%%%%%%%%%%%%%%%%%%%%%%%%%%%%%
\begin{figure}
\centering
\includegraphics[height=6cm]{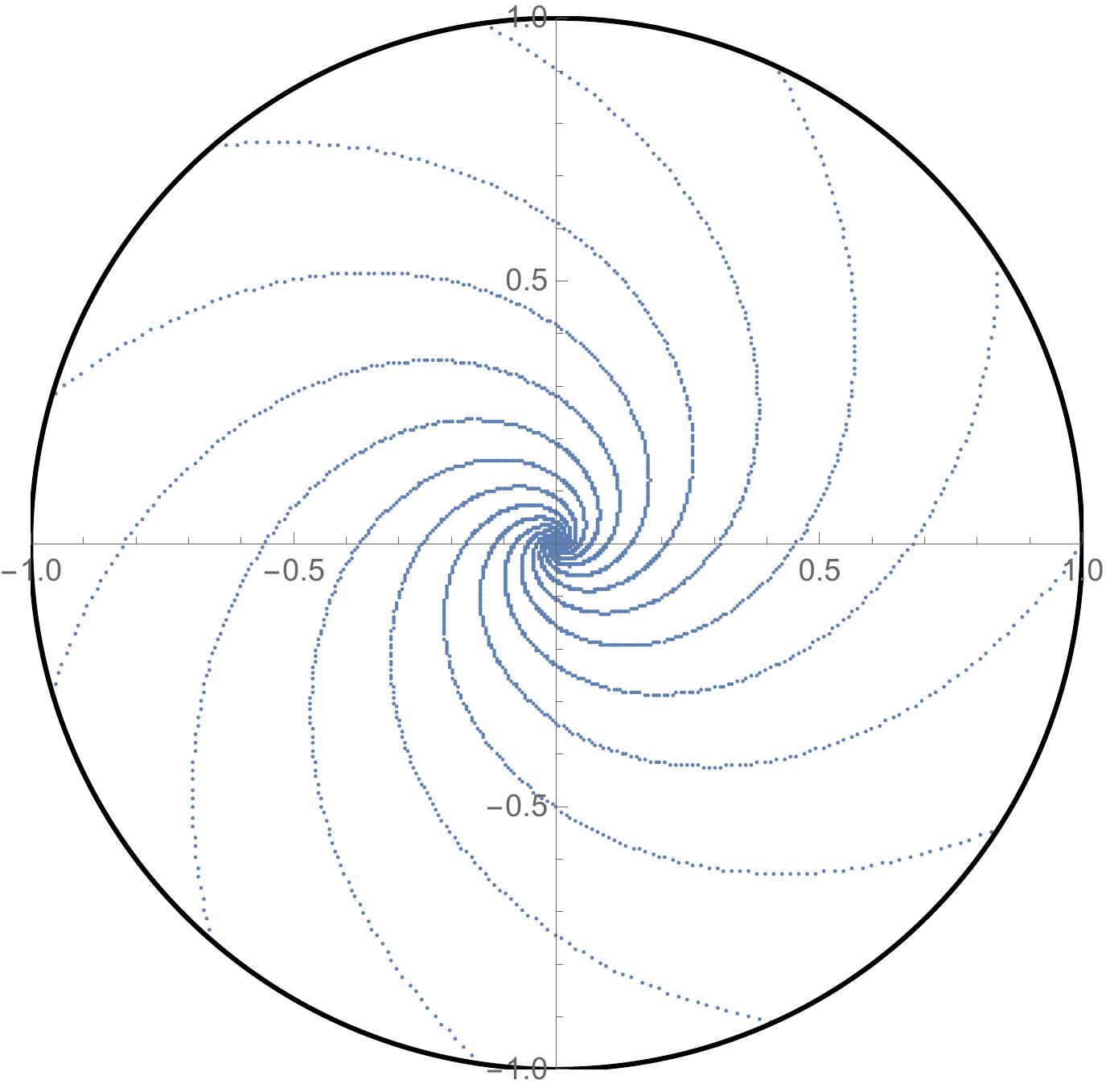}
\hspace{.6cm}
\includegraphics[height=5.6cm]{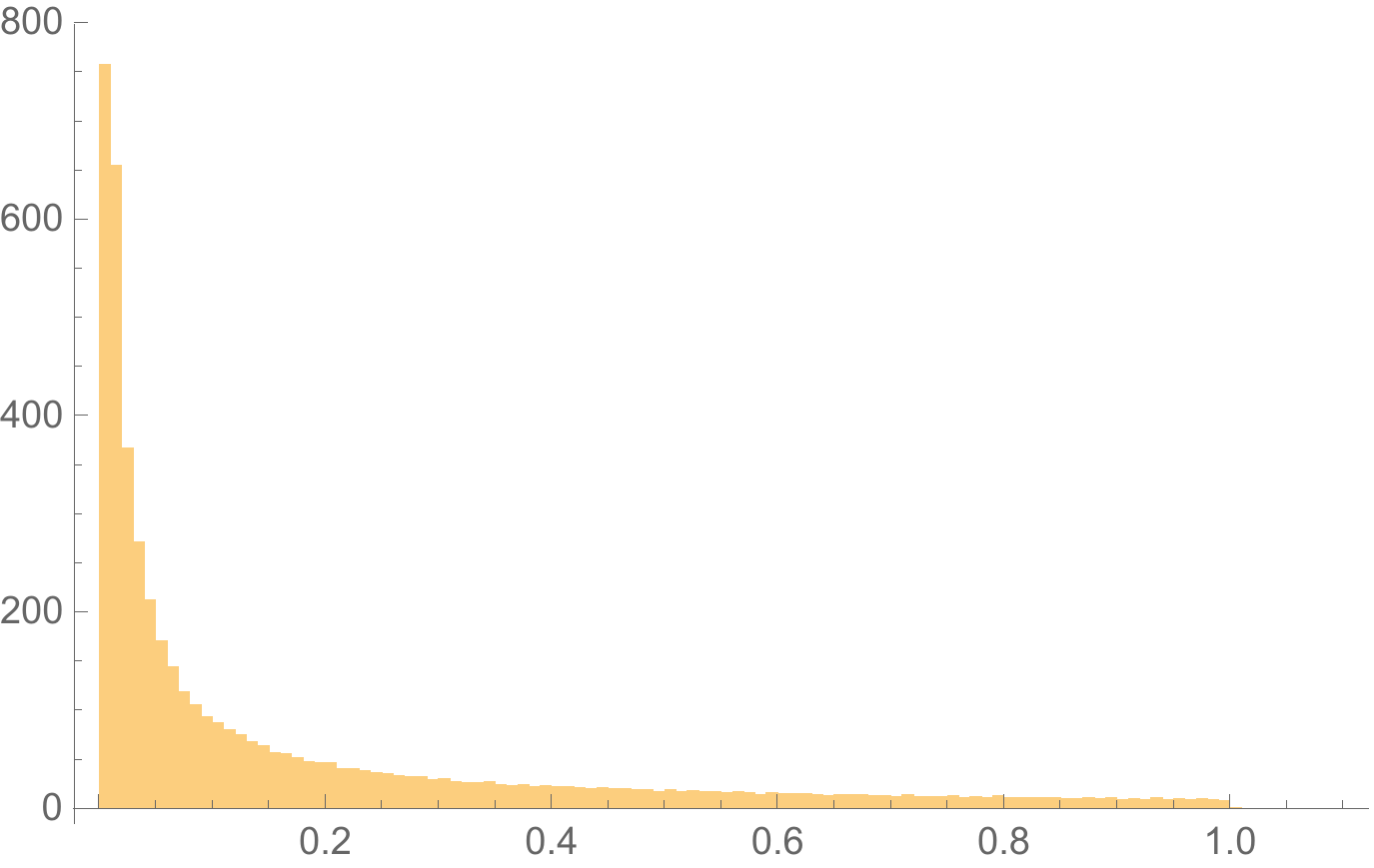}
\caption{\sl 
Plot of an orbit $(z_n)_{n=1,\ldots,N}$ in $\DM$ with $N=5\cdot 10^3$ iterations of the random model described in {\rm Section~\ref{sec-Anderson}}.  The parameters are  $\epsilon=10^{-4}$ and $\delta=10^{-3}$ so that $\epsilon=o(\delta)$, and the initial condition is $z_0=1$. The random variable $\eta_{\sigma}\equiv -2$ is a constant. The histogram shows the distribution of the radii~$|z_n|$. The tail of the distribution is merely due to the thermalization and does not occur if~$z_0=0$.
}
\label{fig-Spec1}
\end{figure}
%%%%%%%%%%%%%%%%%%%%%%%%%%%%%%%

\vspace{2mm}

For $\delta=0$ and arbitrary $\epsilon$, the M\"obius action of $T^{\epsilon,0}_\sigma\in\mbox{\rm SU}(1,1)$ leaves the unit circle $\partial\DM=\SM^1$ invariant.  Hence, there is a $(T^{\epsilon,0}_{\sigma}\cdot)$-invariant measure supported on $\SM^1$. Rigorous perturbation theory for the Lyapunov exponent has been carried out in \cite{JSS,SSS} under the two assumptions $\EM(e^{2\imath \eta_\sigma})\not=1$ and $\EM(e^{4\imath \eta_\sigma})\not=1$, showing that
\begin{equation}
\label{eq-GammaIntro}
\gamma^{\epsilon,0}
\;=\;
\mathcal{D}\,\epsilon^2\;+\;\Oo(\epsilon^3)
\;,
\end{equation}
where $\mathcal{D}\geq 0$ is a constant, the definition of which will be recalled in \eqref{definition-D} below. The property $\mathcal{D}>0$ can be characterized (see Proposition~\ref{Schrader-result} below). If the two assumptions $\EM(e^{2\imath \eta_\sigma})\not=1$ and $\EM(e^{4\imath \eta_\sigma})\not=1$ do not hold, one has to deal with an anomaly. Nevertheless, a more involved analysis still leads to a quadratic behavior in $\epsilon$ as in \eqref{eq-GammaIntro}, but a less explicit formula for the constant $\mathcal{D}$ \cite{SB,SS}.

\vspace{.2cm}

Now let us assume (viii) and (ix) and come to the novel part of this paper, namely the case $\delta>0$. We first focus on the simpler case $\epsilon=0$. Under suitable (weak) conditions on $Q_{\sigma}$, the M\"obius action $T^{0,\delta}_\sigma\cdot$ then drifts towards the center of $\mathbb{D}$ and has a deterministically attracting region around the origin. This implies that the support of $\mu^{0,\delta}$ is a compact subset of $\mathbb{D}$ (see Proposition~\ref{center-support} below). As now $\epsilon$ is increased, this behavior is maintained as long as $\epsilon =o(\delta)$. This is illustrated in Figure~\ref{fig-Spec1} by a numerical experiment for a generic model with the properties described above. Figure~\ref{fig-Spec1} shows that, starting out with an initial condition at $z_0=1$, the orbit rotates with a fixed (thus here non-random) angle, while it spirals towards the attracting region in the center because $\delta>0$ induces this drift. The concrete form of the random family $(T^{\epsilon,\delta}_\sigma)_{\sigma\in\Sigma}$ stems from the Anderson model and is described in detail in Section~\ref{sec-Anderson}. As $\epsilon$ is increased further, the support of the Furstenberg measure generically grows.  However, as long as $\epsilon=o(\delta^{\frac{1}{2}})$, the main weight of $\mu^{\epsilon,\delta}$ is close to the center. This is consistent with the bound \eqref{Furstenberg-result-statement-2} in Theorem~\ref{Furstenberg-result}.

\vspace{.2cm}

\begin{figure}
\centering
\includegraphics[height=6cm]{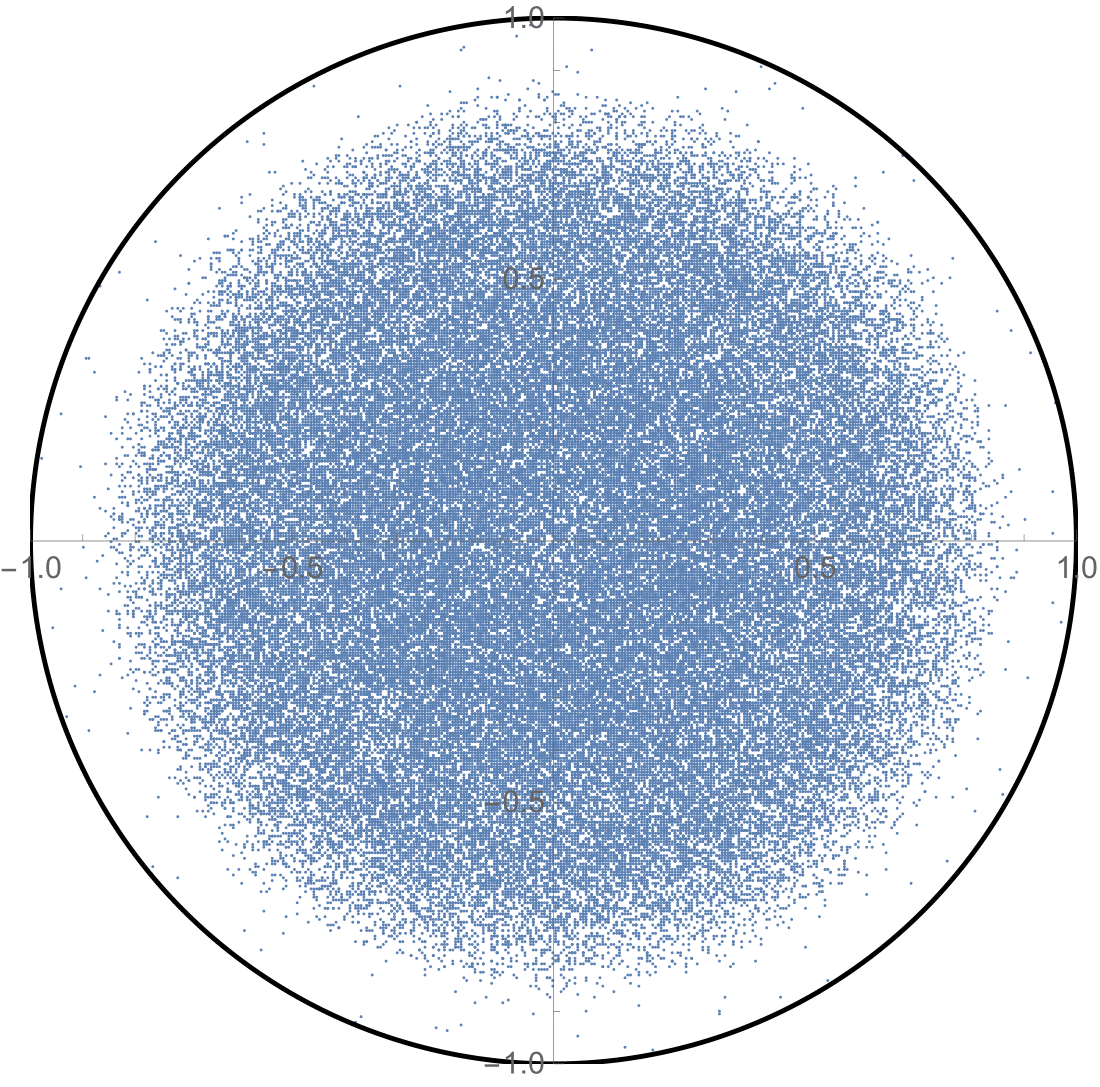}
\hspace{.6cm}
\includegraphics[height=5.6cm]{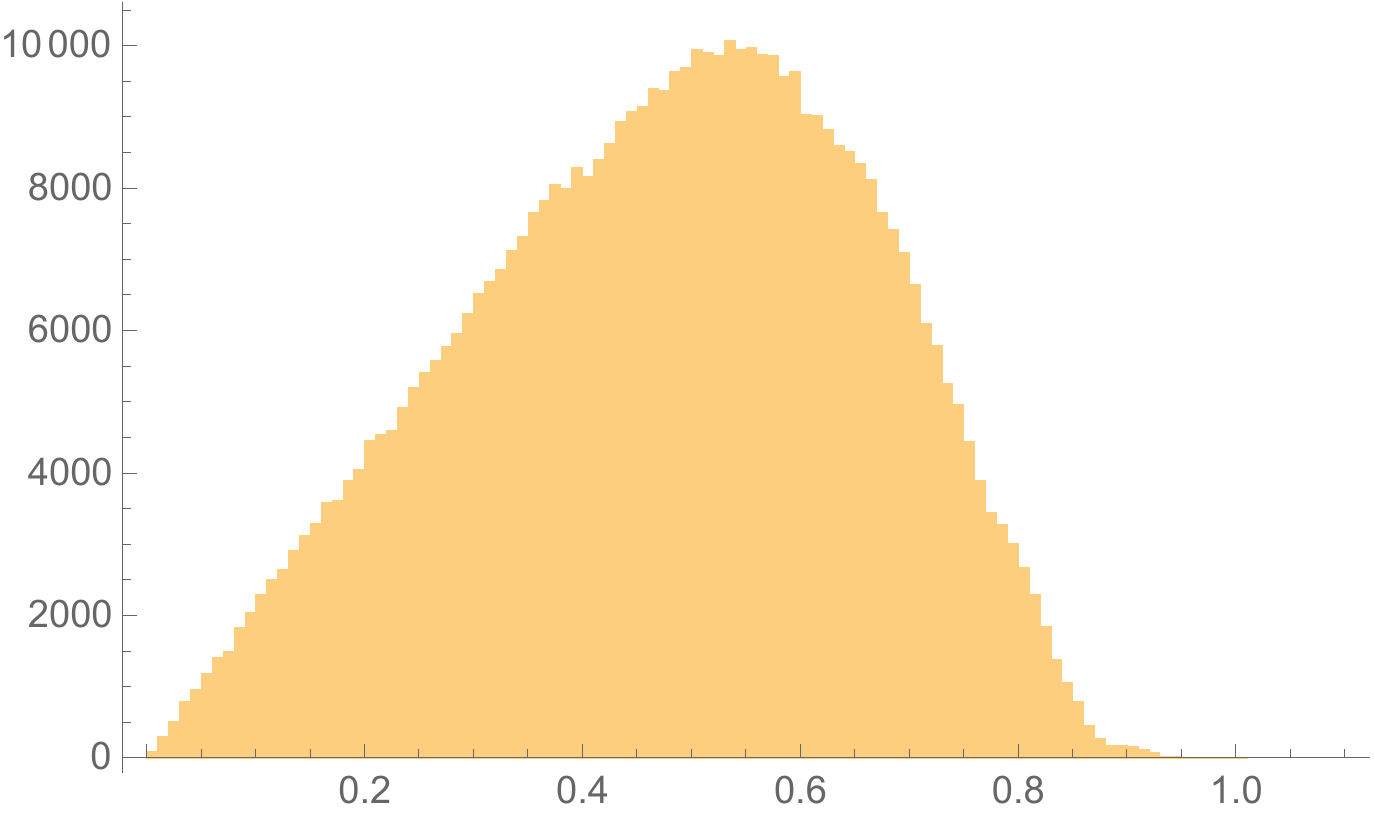}
\caption{\sl
Same plot and histogram as in {\rm Figure~\ref{fig-Spec1}}, but with $\epsilon=0.1$ and $\delta=10^{-3}$. For the plot on the left, $10^5$ iterations were run; for the histogram on the right, $5\cdot 10^5$ iterations were run. As in Figure~\ref{fig-Spec1}, the initial condition is $z_0=1$. If the system starts with $z_0=0$, the plot and the histogram look very similar.
}
\label{fig-Spec2}
\end{figure}

Once $\epsilon^2$ and $\delta$ are of the same order of magnitude, one reaches a crossover regime and the weight of the invariant measure is spread out over the whole closed unit disc. A plot of a typical orbit in that intermediate regime is shown in Figure~\ref{fig-Spec2}. It is one of the main results of the paper to determine the radial distribution of $\mu^{\epsilon,\delta}$ in a weak perturbative sense and to show how it depends on the ratio of $\frac{\delta}{\epsilon^2}$ (see statement~\eqref{Furstenberg-result-density} of Theorem~\ref{Furstenberg-result} below). Actually, as the ratio decreases, the radial distribution is more and more concentrated at the boundary.

%%%%%%%%%%%%%%%%%%%%%%%%%%%%%%%
%%%%%%%%%%%%%%%%%%%%%%%%%%%%%%%

\vspace{.2cm}

\begin{figure}
\centering
\includegraphics[height=6cm]{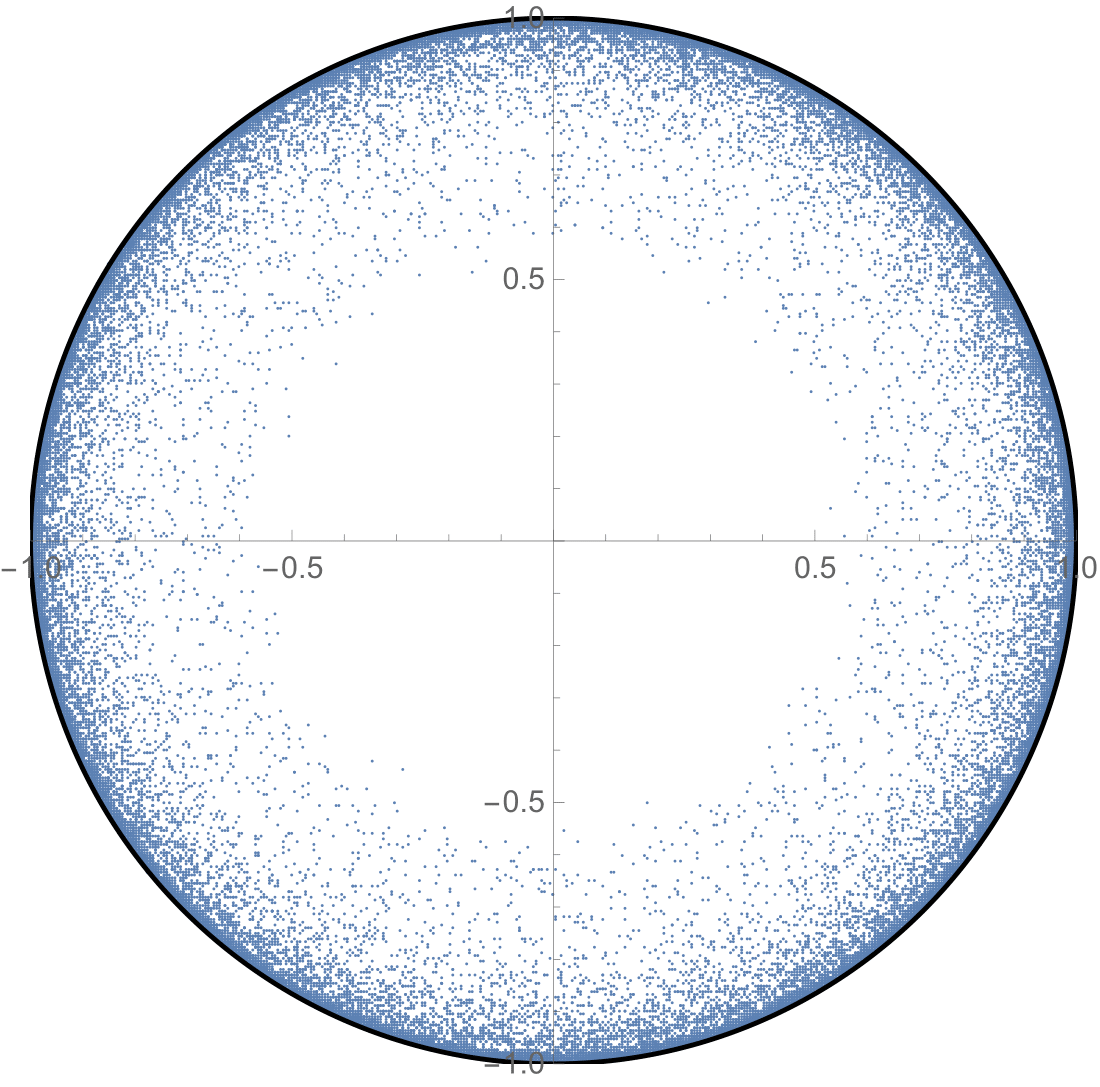}
\hspace{.6cm}
\includegraphics[height=5.6cm]{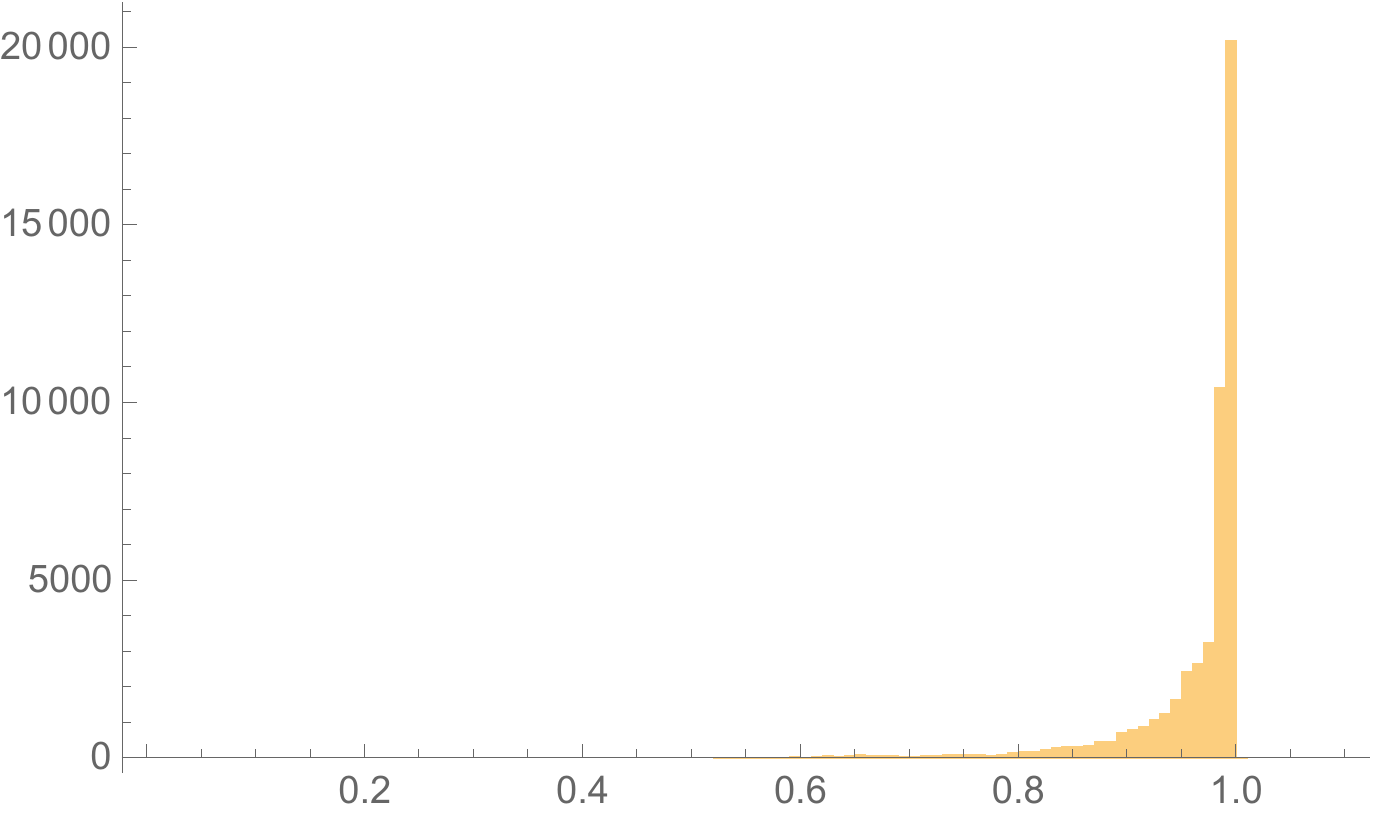}
\caption{\sl
Same plot and histogram as in {\rm Figure~\ref{fig-Spec1}}, but with $\epsilon=0.1$ and $\delta=10^{-5}$ so that $\delta=o(\epsilon^2)$. The number of iterations is $5\cdot 10^{4}$. The initial condition was $z_0=1$. If one chooses $z_0=0$ as initial condition, the orbit takes several hundreds of iterations to attain the boundary, but the histogram after a large number of iterations essentially looks the same.
}
\label{fig-Spec3}
\end{figure}

As $\epsilon$ grows further so that $\delta=o(\epsilon^2)$, the numerical experiments exhibit a striking feature (see Figure~\ref{fig-Spec3}): even though the iterative dynamics $(T^{\epsilon,\delta}_\sigma)^N\cdot z_0$ for each fixed $\sigma$ converges as $N\to\infty$ to a globally attractive fixed point well inside the unit disc (for $\delta>0$), the random orbit sticks to the boundary so that the Furstenberg measure apparently has very little weight outside of a small ring touching $\partial\DM$. This drift to $\partial\DM$ results from the positive Lyapunov exponent (of order $\epsilon^2$ uniformly in $\delta\ll \epsilon^2$) and is consistent with the bound \eqref{Furstenberg-result-statement-1} in Theorem~\ref{Furstenberg-result}. Indeed, presumably the measure $\mu^{\epsilon,\delta}$ converges for $\delta\rightarrow 0$ weakly to the Furstenberg measure $\mu^{\epsilon,0}$ supported on $\mathbb{S}^1$ described above.

%%%%%%%%%%%%%%%%%%%%%%%%%%%%%%%
%%%%%%%%%%%%%%%%%%%%%%%%%%%%%%%

%\vspace{1cm}

%%%%%%%%%%%%%%%%%%%%%%%%%%%
\subsection{Main results on the invariant measures}
\label{sec-InvMeas}

Let us now state the main result on the Furstenberg measure. Its first two items confirm the numerical results of Figures~\ref{fig-Spec1} and \ref{fig-Spec3} in a rather weak form, respectively. More generally, the third item provides an approximation of the radial distribution of $\mu^{\epsilon,\delta}$. To state the results and further ones below, let us introduce the following basis of su$(1,1)$:
\begin{align*}
B_1\;=\;\begin{pmatrix}0&1\\1&0\end{pmatrix}
\;,
\qquad 
B_2\;=\;
\begin{pmatrix}0&\imath\\-\imath&0\end{pmatrix}
\;,
\qquad 
B_3\;=\;
\begin{pmatrix}\imath&0\\0&-\imath\end{pmatrix}
\;.
\end{align*}
The expression for $\mathcal{D}$ in terms of $\eta_{\sigma}$ and the first order term $P_\sigma$~in~\eqref{first-basis-change} is
\begin{align}
\label{definition-D}
\mathcal{D}
\;=\;
\frac{1}{2}\,
\mathbb{E}(|\beta_{\sigma}|^2)
\;+\;
\Re\mathfrak{e}\left(\frac{\mathbb{E}(\beta_{\sigma})\mathbb{E}\left(e^{2\imath\eta_{\sigma}}\overline{\beta_{\sigma}}\right)}{1-\mathbb{E}(e^{2\imath\eta_{\sigma}})}\right)
\;,
\end{align}
where
\begin{align}\label{definition-of-beta}
\beta_{\sigma}\;=\;\frac{1}{2}\,\Tr\big((B_1-\imath B_2)P_{\sigma}\big)
\;.
\end{align}
Let us also introduce a notation for a further constant that turns out to be relevant in the following:
$$
\mathcal{C}
\;=\;
\frac{1}{2}\,\mathbb{E}\big(\Tr(B_3^*Q_{\sigma})\big)\,.
$$
In Remark~\ref{non-negativity-of-q_3} below, it is shown that assumptions (iii), (iv) and (v) imply $\mathcal{C}\geq 0$. If both $\mathcal{C}>0$ and $\mathcal{D}>0$, it is furthermore convenient to use
\begin{equation}
\label{eq-LambdaDef}
\lambda
\;=\;
2\,\frac{\mathcal{C}}{\mathcal{D}}\,\frac{\delta}{\epsilon^2}
\end{equation}
as a measure of the (crucial) relative size of $\delta$ and $\epsilon^2$.

\vspace{.2cm}

Furthermore, we denote by $\Ccont^1([0,1])$ differentiable functions where the derivatives at the boundary points (and only there) are taken one-sided, and this derivative is a continuous function on $[0,1]$. The classes $\Ccont^k([0,1])$  for $k\in\NM$ are then defined by iteration in $k$. 

%%%%%%%%%%%%%%%%%%%%%%%%%%%%%%%
\begin{theo}
\label{Furstenberg-result}
Assume \textnormal{(i)-(ix)} as well as $\mathbb{E}(e^{2\imath  \eta_{\sigma}})\neq 1$ and $\mathbb{E}(e^{4\imath  \eta_{\sigma}})\neq 1$. If $\mathcal{C}>0$, one has
\begin{equation}
\label{Furstenberg-result-statement-2}
\int_{\overline{\mathbb{D}}}  \mu^{\epsilon,\delta}(\textnormal{d}z)\;|z|^2
\;=\;
\mathcal{O}(\delta,\epsilon,\epsilon^2\delta^{-1})
\;,
\end{equation}
and, if $\mathcal{D}>0$, one has
\begin{equation}
\label{Furstenberg-result-statement-1}
\int_{\overline{\mathbb{D}}}  \mu^{\epsilon,\delta}(\textnormal{d}z)\;|z|^2
\;=\;
1\;+\;
\mathcal{O}(\epsilon^{\frac{1}{2}},\delta^{\frac{1}{2}}\epsilon^{-1})\,.
\end{equation}
Further, if $\mathcal{C}>0$ and $\mathcal{D}>0$, the radial distribution of $\mu^{\epsilon,\delta}$ is approximated in a weak sense by the radial density
\begin{align}
\label{Furstenberg-radial-density}
\varrho_{\lambda}(s)
\;=\;
\frac{\lambda}{(1-s)^2}\,\exp\left[-\,\frac{\lambda\,s}{1-s}\right]\,,
\end{align}
with $\lambda$ given by \eqref{eq-LambdaDef}, namely more precisely, for all $h\in\Ccont^2([0,1])$, one has
\begin{align}\label{Furstenberg-result-density}
\int_{\overline{\mathbb{D}}}\mu^{\epsilon,\delta}(\textnormal{d}z)\textnormal{ }{h}(|z|^2)
\;=\;
\int_0^1\textnormal{d}s\,\varrho_{\lambda}(s)\,h(s)\;+\;\mathcal{O}(\epsilon,\epsilon^{-1}\delta)\,.
\end{align}
\end{theo}
%%%%%%%%%%%%%%%%%%%%%%%%%%%%%%%

\begin{figure}
\centering
\includegraphics[height=3.4cm]{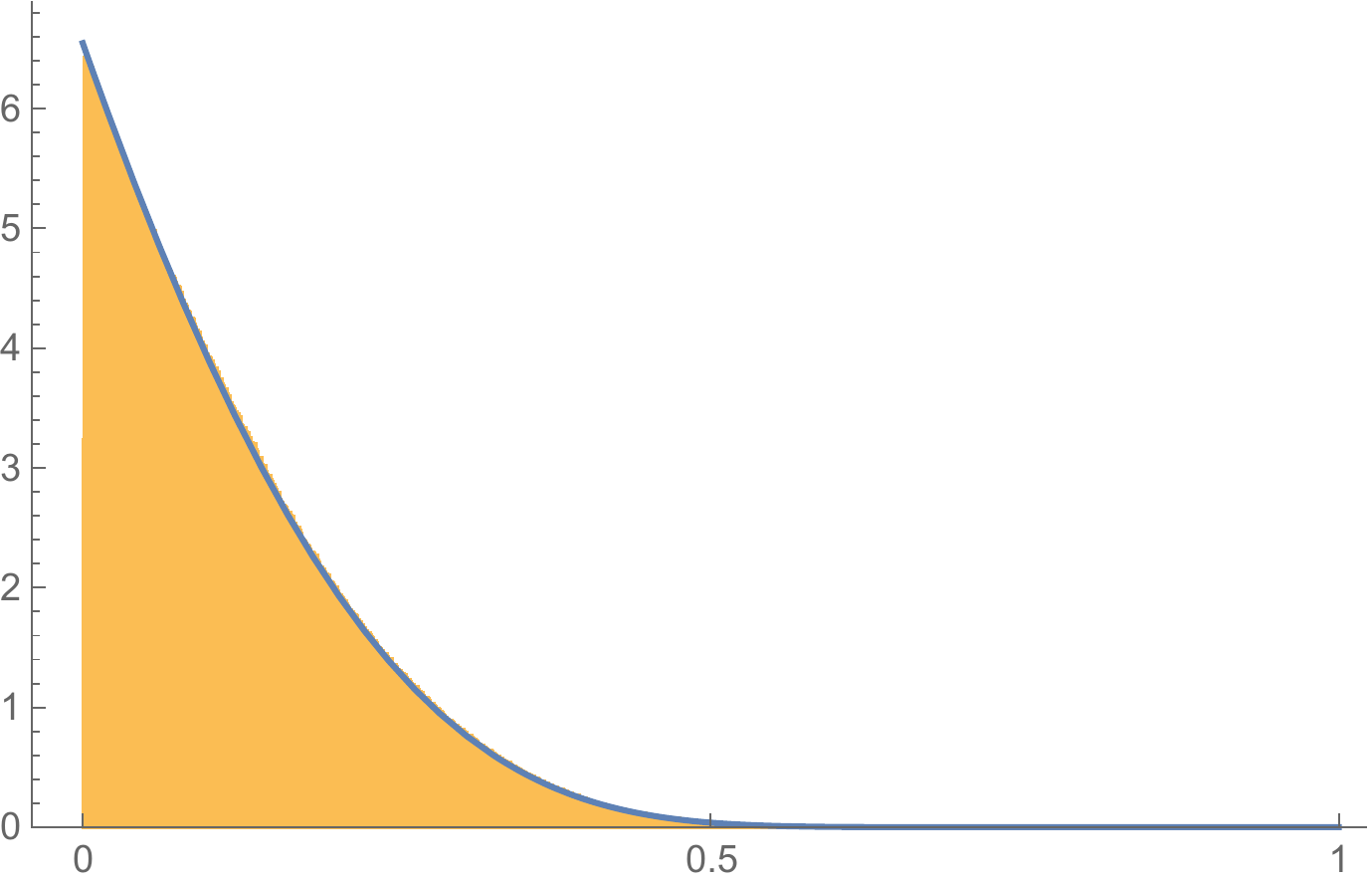}
\includegraphics[height=3.4cm]{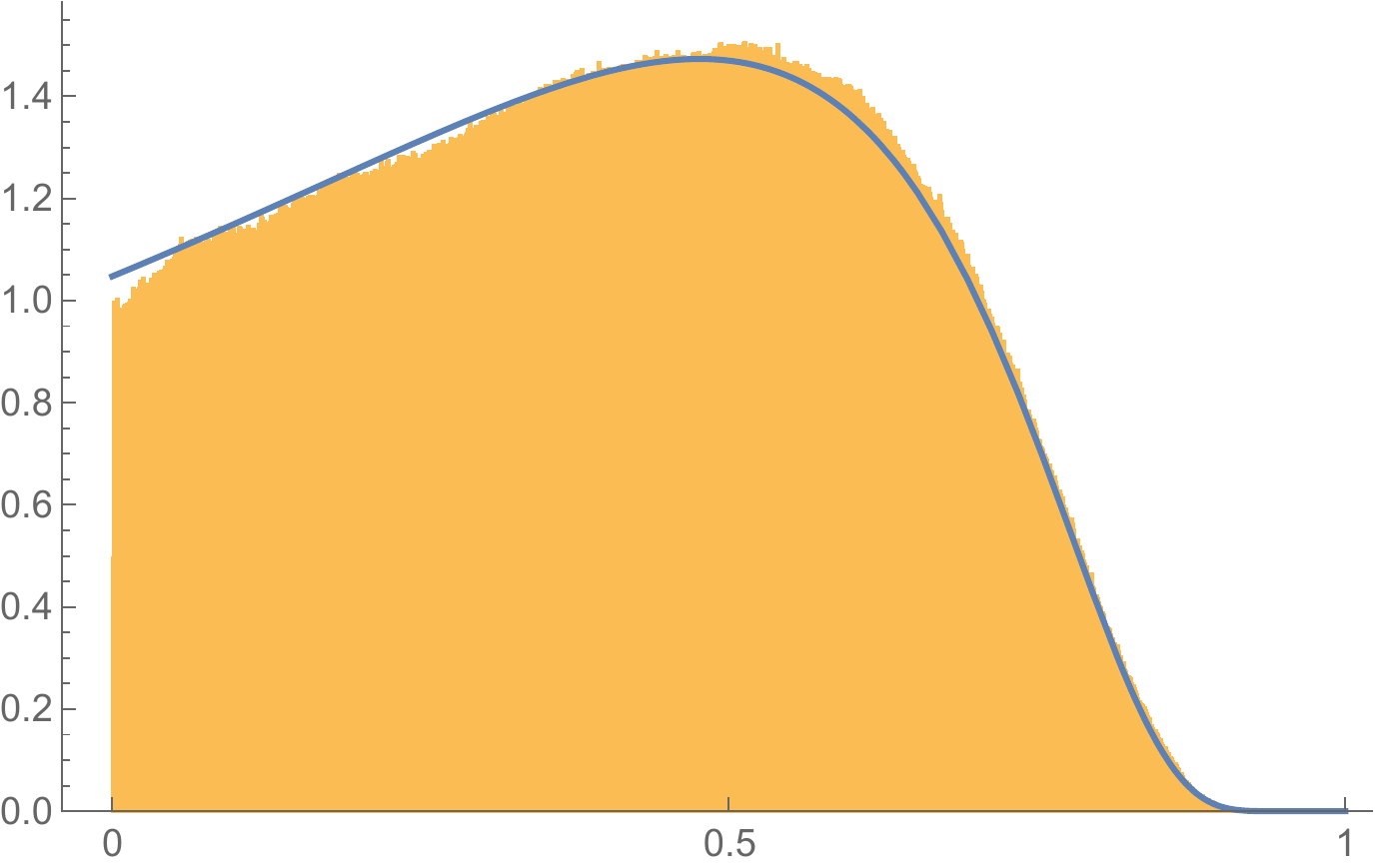}
\includegraphics[height=3.4cm]{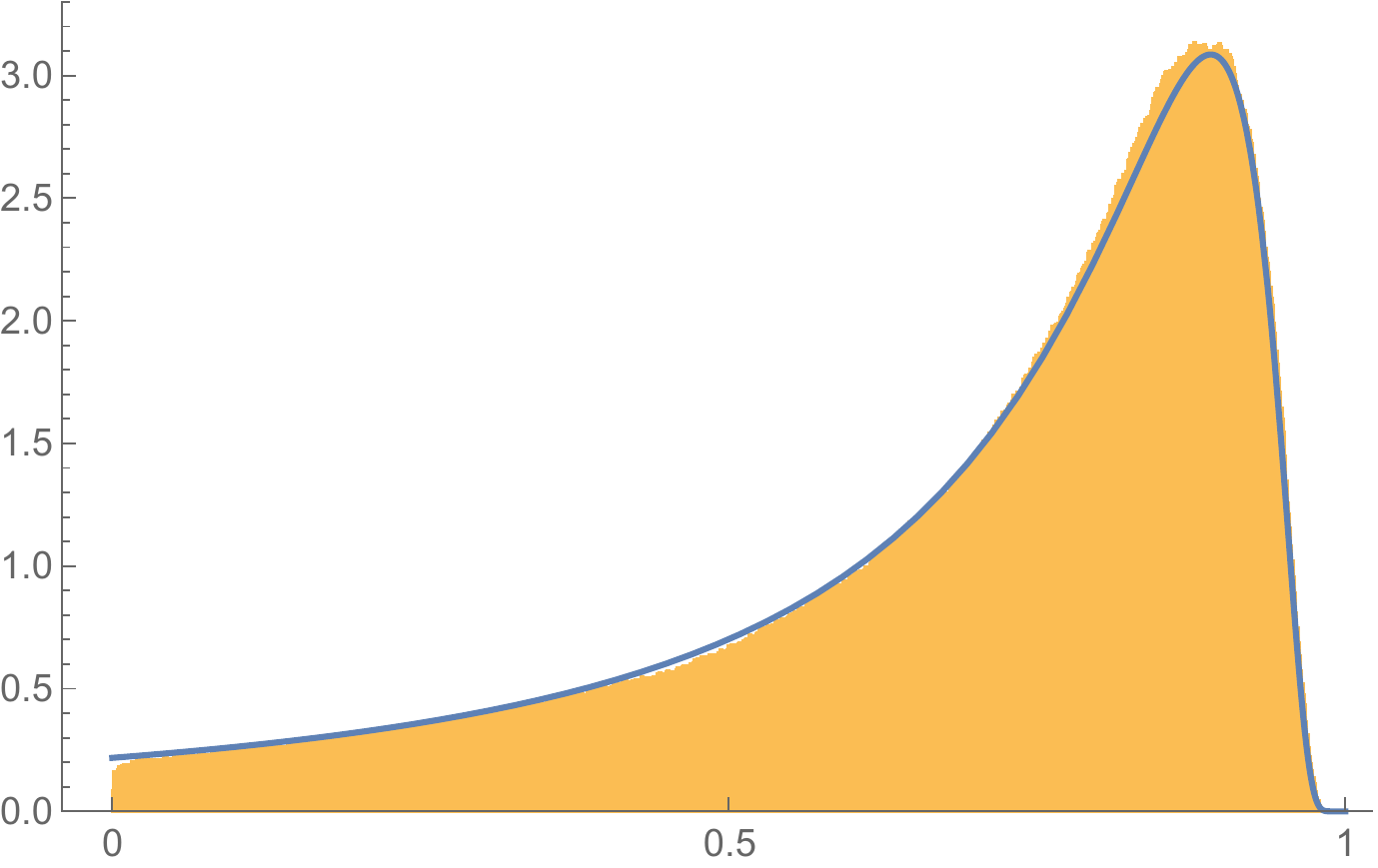}
\caption{\sl Approximate radial density $\varrho_{\lambda}$ {\rm (}blue{\rm )} given by~\eqref{Furstenberg-radial-density} and numerical histogram of values of $|z_n|^2$ obtained after $2\cdot 10^7$ iterations {\rm (}yellow{\rm )}.
The values are $(\epsilon,\delta)=(0.05,7.5\cdot 10^{-4})$ on the left, $(\epsilon,\delta)=(0.05,1.2\cdot 10^{-4})$ in the middle and $(\epsilon,\delta)=(0.05,2.5\cdot 10^{-5})$ on the right.
}
\label{fig-Spec4}
\end{figure}

\begin{rem}
{\rm 
Using the bijection  $s\in[0,1)\mapsto x=\frac{s}{1-s}\in[0,\infty)$ the distribution \eqref{Furstenberg-radial-density} becomes the exponential distribution $\lambda e^{-\lambda x}$ on $[0,\infty)$. Inserting into~\eqref{Furstenberg-result-density} a smooth approximation $h$ of $\chi_{[0,s]}$ yields an approximation (no claim is made on the error bounds which depend on $h$) to the cumulative radial distribution:
$$
\mu^{\epsilon,\delta}\left(\left\{z\in\overline{\mathbb{D}}\,:\, |z|^2\leq s\right\}\right)
\;\approx\;
1\,-\,\exp\left[\frac{-\lambda\,s}{1-s}\right]\,.
%\;+\;
%\mathcal{O}(\epsilon,\epsilon^{-1}\delta,\epsilon^{-2}\delta^2)\,.
$$
Hence for $\lambda\to 0$  one has $\mu^{\epsilon,\delta}\to \delta_1$, while for $\lambda\to \infty$ rather $\mu^{\epsilon,\delta}\to \delta_0$, both up to (uncontrolled) error terms. Thus \eqref{Furstenberg-result-statement-2} and \eqref{Furstenberg-result-statement-1} are consistent  with \eqref{Furstenberg-result-density}. Figure~\ref{fig-Spec4} shows a histogram of the values of $|z|^2$ along an orbit  for the models stated as well as the density $\varrho_{\lambda}$, properly scaled. The agreement is excellent.

\vspace{.1cm}

Let us note that the histograms in Figures~\ref{fig-Spec1},~\ref{fig-Spec2} and~\ref{fig-Spec3} show the distribution of $r=|z|$ (comparable with the orbit plots). Of course, the approximate distribution of $r$ is given by $2r \varrho_\lambda(r^2) \textnormal{d}r$ and it vanishes linearly as $r\to 0$.
}
\hfill $\diamond$
\end{rem}

\begin{rem}
{\rm 
If $f(re^{\imath \theta})=\sum_{j=-J}^Jr^je^{\imath j\theta} f_j(r^2)$ is a trigonometric polynomial in the angle with functions $f_0\in C^2([0,1])$ and $f_j\in\Ccont^1([0,1])$ for $j\in\{\pm 1\,\dots,\pm J\}$, then the techniques below show that 
\begin{align*}
\int_{\overline{\mathbb{D}}}\mu^{\epsilon,\delta}(\textnormal{d}z)\textnormal{ }{f}(z)
&\;=\;
\int_{\overline{\mathbb{D}}}\mu^{\epsilon,\delta}(\textnormal{d}z)\,f_0(|z|^2)\;+\;\mathcal{O}(\epsilon,\delta)\\
&\;=\;\int_0^1\textnormal{d}s\,\varrho_{\lambda}(s)\,f_0(s)\;+\;\mathcal{O}(\epsilon,\epsilon^{-1}\delta,\epsilon^{-2}\delta^2)\,,
\end{align*}
provided that $\EM(e^{2\imath j\eta_\sigma})\not= 1$ for $j=-J,\ldots,J$ (see in particular Lemma~\ref{oscillation-argument}).
}
\hfill $\diamond$
\end{rem}

%%%%%%%%%%%%%%%%%%%%%%%%%%%
\subsection{Expansion of the Lyapunov exponent}
\label{sec-LyapResult}

The second main result is an expansion of the Lyapunov exponent up to the order $\mathcal{O}(\epsilon^3,\epsilon\delta,\delta^2)$:

%%%%%%%%%%%%%%%%%%%%%%%%%%%%%%%
\begin{theo}
\label{main-result}
Assume \textnormal{(i)-(ix)} and $\mathbb{E}(e^{2\imath j \eta_{\sigma}})\neq 1$ for $j=1,2$. Then, one has
\begin{align}
\label{main-result-statement}
\gamma^{\epsilon,\delta}
\;=\;
\mathcal{C}\,\delta\;+\;\mathcal{D}\,\epsilon^2\;+\;\mathcal{O}(\epsilon^3,\epsilon\delta,\delta^2)\,.
\end{align}
\end{theo}
%%%%%%%%%%%%%%%%%%%%%%%%%%%%%%%

The perturbative formula \eqref{main-result-statement} generalizes the result of \cite{SSS} which considered the case $\delta=0$. 
Proposition~\ref{Schrader-result} characterizes the positivity of $\mathcal{D}$ that is crucial for many of the results above.

%%%%%%%%%%%%%%%%%%%%%%%%%%%%%%%
\begin{proposi}[\cite{SSS}]
\label{Schrader-result}
One has $\mathcal{D}\geq 0$ and $\mathcal{D}=0$ if and only if one of the following two mutually excluding cases occurs:
\begin{itemize}
\item[{\rm (i)}] Both $e^{2\imath\eta_{\sigma}}$ and $\beta_{\sigma}$ are $\mathbb{P}$-a.s. constant.
\item[{\rm (ii)}] $\mathbb{E}(e^{2\imath\eta_{\sigma}})=0$ and $\beta_{\sigma}$ is a constant multiple of $1-e^{2\imath\eta_{\sigma}}$.
\end{itemize}
\end{proposi}
%%%%%%%%%%%%%%%%%%%%%%%%%%%%%%%

\begin{rem}
{\rm 
By pushing the techniques of this paper, it is possible to compute also higher order terms in the expansion in $\epsilon$ and $\delta$. This would require to carry out even more cumbersome Taylor expansions in Section~\ref{sec-AlgPrep}, and we refrained from doing so. A more challenging, but presumably feasible extension is a perturbative formula for the variance in the central limit theorem for the Lyapunov exponent. For the case of real $2\times 2$ matrices this was achieved in \cite{SSS} by techniques that would have to be adapted to complex matrices.
}
\hfill $\diamond$
\end{rem}

%%%%%%%%%%%%%%%%%%%%%%%%%%%%%%%%%
\subsection{Behavior without monotonicity assumption}
\label{sec-Drop(v)}

\begin{figure}[H]
\centering
\includegraphics[height=4.4cm]{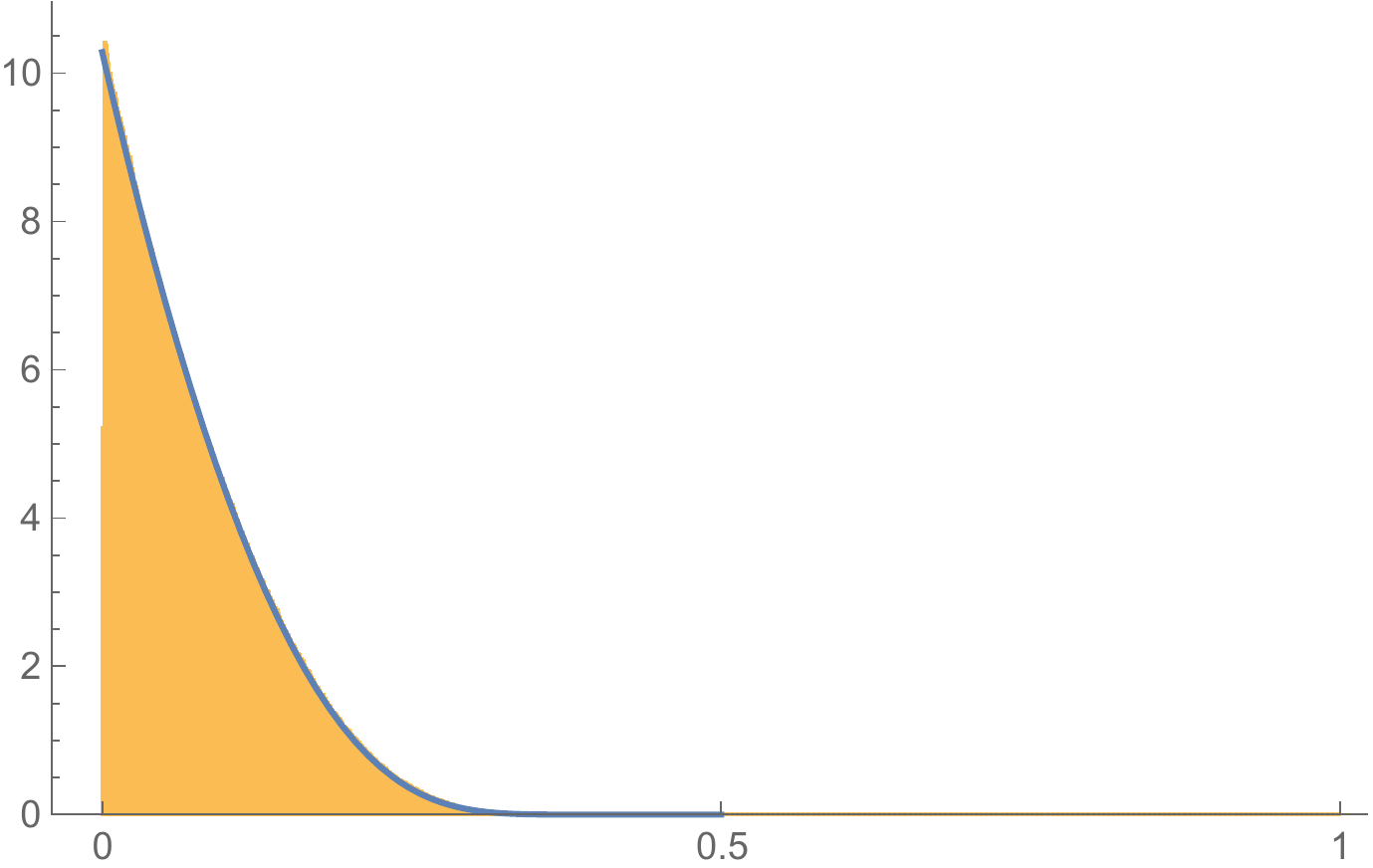}
\hspace{10mm}
\includegraphics[height=4.4cm]{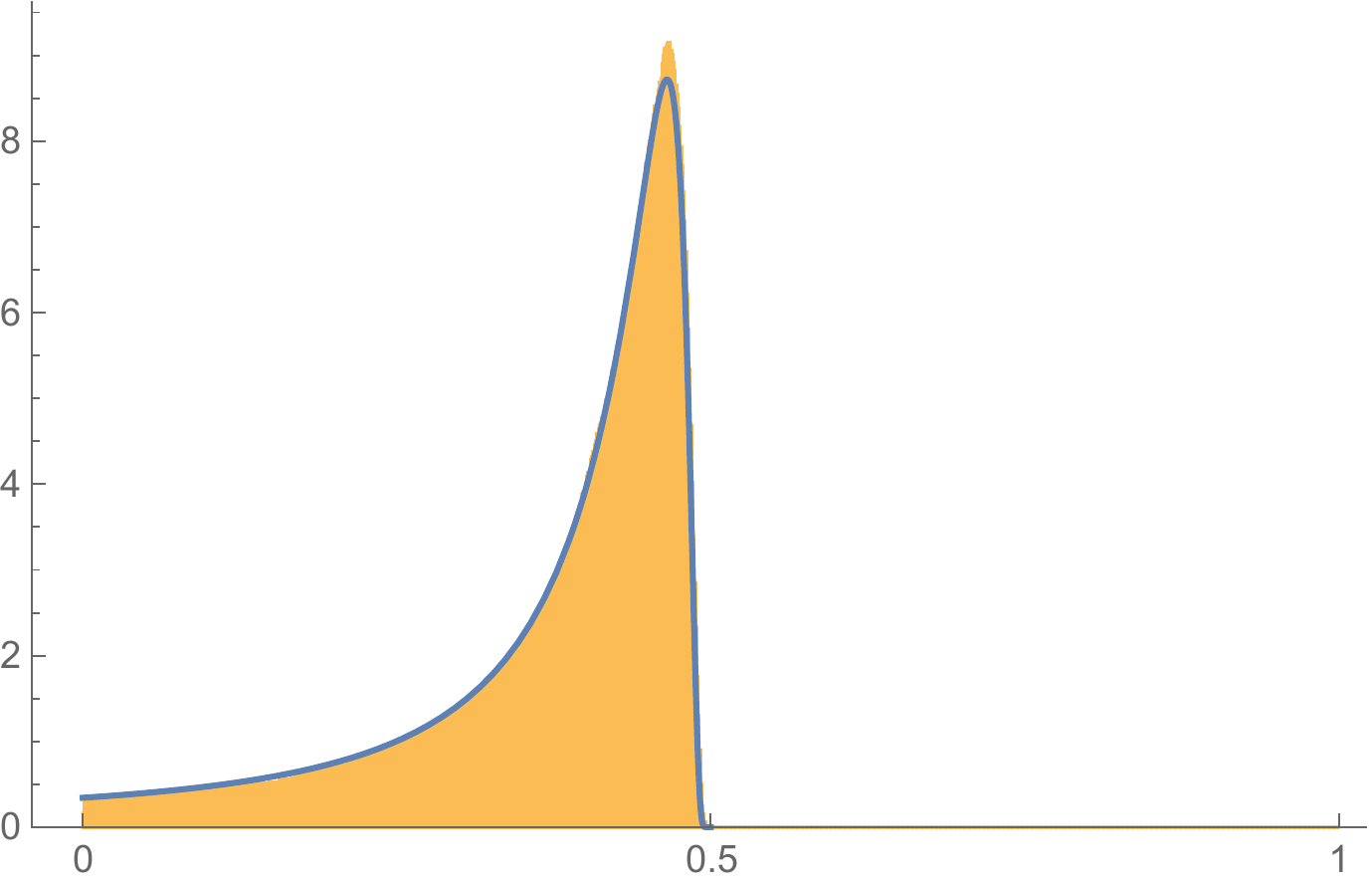}
\caption{\sl 
Numerical histograms of the distribution of $\frac{2}{\pi}\,\arctan(|z_n|^2)\in[0,1]$ after $2\cdot 10^7$ iterations (yellow) and suitably rescaled approximate radial density (blue) with $\mathcal{C}>0$ and $\mathcal{D}>0$ in violation of~{\rm (v)}. The values are $(\epsilon,\delta)=(0.05,7.5\cdot 10^{-4})$ on the left and $(\epsilon,\delta)=(0.05,2.5\cdot 10^{-5})$ on the right. The model is described in detail in {\rm Section~\ref{sec-Anderson}}.
}
\label{fig-drop-of-v}
\end{figure}

As shown in estimate~\eqref{invariance-of-disc}, the monotonicity assumption (v) guarantees the invariance of the unit disc under the M{\"o}bius action of $T^{\epsilon,\delta}_{\sigma}$. If (v) is dropped, the generated random dynamical system generically explores the whole Riemann sphere. While adapting the methods of this paper may allow to deal with this situation, this goes beyond the scope of the present study. Let us merely provide some numerical and intuitive insight on what to expect. For that purpose, we study the same model as in Figure~\ref{fig-Spec1} to \ref{fig-Spec4} but replace $\delta$ with $d_n\delta$, where $(d_n)_{n\geq 1}$ is a sequence of i.i.d. random variables whose elements $d_n$ take both positive and negative values so that assumption (v) is broken (see Section~\ref{sec-Anderson} for a more detailed description of the model). If the distribution of the $d_n$ is such that nevertheless $\mathcal{C}>0$, the numerics in Figure~\ref{fig-drop-of-v} provide evidence that \eqref{Furstenberg-result-density} withstands at least in an approximate manner, namely there is very little weight outside of the unit disc. This is intuitively reasonable as $\mathcal{C}>0$ assures that in average there is a drift to the inside of the unit disc.

%%%%%%%%%%%%%%%%%%%%%%%%%%%%%%%
%%%%%%%%%%%%%%%%%%%%%%%%%%%%%%%

\begin{figure}[H]
\centering
\includegraphics[height=4.4cm]{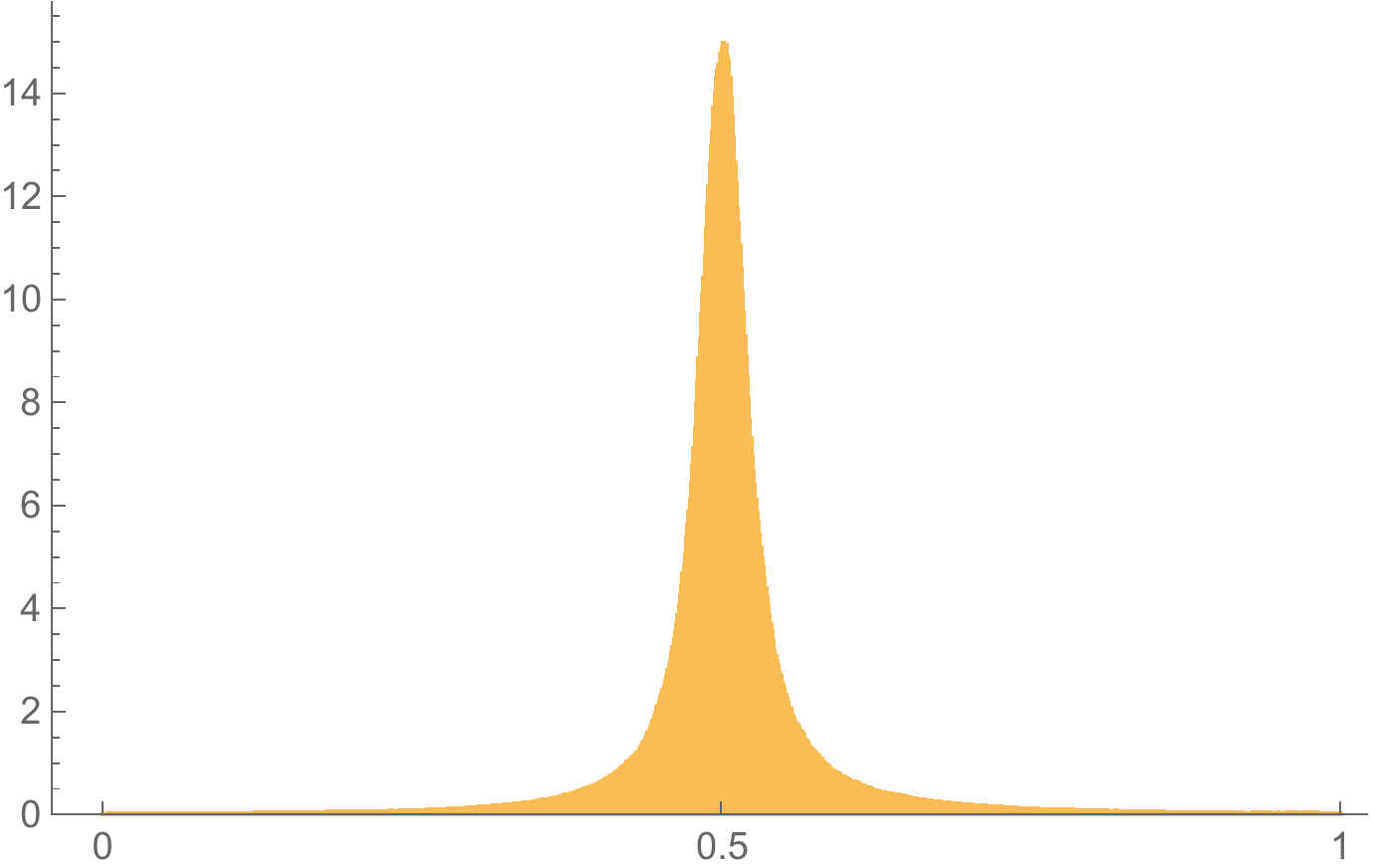}
\hspace{10mm}
\includegraphics[height=4.4cm]{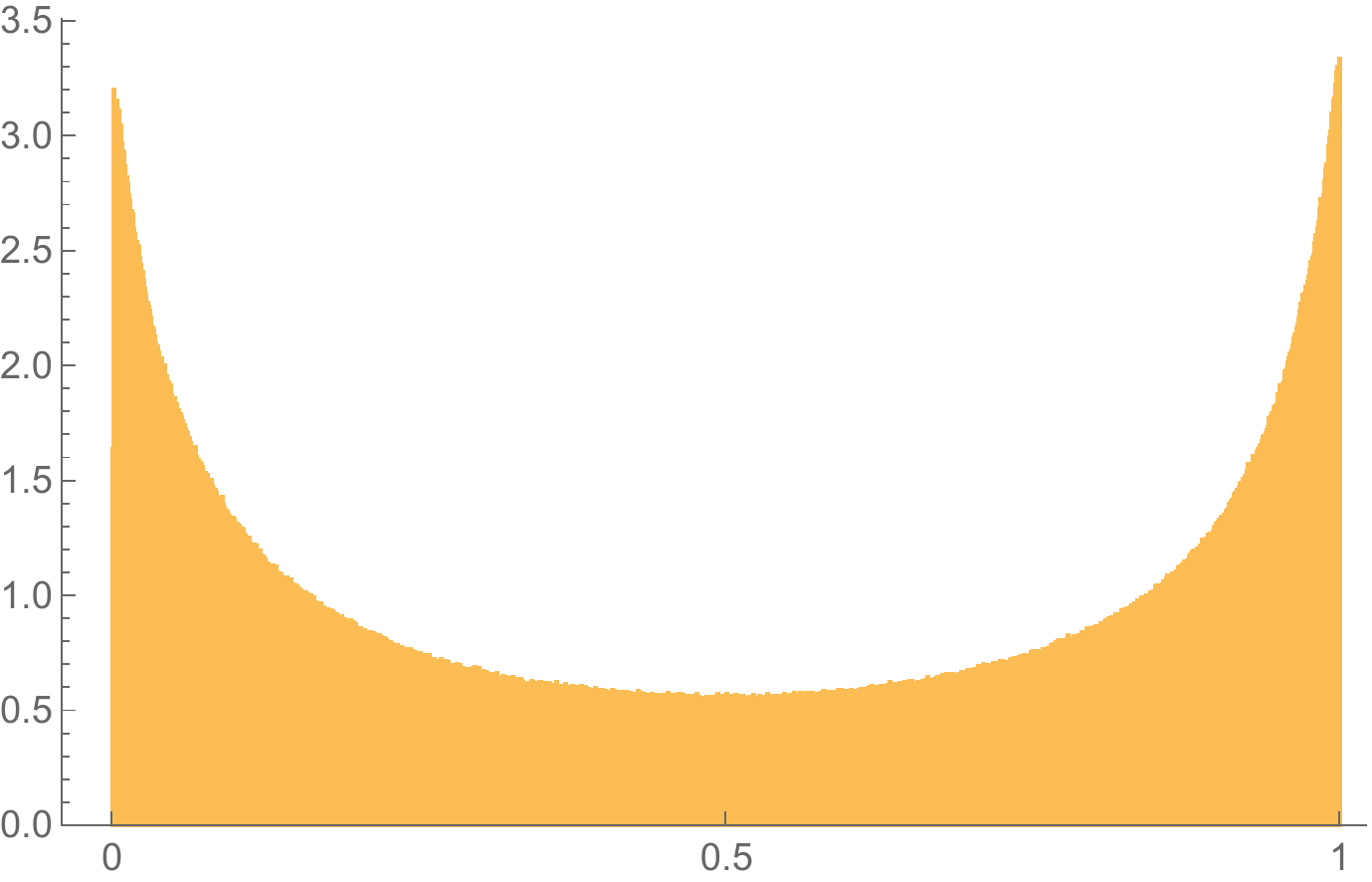}
\caption{\sl 
Numerical histograms of the distribution of $\frac{2}{\pi}\,\arctan(|z_n|^2)\in [0,1]$ after $2\cdot 10^7$ iterations for the same model as in {\rm Figure~\ref{fig-drop-of-v}}, but with $\mathcal{C}=0$ and $\mathcal{D}>0$. The values are $(\epsilon,\delta)=(0.05,5\cdot 10^{-4})$ on the left and $(\epsilon,\delta)=(2\cdot 10^{-4},0.05)$ on the right, further details are again found in {\rm Section~\ref{sec-Anderson}}.
}
\label{fig-drop-of-v-and-C=0}
\end{figure}

A fundamentally different behavior is observable at $\mathcal{C}=0$. Now there is no longer a drift to the inside of the unit disc, rather the $\delta$ term leads to fluctuations in the radial variable. If, however, $\delta\ll\epsilon^2$ one may expect that the positivity of the Lyapunov exponent again forces the random dynamics to stay close to the unit circle, similar as in Figure~\ref{fig-Spec3}. This is confirmed in the first plot in Figure~\ref{fig-drop-of-v-and-C=0}. The second plot shows that for $\delta\gg \epsilon$ the weight is rather pushed to radii $0$ and $\infty$, which results from a diffusive force between these regions. Providing an analytical expression for both of these distributions is an open problem.

%%%%%%%%%%%%%%%%%%%%%%%%%%%%%%%
%%%%%%%%%%%%%%%%%%%%%%%%%%%%%%%

%%%%%%%%%%%%%%%%%%%%%%%%%%%%%%%%%%%%%%%%%%
\section{Analysis of the Furstenberg measure}

From now on the framework described in Section~\ref{sec-2para} is supposed to hold. Therefore the random dynamical system~\eqref{eq-RDS} reads more explicitly
\begin{align}\label{explicit-definition-of-the-random-dynamical-system}
z_n\;=\;T^{\epsilon,\delta}_{\sigma_n}\cdot z_{n-1}\,,
\end{align}
where $(\sigma_n)_{n\in\mathbb{N}}$ is a random sequence of independent and identically distributed copies of $\sigma$. The sequence $(\sigma_n)_{n\in\mathbb{N}}$ is hence distributed according to $\mathbf{P}=\mathbb{P}^{\otimes \mathbb{N}}$. The average w.r.t. $\mathbf{P}$ is denoted by $\mathbf{E}$. Moreover, any random variable of the form $X_{\sigma_n}$ will simply be denoted by $X_n$. By definition, the Furstenberg measure satisfies
\begin{align}\label{Furstenberg-definition}
\int_{\overline{\mathbb{D}}}\mu^{\epsilon,\delta}(\textnormal{d}z)\;f(z)
\;=\;
\int_{\overline{\mathbb{D}}}\mu^{\epsilon,\delta}(\textnormal{d}z)\;
\;\EM\;f(T^{\epsilon,\delta}_{\sigma}\cdot z)
\;
\end{align}
for all continuous functions $f$. Iteration and averaging then shows that all $N\in\mathbb{N}$ obey
\begin{align}\label{Furstenberg-property}
\int_{\overline{\mathbb{D}}}\mu^{\epsilon,\delta}(\textnormal{d}z)\;f(z)
\;=\;
\int_{\overline{\mathbb{D}}}\mu^{\epsilon,\delta}(\textnormal{d}z)\;
\EE\;\frac{1}{N}\sum\limits_{n=0}^{N-1}
\;f(z_n)
\;,
\end{align}
where $z=z_0$ is the initial condition in the definition~\eqref{explicit-definition-of-the-random-dynamical-system} of the random dynamics. Theorem~\ref{Furstenberg-result} will be proved by analyzing the Birkhoff sum
$$
\EE\;\frac{1}{N}\sum\limits_{n=0}^{N-1}\;f(z_n)\,.
$$
%

%%%%%%%%%%%%%%%%%%%%%%%%%%%%%%%
\subsection{Algebraic preparations}
\label{sec-AlgPrep}

Let us introduce the real-valued random variables $(p_{j,\sigma})_{j=1}^3$, $(p^{\prime}_{j,\sigma})_{j=1}^3$ and $(q_{j,\sigma})_{j=1}^3$ by
\begin{align*}
{P}_{\sigma}\;=\;\sum\limits_{j=1}^3p_{j,\sigma}B_j\,,\qquad {P}^{\prime}_{\sigma}\;=\;\sum\limits_{j=1}^3p^{\prime}_{j,\sigma}B_j\,,\qquad {Q}_{\sigma}\;=\;\sum\limits_{j=1}^3q_{j,\sigma}B_j\,.
\end{align*}
The Baker-Campbell-Hausdorff formula implies the identity
\begin{align}
\label{Baker}
T^{\epsilon,\delta}_{\sigma}\;=\;{R}_{\eta_{\sigma}}e^{\epsilon (p_{3,\sigma}+\epsilon \tilde{p}_{3,\sigma})B_3}e^{\epsilon (p_{2,\sigma}+\epsilon \tilde{p}_{2,\sigma})B_2}e^{\epsilon (p_{1,\sigma}+\epsilon \tilde{p}_{1,\sigma})B_1}e^{\imath\delta q_{3,\sigma}B_3}e^{\imath\delta q_{2,\sigma}B_2}e^{\imath\delta q_{1,\sigma}B_1}\;+\;\mathcal{O}(\epsilon^3,\epsilon\delta,\delta^2)\,,
\end{align}
where $(\tilde{p}_{j,\sigma})_{j=1}^3$ are real-valued random variables containing the coefficients of $P^{\prime}_{\sigma}$ and 
the commutators of the terms of the order $\mathcal{O}(\epsilon)$. Due to the assumptions (vi) and (vii), the variables $({p}_{j,\sigma})_{j=1}^3$, $(\tilde{p}_{j,\sigma})_{j=1}^3$, $({q}_{j,\sigma})_{j=1}^3$ and the terms of order $\mathcal{O}(\epsilon^3,\delta^2,\epsilon\delta)$ in~\eqref{Baker} have compact support. One is thus led to compute the exponentials of $tB_1$, $tB_2$, $tB_3$, $\imath tB_1$, $\imath tB_2$, $\imath tB_3$ for all $t\in\mathbb{R}$:
\begin{align}\label{exponentials}
\begin{split}
 e^{tB_1}&\;=\;\begin{pmatrix}\cosh(t)&\sinh(t)\\\sinh(t)&\cosh(t)\end{pmatrix}\,,
\qquad e^{tB_2}\;=\;\begin{pmatrix}\cosh(t)&\imath \sinh(t)\\-\imath\sinh(t)&\cosh(t)\end{pmatrix}\,,
 \quad e^{tB_3}\;=\;\begin{pmatrix}e^{\imath t}&0\\0&e^{-\imath t}\end{pmatrix}\,,\\
 e^{\imath tB_1}&\;=\;\begin{pmatrix}\cos(t)&\imath \sin(t)\\\imath\sin(t)&\cos(t)\end{pmatrix}\,,
 \qquad\hspace{0.5mm} e^{\imath tB_2}\;=\;\begin{pmatrix}\cos(t)&-\sin(t)\\\sin(t)&\cos(t)\end{pmatrix}\,,
 \qquad \hspace{2.4mm} e^{\imath tB_3}\;=\;\begin{pmatrix}e^{-t}&0\\0&e^{t}\end{pmatrix}\,.
\end{split}
\end{align}

%%%%%%%%%%%%%%%%%%%%%%%%%%%%%%%
\begin{rem}\label{non-negativity-of-q_3}
{\rm 
The assumption \textnormal{(v)} guarantees $q_{3,\sigma}\geq 0$ for all $\sigma\in\Sigma$. Thus, $\mathcal{C}=\mathbb{E}(q_{3,\sigma})\geq 0$ .~\hfill$\diamond$
}
\end{rem}
%%%%%%%%%%%%%%%%%%%%%%%%%%%%%%%

Lemma~\ref{action-on-z-explicit} summarizes explicit expansions of the action $T^{\epsilon,\delta}_{\sigma}\cdot$ up to the order $\mathcal{O}(\epsilon^3,\epsilon\delta,\delta^2)$. For this purpose, it will be helpful to introduce further random variables by
\begin{align*}
\beta_{\sigma}\;=\;p_{1,\sigma}-\imath p_{2,\sigma}\,,
\qquad
\beta^{\prime}_{\sigma}\;=\;p_{1,\sigma}^{\prime}-\imath p_{2,\sigma}^{\prime} \,,
\qquad
\tilde{\beta}_{\sigma}
\;=\;
\tilde{p}_{1,\sigma}-\imath\tilde{p}_{2,\sigma}
\,,
\qquad
\xi_{\sigma}\;=\;q_{1,\sigma}-\imath q_{2,\sigma}\,.
\end{align*}
Note that $\beta_{\sigma}$ is a rewriting of the definition~\eqref{definition-of-beta}.

%%%%%%%%%%%%%%%%%%%%%%%%%%%%%%%
\begin{lemma}\label{action-on-z-explicit}
All $z\in\overline{\mathbb{D}}$ satisfy
\begin{align}\label{action-on-z-explicit-statement-1}
T^{\epsilon,\delta}_{\sigma}\cdot z
\;=\; 
& 
\,
e^{2\imath\eta_{\sigma}}\bigg[z+\epsilon\big(\overline{\beta}_{\sigma}+2\imath p_{3,\sigma}z-\beta_{\sigma} z^2\big)+\delta\big(\imath\overline{\xi}_{\sigma}-2q_{3,\sigma}z-\imath \xi_{\sigma} z^2\big)
\nonumber
\\
&\;+\;\epsilon^2
\Big(\overline{\tilde{\beta}}_{\sigma}-\tilde{\beta}_{\sigma} z^2+2\imath p_{3,\sigma}\big(\overline{\beta}_{\sigma}-{\beta}_{\sigma}z^2\big)
-
\big(|\beta_{\sigma}|^2+\imath\,\Im\mathfrak{m}(\beta_{\sigma}^2)-\imath \tilde{p}_{3,\sigma}+2p_{3,\sigma}^2\big)z+\beta_{\sigma}^2z^3\Big)
\bigg]
\nonumber
\\
&
\;+\;\mathcal{O}(\epsilon^3,\epsilon\delta,\delta^2)
\end{align}
and
\begin{align}
|T^{\epsilon,\delta}_{\sigma}\cdot z|^2
\;=\;
&
|z|^2\;+\;2\epsilon\,\Re\mathfrak{e}(\beta_{\sigma}z)(1-|z|^2)\;+\;\epsilon^2\big(|\beta_{\sigma}|^2(1-|z|^2)
\;+\;2\,\Re\mathfrak{e}\big(\tilde{\beta}_{\sigma}z-\beta^2_{\sigma}z^2\big)\big)(1-|z|^2)
\nonumber
\\
&
\;+\;2\delta\big[\Im \mathfrak{m}(\xi_{\sigma} z)[1+|z|^2]-2 q_{3,\sigma} |z|^2\big]
\;+\;
\mathcal{O}(\epsilon^3,\epsilon\delta,\delta^2)\,.
\label{action-on-z-explicit-statement-2}
\end{align}
Moroever, all $g\in\Ccont^2([0,1])$ and all $z\in\overline{\mathbb{D}}$ satisfy
\begin{align}\label{action-on-z-explicit-statement-2-b}
g\left(|T^{\epsilon,\delta}_{\sigma}\cdot z|^2\right)
\;=\; 
&\, g(|z|^2)\;+\;g^{\prime}(|z|^2)\bigg[2\,\epsilon\,\Re\mathfrak{e}(\beta_{\sigma}z)(1-|z|^2)\,+\,2\,\delta\,\left[\Im \mathfrak{m}(\xi_{\sigma} z)[1+|z|^2]\,-\,2 q_{3,\sigma} |z|^2\right]
\nonumber
\\
&
\;+\;\epsilon^2\,\big[|\beta_{\sigma}|^2(1-|z|^2)+2\,\Re\mathfrak{e}\big(\tilde{\beta}_{\sigma}z-\beta^2_{\sigma}z^2\big)\big](1-|z|^2)\bigg]
\nonumber
\\
&\;+\;{\epsilon^2}\,g^{\prime\prime}(|z|^2)\,\left[\Re\mathfrak{e}(\beta_{\sigma}^2z^2)\;+\;|\beta_{\sigma}|^2|z|^2\right](1-|z|^2)^2\;+\;\mathcal{O}(\epsilon^3,\epsilon\delta,\delta^2)\,.
\end{align}
\end{lemma}
%%%%%%%%%%%%%%%%%%%%%%%%%%%%%%%

\noindent \textbf{Proof.} For $z\in\overline{\mathbb{D}}$, let us begin by computing the identities
\begin{align}
e^{\epsilon(p_{1,\sigma}+\epsilon \tilde{p}_{1,\sigma})B_1}\cdot z&\;=\;z\;+\;\epsilon[1-z^2]p_{1,\sigma}\;+\;\epsilon^2[1-z^2][\tilde{p}_{1,\sigma}-zp_{1,\sigma}^2]\;+\;\mathcal{O}(\epsilon^3)\,,\label{action-on-z-explicit-1}\\
e^{\epsilon(p_{2,\sigma}+\epsilon \tilde{p}_{2,\sigma})B_2}\cdot z&\;=\;z\;+\;\imath \epsilon[1+z^2]p_{2,\sigma}\;+\;\epsilon^2[1+z^2][\imath \tilde{p}_{2,\sigma}-zp_{2,\sigma}^2]\;+\;\mathcal{O}(\epsilon^3)\,,\label{action-on-z-explicit-2}\\
e^{\epsilon(p_{3,\sigma}+\epsilon \tilde{p}_{3,\sigma})B_1}\cdot z&\;=\;z\;+\;2\imath \epsilon p_{3,\sigma}z\;+\;2\epsilon^2[\imath \tilde{p}_{3,\sigma}-p_{3,\sigma}^2]z\;+\;\mathcal{O}(\epsilon^3)\,,\label{action-on-z-explicit-3}\\
e^{\imath\delta q_{3,\sigma}B_3}e^{\imath\delta q_{2,\sigma}B_2}e^{\imath\delta q_{1,\sigma}B_1}\cdot z&\;=\;z\;+\;\delta\left[\imath\overline{\xi}_{\sigma}-2q_{3,\sigma}z-\imath \xi_{\sigma} z^2\right]\;+\;\mathcal{O}(\delta^2)\,.\label{action-on-z-explicit-4}
\end{align}
Next, by combining~\eqref{Baker},~\eqref{action-on-z-explicit-1},~\eqref{action-on-z-explicit-2} and~\eqref{action-on-z-explicit-4}, one obtains for all $z\in\overline{\mathbb{D}}$ the equation
\begin{align}\label{action-on-z-explicit-5}
\begin{split}
&e^{-\epsilon(p_{3,\sigma}+\epsilon \tilde{p}_{3,\sigma})B_1}R_{\eta_{\sigma}}^{-1}T^{\epsilon,\delta}_{\sigma}\cdot z
\;=\;e^{\epsilon(p_{2,\sigma}+\epsilon \tilde{p}_{2,\sigma})B_2}e^{\epsilon(p_{1,\sigma}+\epsilon \tilde{p}_{1,\sigma})B_1}e^{\imath\delta q_{3,\sigma}B_3}e^{\imath\delta q_{2,\sigma}B_2}e^{\imath\delta q_{1,\sigma}B_1}\cdot z\\
&\qquad \;=\;z\;+\;\epsilon[1-z^2]p_{1,\sigma}\;+\;\epsilon^2[1-z^2][\tilde{p}_{1,\sigma}-zp_{1,\sigma}^2]\;+\;\imath \epsilon\big[1+z^2+2\epsilon p_{1,\sigma}z[1-z^2]\big]p_{2,\sigma}\\
&\qquad\hspace{6mm}\;+\;\epsilon^2[1+z^2][\imath\, \tilde{p}_{2,\sigma}-zp_{2,\sigma}^2]\;+\;\delta\left[\imath\,\overline{\xi}_{\sigma}-2q_{3,\sigma}z-\imath \,\xi_{\sigma} z^2\right]\;+\;\mathcal{O}(\epsilon^3,\epsilon\delta,\delta^2)\\
&\qquad \;=\;z\;+\;\epsilon\Big[\overline{\beta}_{\sigma}-\beta_{\sigma} z^2\Big]\;+\;\epsilon^2\Big[\overline{\tilde{\beta}}_{\sigma}-[|\beta_{\sigma}|^2\;+\;\imath\,\Im\mathfrak{m}(\beta_{\sigma}^2)]z-\tilde{\beta}_{\sigma}z^2+\beta_{\sigma}^2z^3\Big]\\
&\qquad\hspace{6mm}\;+\;\delta\left[\imath\,\overline{\xi}_{\sigma}-2q_{3,\sigma}z-\imath\, \xi_{\sigma} z^2\right]\;+\;\mathcal{O}(\epsilon^3,\epsilon\delta,\delta^2)\,.
\end{split}
\end{align}
As the action $e^{-\epsilon(p_{3,\sigma}+\epsilon \tilde{p}_{3,\sigma})B_1}R_{\eta_{\sigma}}^{-1}\cdot $ preserves the modulus,~\eqref{action-on-z-explicit-5} implies that all $z\in\overline{\mathbb{D}}$ satisfy
\begin{align*}
|T^{\epsilon,\delta}_{\sigma}\cdot z|^2&\;=\;|e^{-\epsilon(p_{3,\sigma}\;+\;\epsilon \tilde{p}_{3,\sigma})B_1}R_{\eta_{\sigma}}^{-1}T^{\epsilon,\delta}_{\sigma}\cdot z|^2\\
&\;=\;|z|^2\;+\;2\epsilon\,\Re\mathfrak{e}(\beta_{\sigma}z)(1-|z|^2)\;+\;\epsilon^2\big[|\beta_{\sigma}|^2(1-|z|^2)\;+\;2\,\Re\mathfrak{e}\big(\tilde{\beta}_{\sigma}z-\beta^2_{\sigma}z^2\big)\big](1-|z|^2)\\
&\;\;\;\;\;\;\;+\;2\delta\left[\Im \mathfrak{m}(\xi_{\sigma} z)[1+|z|^2]-2 q_{3,\sigma} |z|^2\right]\;+\;\mathcal{O}(\epsilon^3,\epsilon\delta,\delta^2)\,
\end{align*}
which proves~\eqref{action-on-z-explicit-statement-2}. Moreover, combining~\eqref{action-on-z-explicit-3} with~\eqref{action-on-z-explicit-5} yields
\begin{align*}
R_{\eta_{\sigma}}^{-1}T^{\epsilon,\delta}_{\sigma}\cdot z&\;=\;z\;+\;\epsilon\Big[\overline{\beta}_{\sigma}-\beta_{\sigma} z^2\Big]\;+\;\epsilon^2\Big[\overline{\tilde{\beta}}_{\sigma}-[|\beta_{\sigma}|^2\;+\;\imath\,\Im\mathfrak{m}(\beta_{\sigma}^2)]z-\tilde{\beta}_{\sigma}z^2\;+\;\beta_{\sigma}^2z^3\Big]\\
&\hspace{7.5mm}\;+\;\delta\left[\imath\overline{\xi}_{\sigma}-2q_{3,\sigma}z-\imath \xi_{\sigma} z^2\right]\;+\;2\imath \epsilon p_{3,\sigma}\left(z\;+\;\epsilon\big[\textnormal{ }\overline{\beta}_{\sigma}-\beta_{\sigma} z^2\big]\right)\;+\;2\epsilon^2[\imath \tilde{p}_{3,\sigma}-p_{3,\sigma}^2]z\\
&\hspace{7.5mm}\;+\;\mathcal{O}(\epsilon^3,\epsilon\delta,\delta^2)\\
&\;=\;z\;+\;\epsilon\Big[\overline{\beta}_{\sigma}\;+\;2\imath p_{3,\sigma}z-\beta_{\sigma} z^2\Big]\;+\;\delta\Big[\imath\,\overline{\xi}_{\sigma}-2q_{3,\sigma}z-\imath\, \xi_{\sigma} z^2\Big]\\
&\hspace{7.5mm}\;+\;\epsilon^2\Big[\overline{\tilde{\beta}}_{\sigma}-\tilde{\beta}_{\sigma} z^2\;+\;2\imath \, p_{3,\sigma}\big(\overline{\beta}_{\sigma}-{\beta}_{\sigma}z^2\big)\Big]
\\
& \hspace{7.5mm}\;-\;\epsilon^2\Big[(|\beta_{\sigma}|^2\;+\;\imath\,\Im\mathfrak{m}(\beta_{\sigma}^2)-\imath\, \tilde{p}_{3,\sigma}\;+\;2p_{3,\sigma}^2)z+\beta_{\sigma}^2z^3\Big]
\;+\;\mathcal{O}(\epsilon^3,\epsilon\delta,\delta^2)
\end{align*}
for all $z\in\overline{\mathbb{D}}$, which implies~\eqref{action-on-z-explicit-statement-1} due to $R_{\eta_{\sigma}}\cdot z=e^{2\imath\eta_{\sigma}}z$.
As for~\eqref{action-on-z-explicit-statement-2-b}, let us use the identity~\eqref{action-on-z-explicit-statement-2} and Taylor's theorem in the first and second step, respectively, to~obtain
\begin{align}\label{action-on-z-explicit-6}
g\left(|T^{\epsilon,\delta}_{\sigma}\cdot z|^2\right)&\;=\;g\left(|z|^2\;+\;\epsilon\AT\;+\;\epsilon^2\BT\;+\;\delta\CT\;+\;\mathcal{O}(\epsilon^3,\epsilon\delta,\delta^2)\right)
\nonumber 
\\
&\;=\;g(|z|^2)\;+\;\left[\epsilon\AT\;+\;\epsilon^2\BT\;+\;\delta\CT\right]\,g^{\prime}(|z|^2)\;+\;\frac{\epsilon^2}{2}\AT^2\,g^{\prime\prime}(|z|^2)\;+\;\mathcal{O}(\epsilon^3,\epsilon\delta,\delta^2)\,,
\end{align}
where
\begin{align}
\AT&\;=\;2\,\Re\mathfrak{e}(\beta_{\sigma}z)(1-|z|^2)\,,\label{action-on-z-explicit-7}\\
\BT&\;=\;\big[|\beta_{\sigma}|^2(1-|z|^2)\;+\;2\,\Re\mathfrak{e}\big(\tilde{\beta}_{\sigma}z-\beta^2_{\sigma}z^2\big)\big](1-|z|^2)\,,\label{action-on-z-explicit-8}\\
\CT&\;=\;2\big[\Im \mathfrak{m}(\xi_{\sigma} z)(1+|z|^2)\;-\;2 q_{3,\sigma} |z|^2\big]\,.\label{action-on-z-explicit-9}
\end{align}
Inserting equations~\eqref{action-on-z-explicit-7},~\eqref{action-on-z-explicit-8},~\eqref{action-on-z-explicit-9} and
\begin{align*}
\AT^2\;=\;4\,\Re\mathfrak{e}(\beta_{\sigma} z)^2(1-|z|^2)^2\;=\;2\left[\Re\mathfrak{e}(\beta_{\sigma}^2z^2)\;+\;|\beta_{\sigma}|^2|z|^2\right](1-|z|^2)^2
\end{align*}
into~\eqref{action-on-z-explicit-6} yields the desired identity~\eqref{action-on-z-explicit-statement-2-b}.
\hfill $\Box$

%%%%%%%%%%%%%%%%%%%%%%%%%%%%%%%
\subsection{Oscillatory phase argument to lowest order}
\label{osci-lowest-order}

At $\epsilon=\delta=0$, the dynamics is simply a rotation around the origin by the random angle $\eta_{\sigma}$. Due to the assumption $\mathbb{E}(e^{2\imath\eta_{\sigma}})\neq 1$, there is a proper (non-trivial) average rotation. Hence Birkhoff sums like
\begin{align*}
\mathbf{E}\frac{1}{N}\sum\limits_{n=0}^{N-1}z_n\,g(|z_n|^2)
\end{align*}
for functions $g:[0,1]\to \CM$ tend to zero for large $N$ because the phases $(\theta_n)_{n\geq 1}$ of $(z_n)_{n\geq 1}=(r_ne^{\imath\theta_n})_{n\geq 1}$ lead to oscillatory summands with constant moduli $(r_n)_{n\geq 1}$. If $\epsilon$ and $\delta$ are non-zero, the same behavior still holds approximately, as stated by Lemma~\ref{oscillation-argument} for sufficiently smooth functions $g$. The basic idea of the argument leading to the following statement goes back to Pastur and Figotin~\cite{PF} and was applied \textit{e.g.} in~\cite{JSS} and~\cite{SSS}.

%%%%%%%%%%%%%%%%%%%%%%%%%%%%%%%%%%%%%%%%%%%%%%%
\begin{lemma}
\label{oscillation-argument}
Let $ j\in\mathbb{Z}\setminus\{0\}$ and $g\in\Ccont^1([0,1])$. If $\mathbb{E}(e^{2\imath j\eta_{\sigma}})\neq 1$, then one has
\begin{align}\label{oscillation-argument-statement-1}
\mathbf{E}\frac{1}{N}\sum\limits_{n=0}^{N-1}z^{ j}_n\,g(|z_n|^2)
\;=\;
\mathcal{O}(\epsilon,\delta,N^{-1})
\end{align}
and 
\begin{align}\label{oscilllation-argument-statement-2}
\int_{\overline{\mathbb{D}}}\mu^{\epsilon,\delta}(\textnormal{d}z)\textnormal{ }z^{ j}g(|z|^2)
\;=\;
\mathcal{O}(\epsilon,\delta)\,,
\end{align}
where the constants of the error bounds depend on $j$ and $g$.
\end{lemma}
%%%%%%%%%%%%%%%%%%%%%%%%%%%%%%%%%%%%%%%%%%%%%%%

\noindent\textbf{Proof.} As $T^{\epsilon,\delta}_{\sigma}\cdot z=e^{2\imath\eta_{\sigma}}z\;+\;\mathcal{O}(\epsilon,\delta)$, Taylor's theorem implies for  all $n\in\{0,\dots,N-1\}$ 
\begin{align*}
z_{n+1}^{ j}g(|z_{n+1}|^2)\;=\;e^{2\imath j\eta_{n+1}}z_n^{ j}g(|z_{n}|^2)\;+\;\mathcal{O}(\epsilon,\delta)
\end{align*}
and hence
\begin{align}\label{osci-1}
\mathbf{E}\;z_{n+1}^{ j}g(|z_{n+1}|^2)\;=\;\mathbb{E}(e^{2\imath j\eta_{\sigma}})\textnormal{ }\mathbf{E}\;z_n^{ j}g(|z_{n}|^2)\;+\;\mathcal{O}(\epsilon,\delta)\,.
\end{align}
Averaging~\eqref{osci-1} from $n=0$ to $N-1$ yields
\begin{align}\label{osci-2}
\mathbf{E}\,\frac{1}{N}\sum\limits_{n=1}^{N}z_{n}^{ j}g(|z_{n}|^2)&\;=\;\mathbb{E}(e^{2\imath j\eta_{\sigma}})\textnormal{ }\mathbf{E}\,\frac{1}{N}\sum\limits_{n=0}^{N-1}z_n^{ j}g(|z_{n}|^2)\;+\;\mathcal{O}(\epsilon,\delta)\,,
\end{align}
which is equivalent to 
\begin{align*}
\mathbf{E}\frac{1}{N}\sum\limits_{n=0}^{N-1}z_n^{ j}g(|z_{n}|^2)&\;=\;\frac{z_0^{ j}g(|z_0|^2)-\mathbf{E}z_N^{ j}g(|z_N|^2)}{\left[1-\mathbb{E}(e^{2\imath j\eta_{\sigma}})\right]N}\;+\;\mathcal{O}(\epsilon,\delta)\;=\;\mathcal{O}(\epsilon,\delta,N^{-1})\,.
\end{align*}
This proves~\eqref{oscillation-argument-statement-1}, which, in turn, implies~\eqref{oscilllation-argument-statement-2} by using~\eqref{Furstenberg-property} in the limit $N\rightarrow\infty$.
\hfill $\Box$

\vspace{.2cm}

In fact, Birkhoff sums like~\eqref{oscillation-argument-statement-1} can be computed more precisely by analyzing the terms of the order $\mathcal{O}(\epsilon,\delta)$ in~\eqref{osci-2}. Lemma~\ref{oscillation-argument-first-order} treats the case $j=1$ with $g\in\Ccont^2([0,1])$.

%%%%%%%%%%%%%%%%%%%%%%%%%%%%%%%
\begin{lemma}\label{oscillation-argument-first-order}
Suppose that $\mathbb{E}(e^{2\imath\eta_{\sigma}})\neq 1$ and $\mathbb{E}(e^{4\imath\eta_{\sigma}})\neq 1$. Then for all $g\in \Ccont^2([0,1])$ and $N\in\mathbb{N}$
\begin{align}
\mathbf{E}\frac{1}{N}\sum\limits_{n=0}^{N-1}z_n\,g(|z_n|^2)
\;=\;
& \,
\epsilon\, \frac{\mathbb{E}\left(e^{2\imath\eta_{\sigma}}\overline{\beta_{\sigma}}\right)}{1-\mathbb{E}(e^{2\imath\eta_{\sigma}})}\,\mathbf{E}\,
\frac{1}{N}\sum\limits_{n=0}^{N-1}\Big[g(|z_n|^2)+|z_n|^2(1-|z_n|^2)g^{\prime}(|z_n|^2)\Big]
\nonumber
\\
& \;+\;\mathcal{O}(\epsilon^2,\delta,N^{-1})
\;.
\label{oscillation-argument-first-order-statement-1}
\end{align}
Moreover, all $g\in \Ccont^2([0,1])$ satisfy
\begin{align}\label{oscillation-argument-first-order-statement-2}
\int_{\overline{\mathbb{D}}}\mu^{\epsilon,\delta}(\textnormal{d}z)\,z\,g(|z|^2)
\;=\;
\epsilon\, \frac{\mathbb{E}\left(e^{2\imath\eta_{\sigma}}\overline{\beta_{\sigma}}\right)}{1-\mathbb{E}(e^{2\imath\eta_{\sigma}})}\int_{\overline{\mathbb{D}}}\mu^{\epsilon,\delta}(\textnormal{d}z)\textnormal{ }\Big[g(|z|^2)+|z|^2(1-|z|^2)g^{\prime}(|z|^2)\Big]\,+\,\mathcal{O}(\epsilon^2,\delta)\,.
\end{align}
\end{lemma}
%%%%%%%%%%%%%%%%%%%%%%%%%%%%%%%

\noindent\textbf{Proof.} The identity~\eqref{action-on-z-explicit-statement-1} implies for all $n\in\{0,\dots,N-1\}$ 
\begin{align}\label{oscillation-argument-first-order-1}
z_{n+1}
\;=\;
e^{2\imath\eta_{n+1}}\bigg[z_n+\epsilon\Big[\overline{\beta}_{n+1}+2\imath p_{3,n+1}z-\beta_{n+1} z_n^2\Big]\bigg]\;+\;\mathcal{O}(\epsilon^2,\delta)\,.
\end{align}
Moreover, the identity~\eqref{action-on-z-explicit-statement-2-b} imply for all $n\in\{0,\dots,N-1\}$ 
\begin{align}\label{oscillation-argument-first-order-2}
g\left(|z_{n+1}|^2\right)&
\;=\;
g(|z_n|^2)+2\,\epsilon\,\Re\mathfrak{e}(\beta_{n+1}z_n)\,(1-|z_n|^2)\,g^{\prime}(|z_n|^2)\;+\;\mathcal{O}(\epsilon^2,\delta)\,.
\end{align}
Combining~\eqref{oscillation-argument-first-order-1} and~\eqref{oscillation-argument-first-order-2} yields for all $n\in\{0,\dots,N-1\}$ the identity
\begin{align*}
%\label{oscillation-argument-first-order-3}
z_{n+1}\,g\left(|z_{n+1}|^2\right)
\;=\; &
e^{2\imath\eta_{n+1}}\bigg[z_{n}\,g\left(|z_{n}|^2\right)+\epsilon\bigg(\Big[\overline{\beta}_{n+1}+2\imath p_{3,n+1}z_n-\beta_{n+1} z_n^2\Big]\,g\left(|z_{n}|^2\right)\\
&\quad+\Big[\beta_{n+1}z_n^2+\overline{\beta_{n+1}}|z_n|^2\Big]\,(1-|z_n|^2)\,g^{\prime}(|z_n|^2)\bigg)\bigg]\;+\;\mathcal{O}(\epsilon^2,\delta)\\
\;=\; &
e^{2\imath\eta_{n+1}}\,z_{n}\,g\left(|z_{n}|^2\right)+\epsilon\,e^{2\imath\eta_{n+1}}\overline{\beta_{n+1}}\,\Big[g(|z_n|^2)+|z_n|^2(1-|z_n|^2)\,g^{\prime}(|z_n|^2)\Big]\\
&\quad +\epsilon\,e^{2\imath\eta_{n+1}}\beta_{n+1}\,z_n^2\,\Big[|z_n|^2(1-|z_n|^2)\,g^{\prime}(|z_n|^2)-g(|z_n|^2)\Big]\\
&\quad+2\imath\,\epsilon\,e^{2\imath\eta_{n+1}}p_{3,n+1}\,z_n\,g(|z_n|^2)\;+\;\mathcal{O}(\epsilon^2,\delta)\,.
\end{align*}
Averaging this
%~\eqref{oscillation-argument-first-order-3} 
from $n=0$ to $N-1$ yields
\begin{align*}
%\label{oscillation-argument-first-order-4}
%\begin{split}
\mathbf{E}\,\frac{1}{N}\sum\limits_{n=0}^{N-1}z_{n+1}\,g\left(|z_{n+1}|^2\right)
\;=\; &
\mathbb{E}(e^{2\imath \eta_{\sigma}})\,\mathbf{E}\,\frac{1}{N}\sum\limits_{n=0}^{N-1}z_{n}\,g\left(|z_{n}|^2\right)\\
&
\;+\;\epsilon\,\mathbb{E}\big(e^{2\imath\eta_{\sigma}}\overline{\beta_{\sigma}}\big)\,\mathbf{E}\,\frac{1}{N}\sum\limits_{n=0}^{N-1}\Big[g(|z_n|^2)+|z_n|^2(1-|z_n|^2)\,g^{\prime}(|z_n|^2)\Big]\\
&\;+\;\epsilon\,\mathbb{E}\big(e^{2\imath\eta_{\sigma}}\beta_{\sigma}\big)\,\mathbf{E}\frac{1}{N}\sum\limits_{n=0}^{N-1}z_n^2\,\Big[|z_n|^2(1-|z_n|^2)\,g^{\prime}(|z_n|^2)-g(|z_n|^2)\Big]\\
&\;+\;2\imath\,\epsilon\,\mathbb{E}\big(e^{2\imath\eta_{\sigma}}p_{3,\sigma}\big)\,\mathbf{E}\,\frac{1}{N}\sum\limits_{n=0}^{N-1}z_n\,g(|z_n|^2)\;+\;\mathcal{O}(\epsilon^2,\delta)\,.
%\end{split}
\end{align*}
According to Lemma~\ref{oscillation-argument}, the third and the fourth summand are of the order $\mathcal{O}(\epsilon^2,\delta,N^{-1}\epsilon)$. Hence
\begin{align*}
\mathbf{E}\frac{1}{N}\sum\limits_{n=0}^{N-1}z_{n+1}\,g\left(|z_{n+1}|^2\right)
\;=\;
&\,
\epsilon\,\frac{\mathbb{E}\left(e^{2\imath\eta_{\sigma}}\overline{\beta_{\sigma}}\right)}{1-\mathbb{E}(e^{2\imath\eta_{\sigma}})}
\,\mathbf{E}\,\frac{1}{N}\sum\limits_{n=0}^{N-1}\Big[g(|z_n|^2)+|z_n|^2(1-|z_n|^2)\,g^{\prime}(|z_n|^2)\Big]\\
&\quad
+\,\frac{z_0\,g\left(|z_{0}|^2\right)-\mathbf{E}\,z_N\,g\left(|z_{N}|^2\right)}{\left[1-\mathbb{E}(e^{2\imath\eta_{\sigma}})\right]N}
\;+\;\mathcal{O}(\epsilon^2,\delta,N^{-1}\epsilon)\\
\;=\;
&\,
\epsilon\,\frac{\mathbb{E}\left(e^{2\imath\eta_{\sigma}}\overline{\beta_{\sigma}}\right)}{1-\mathbb{E}(e^{2\imath\eta_{\sigma}})}
\,\mathbf{E}\,\frac{1}{N}\sum\limits_{n=0}^{N-1}\Big[g(|z_n|^2)+|z_n|^2(1-|z_n|^2)\,g^{\prime}(|z_n|^2)\Big]\\
&\quad\;+\;\mathcal{O}(\epsilon^2,\delta,N^{-1})\,,
\end{align*}
which  implies~\eqref{oscillation-argument-first-order-statement-1}, which, in turn, implies~\eqref{oscillation-argument-first-order-statement-2} by using~\eqref{Furstenberg-property} in the limit $N\rightarrow\infty$.
\hfill $\Box$

\vspace{4mm}

For the computation of the Lyapunov exponent, the special case $g(s)=(1+s)^{-1}$ in \eqref{oscillation-argument-first-order-statement-2} will be relevant, and it is also one of the elements in the proof of Theorem~\ref{main-result}:

%%%%%%%%%%%%%%%%%%%%%%%%%%%%%%%%%%
\begin{coro}
The Furstenberg measure $\mu^{\epsilon,\delta}$ satisfies 
\begin{align}\label{oscillation-argument-first-order-corollary-statement}
\int_{\overline{\mathbb{D}}}\mu^{\epsilon,\delta}(\textnormal{d}z)\textnormal{ }\frac{z}{1+|z|^2}\;=\;\epsilon\, \frac{\mathbb{E}\left(e^{2\imath\eta_{\sigma}}\overline{\beta_{\sigma}}\right)}{1-\mathbb{E}(e^{2\imath\eta_{\sigma}})}\int_{\overline{\mathbb{D}}}\mu^{\epsilon,\delta}(\textnormal{d}z)\textnormal{ }\frac{1+|z|^4}{(1+|z|^2)^2}\;+\;\mathcal{O}(\epsilon^2,\delta)\,.
\end{align}
\end{coro}
%%%%%%%%%%%%%%%%%%%%%%%%%%%%%%%%%%

%%%%%%%%%%%%%%%%%%%%%%%%%%%%%%%%%%
\subsection{Oscillatory phase argument to second order}
\label{sec-osci-ph-arg-2nd-order}

The section applies the oscillation argument to functions that only depend on the modulus of their argument. This allows to complete the proof of Theorem~\ref{Furstenberg-result}.

%%%%%%%%%%%%%%%%%%%%%%%%%%%%%%%%%%
\begin{lemma}\label{overall-oscillation-argument}
Suppose that $\mathbb{E}(e^{2\imath\eta_{\sigma}})\neq 1$ and $\mathbb{E}(e^{4\imath\eta_{\sigma}})\neq 1$ hold. Then all $g\in\Ccont^3([0,1])$ satisfy 
\begin{align}\label{overall-oscillation-argument-statement-1}
%\begin{split}
2\,\mathcal{C}\,\delta\,\mathbf{E}\frac{1}{N} & \sum\limits_{n=0}^{N-1}|z_n|^2\,g^{\prime}(|z_{n}|^2)
\nonumber
\\
& =\;\mathcal{D}\,\epsilon^2\,\mathbf{E}\frac{1}{N}\sum\limits_{n=0}^{N-1}(1-|z_n|^2)^2\left(g^{\prime}(|z_n|^2)+|z_n|^2\,g^{\prime\prime}(|z_n|^2)\right)\;+\;\mathcal{O}(\epsilon^3,\epsilon\delta,\delta^2,  N^{-1})
%\end{split}
\end{align}
for all $N\in\mathbb{N}$, and
\begin{align}
\label{overall-oscillation-argument-statement-2}
\int_{\overline{\mathbb{D}}}\mu^{\epsilon,\delta}(\textnormal{d}z)
\Big[
2\,\mathcal{C}\,\delta \, |z|^2\,g^{\prime}(|z|^2)-\mathcal{D}\,\epsilon^2\,(1-|z|^2)^2\left(g^{\prime}(|z|^2)+|z|^2\,g^{\prime\prime}(|z|^2)\right)
\Big]
\;=\;
\mathcal{O}(\epsilon^3,\epsilon\delta,\delta^2)\,.
\end{align}
\end{lemma}
%%%%%%%%%%%%%%%%%%%%%%%%%%%%%%%%%%

\noindent\textbf{Proof.} Let us start out with \eqref{action-on-z-explicit-statement-2-b} for each point of the orbit $(z_n)_{n\in\mathbb{N}}$. Taking the average along the orbit leads to 
\begin{align*}
%\label{overall-oscillation-argument-1}
%\begin{split}
\mathbf{E}\frac{1}{N}\sum\limits_{n=0}^{N-1}g(|z_{n+1}|^2)\;= & \;\mathbf{E}\frac{1}{N}\sum\limits_{n=0}^{N-1}g(|z_{n}|^2)+2\,\epsilon\,\Re\mathfrak{e}\left(\mathbb{E}(\beta_{\sigma})\,\mathbf{E}\frac{1}{N}\sum\limits_{n=0}^{N-1}z_n(1-|z_n|^2)\,g^{\prime}(|z_{n}|^2)\right)\\
&\quad -\,4\,\mathcal{C}\,\delta\,\mathbf{E}\frac{1}{N}\sum\limits_{n=0}^{N-1}|z_n|^2\,g^{\prime}(|z_{n}|^2)+\epsilon^2\,\mathbb{E}(|\beta_{\sigma}|^2)\,\mathbf{E}\frac{1}{N}\sum\limits_{n=0}^{N-1}(1-|z_n|^2)^2\,g^{\prime}(|z_{n}|^2)\\
&\quad +{\epsilon^2}\,\mathbb{E}(|\beta_{\sigma}|^2)\,\mathbf{E}\frac{1}{N}\sum\limits_{n=0}^{N-1}|z_n|^2(1-|z_n|^2)^2\,g^{\prime\prime}(|z_n|^2)\\
&\quad +2\,\delta\,\Im\mathfrak{m}\left(\mathbb{E}(\xi_{\sigma})\,\mathbf{E}\frac{1}{N}\sum\limits_{n=0}^{N-1}z_n[1+|z_n|^2]\,g^{\prime}(|z_{n}|^2)\right)\\
&\quad+2\,\epsilon^2\,\Re\mathfrak{e}\left(\mathbf{E}\frac{1}{N}\sum\limits_{n=0}^{N-1}\big[\mathbb{E}(\tilde{\beta}_{\sigma})z_n-\mathbb{E}(\beta^2_{\sigma})z_n^2\big]\,(1-|z_n|^2)\,g^{\prime}(|z_{n}|^2)\right)\\
&\quad+\epsilon^2\,\Re\mathfrak{e}\left(\mathbb{E}(\beta^2_{\sigma})\,\mathbf{E}\frac{1}{N}\sum\limits_{n=0}^{N-1}z_n^2\,(1-|z_n|^2)^2\,g^{\prime\prime}(|z_{n}|^2)\right)\;+\;\mathcal{O}(\epsilon^3,\epsilon\delta,\delta^2)\,.
%\end{split}
\end{align*}
According to Lemma~\ref{oscillation-argument}, the last three lines are of the order $\mathcal{O}(\epsilon^3,\epsilon\delta,\delta^2, \epsilon^2N^{-1},\delta N^{-1})$. Thus
\begin{align*}
%\label{overall-oscillation-argument-2}
%\begin{split}
&2\,\mathcal{C}\,\delta\,\mathbf{E}\frac{1}{N}\sum\limits_{n=0}^{N-1}|z_n|^2\,g^{\prime}(|z_{n}|^2)\\
&=\;\frac{g(|z_0|^2)-\mathbf{E}\,g(|z_N|^2)}{2N}+\epsilon\,\Re\mathfrak{e}\left(\mathbb{E}(\beta_{\sigma})\,\mathbf{E}\frac{1}{N}\sum\limits_{n=0}^{N-1}z_n(1-|z_n|^2)\,g^{\prime}(|z_{n}|^2)\right)\\
&\quad\;\;+\,\epsilon^2\,\frac{1}{2}\mathbb{E}(|\beta_{\sigma}|^2)\,\mathbf{E}\frac{1}{N}\sum\limits_{n=0}^{N-1}(1-|z_n|^2)^2\,\left[g^{\prime}(|z_n|^2)+|z_n|^2\,g^{\prime\prime}(|z_n|^2)\right]\;+\;\mathcal{O}(\epsilon^3,\epsilon\delta,\delta^2, \epsilon^2N^{-1},\delta N^{-1})\\
&=\;\epsilon\,\Re\mathfrak{e}\left(\mathbb{E}(\beta_{\sigma})\,\mathbf{E}\frac{1}{N}\sum\limits_{n=0}^{N-1}z_n(1-|z_n|^2)\,g^{\prime}(|z_{n}|^2)\right)\\
&\quad\;+\,\epsilon^2\,\frac{1}{2}\mathbb{E}(|\beta_{\sigma}|^2)\,\mathbf{E}\frac{1}{N}\sum\limits_{n=0}^{N-1}(1-|z_n|^2)^2\,\left[g^{\prime}(|z_n|^2)+|z_n|^2\,g^{\prime\prime}(|z_n|^2)\right]\;+\;\mathcal{O}(\epsilon^3,\epsilon\delta,\delta^2,  N^{-1})\,.
%\end{split}
\end{align*}
In view of Lemma~\ref{oscillation-argument-first-order} and due to the identity
$$
(1-s)g'(s)\;+\;s(1-s)\partial_s\big((1-s)g'(s)\big)
\;=\;
(1-s)^2\big(g'(s)+s g''(s)\big)
$$
for $s=|z|^2$, this implies~\eqref{overall-oscillation-argument-statement-1}. In the limit $N\rightarrow\infty$, one infers~\eqref{overall-oscillation-argument-statement-2} by using~\eqref{Furstenberg-property}.
\hfill $\Box$

\vspace{.2cm}

Inserting $g(s)=\log(1+s)$ in~\eqref{overall-oscillation-argument-statement-2} yields an equation used in the proof of Theorem~\ref{main-result}:

%%%%%%%%%%%%%%%%%%%%%%%%%%%%%%%%%%
\begin{coro}\label{Lyapunov-auxiliary-equation}
The Furstenberg measure $\mu^{\epsilon,\delta}$ satisfies the identity
\begin{align}\label{Lyapunov-auxiliary-equation-statement}
2\,\mathcal{C}\,\delta\int_{\overline{\mathbb{D}}}\textnormal{d}\mu^{\epsilon,\delta}(z)\textnormal{ }\frac{|z|^2}{1+|z|^2}
\;-\;
\mathcal{D}\,\epsilon^2\int_{\overline{\mathbb{D}}}\textnormal{d}\mu^{\epsilon,\delta}(z)\textnormal{ }\left[\frac{1-|z|^2}{1+|z|^2}\right]^2
\;=\;
\mathcal{O}(\epsilon^3,\epsilon\delta,\delta^2)\,.
\end{align}
\end{coro}
%%%%%%%%%%%%%%%%%%%%%%%%%%%%%%%%%%

\vspace{.2cm}

Based on the statement~\eqref{overall-oscillation-argument-statement-2} of Lemma~\ref{overall-oscillation-argument}, it is now possible to proceed with the proof of Theorem~\ref{Furstenberg-result}. Before going into technical details, let us give some intuition though, principally based on the general strategy outlined in \cite{SS}. For that purpose, let us set $s=|z|^2$ and suppose that $\mu^{\epsilon,\delta}$ is absolutely continuous on radial functions, namely that there exists a probability density $\varrho^{\epsilon,\delta}:[0,1]\to [0,\infty]$ such that
$$
\int_{\overline{\mathbb{D}}}\mu^{\epsilon,\delta}(\textnormal{d}z)\,g(|z|^2)
\;=\;
\int^1_0 \textnormal{d}s\,\varrho^{\epsilon,\delta}(s)\,g(s)
\;.
$$
Supposing, moreover, that $\Dd>0$, one can rewrite \eqref{overall-oscillation-argument-statement-2} divided by $\Dd \epsilon^2$ as
\begin{equation}
\label{eq-ImageSmall}
\int^1_0 \textnormal{d}s\,\varrho^{\epsilon,\delta}(s)\,
\Big(\lambda\,s g'(s)\,-\,(1-s)^2\big(g'(s)+sg''(s)\big)\Big)
\;=\;
\Oo(\epsilon,\epsilon^{-1}\delta,\epsilon^{-2}\delta^2)
\;,
\end{equation}
with $\lambda$ defined as in \eqref{eq-LambdaDef}. Given this link between $\epsilon$ and $\delta$, let us set $\varrho_\lambda=\varrho^{\epsilon,\delta}$. Therefore it is of interest to define a second-order differential operator $\Ll_\lambda:C^2([0,1])\to C([0,1])$ by
$$
\Ll_\lambda\;=\;\big(\lambda \,-\,(1-s)^2\partial_s\big) s\partial_s
\;=\;
-s(1-s)^2\partial_s^2\,+\,\big(\lambda s-(1-s)^2\big)\partial_s
\;.
$$
Then \eqref{eq-ImageSmall} states that functions in the image of $\Ll_\lambda$ have a small expectation w.r.t. $\varrho_\lambda$.
Furthermore, let us introduce a formal adjoint $\Ll_\lambda^*:C^2([0,1])\to C([0,1])$ of $\Ll$ by
$$
\Ll_\lambda^*
\;=\;
-\partial_s\,s\big(\lambda +\partial_s (1-s)^2\big)
\;.
$$
Supposing that $\varrho_\lambda$ is also in  $C^2([0,1])$, partial integration leads to
$$
\int^1_0 \textnormal{d}s\,\varrho_\lambda(s)\,(\Ll_\lambda g)(s)
\;=\;
\int^1_0 \textnormal{d}s\,(\Ll_\lambda^* \varrho_\lambda)(s)\,g(s)
\;+\;
\lambda\,\varrho_\lambda(1)\,g(1)
\;.
$$
Hence by the above, this is of order $\Oo(\epsilon,\epsilon^{-1}\delta,\epsilon^{-2}\delta^2)$ for all $g\in C^3([0,1])$. This suggests that $\varrho_\lambda(1)=0$. One is thus led to determine the non-negative elements of the kernel of $\Ll_\lambda^*$ which vanish at $1$. The corresponding subspace  contains the normalized function $\varrho_\lambda$ given by \eqref{Furstenberg-radial-density}. It actually already lies in the kernel of the first order operator $\big(\lambda s+\partial_s (1-s)^2\big)$ which is part of $\Ll_\lambda^*$. Also note that $\varrho_\lambda(1)=0$ and that L'H\^opital's rule allows to compute the limits $s\to 0$ and $s\to 1$ of $\varrho_\lambda$ and its derivatives, implying that $\varrho_\lambda\in C^2([0,1])$.  Of course, at this point these formal arguments have to be completed. For example, it is necessary to show that the kernel of $\Ll_\lambda^*$ is one-dimensional. This and other analytical issues have to deal with the fact that both $\Ll_\lambda$ and $\Ll_\lambda^*$ are singular elliptic in the sense that the highest order term $-s(1-s)^2\partial_s^2$ has a coefficient function that vanishes at the boundary points $s=0$ and $s=1$ (this can be dealt with by the techniques of the appendix in \cite{SS}). Here the proof of Theorem~\ref{Furstenberg-result} rather follows a more direct approach.

\vspace{.2cm}

\noindent\textbf{Proof of Theorem~\ref{Furstenberg-result}.} Inserting $g=\mathds{1}$ into~\eqref{overall-oscillation-argument-statement-2} yields
\begin{align}\label{Furstenberg-result-proof-1}
\mathcal{D}\,\epsilon^2\int_{\overline{\mathbb{D}}}\mu^{\epsilon,\delta}(\textnormal{d}z)\textnormal{ }(1-|z|^2)^2-2\,\mathcal{C}\,\delta\int_{\overline{\mathbb{D}}}\mu^{\epsilon,\delta}(\textnormal{d}z)\textnormal{ }{|z|^2}\;=\;\mathcal{O}(\epsilon^3,\epsilon\delta,\delta^2)\,,
\end{align}
which implies the first statement~\eqref{Furstenberg-result-statement-2} if $\mathcal{C}>0$. Moreover, if $\mathcal{D}>0$, the identity~\eqref{Furstenberg-result-proof-1} also implies
\begin{align*}
\int_{\overline{\mathbb{D}}}\mu^{\epsilon,\delta}(\textnormal{d}z)\textnormal{ }(1-|z|^2)^2\;=\;\mathcal{O}(\epsilon,\delta\epsilon^{-2})\,,
\end{align*}
from which one infers the second statement~\eqref{Furstenberg-result-statement-1} by using Jensen's inequality:
\begin{align*}
\int_{\overline{\mathbb{D}}}\mu^{\epsilon,\delta}(\textnormal{d}z)\textnormal{ }|z|^2\;=\;1-\int_{\overline{\mathbb{D}}}\mu^{\epsilon,\delta}(\textnormal{d}z)\textnormal{ }(1-|z|^2)
\;\geq\; 
1-\left(\int_{\overline{\mathbb{D}}}\mu^{\epsilon,\delta}(\textnormal{d}z)\textnormal{ }(1-|z|^2)^2\right)^{\frac{1}{2}}\,.
\end{align*}
Let us now come to the third statement~\eqref{Furstenberg-result-density}. In view of the identity~\eqref{overall-oscillation-argument-statement-2} of Lemma~\ref{overall-oscillation-argument}, the task is to find a function $g\in C^3([0,1])$ that satisfies $\mathcal{L}_{\lambda}\,g=\tilde{h}$, where $\tilde{h}\in C^2([0,1])$ is defined in terms of $h$ and~\eqref{Furstenberg-radial-density}~by
\begin{align}\label{Furstenberg-result-proof-2}
\tilde{h}(s)
\;=\;
h(s)\;-\;\int_{0}^1\textnormal{d}x\,\varrho_{\lambda}(x)\,h(x)\,.
\end{align}
For this purpose, let us first solve the first order differential equation
\begin{align}\label{Furstenberg-result-proof-3}
\big(\lambda-(1-s)^2\partial_s\big) F(s)
\;=\;
\tilde{h}(s)
\end{align}
in the open interval $(0,1)$ by the method of variation of constants:
$$
F(s)
\;=\;
\exp\left[\frac{\lambda}{1-s}\right]\int_{0}^s\textnormal{d}x\,\frac{\tilde{h}(x)}{(1-x)^2}\exp\left[\frac{-\lambda}{1-x}\right]\,.
$$
The function $F:(0,1)\rightarrow\mathbb{R}$ lies in $C^3((0,1))$ and so does ${G}$ given by ${G}(s)=F(s)\,s^{-1}$. Due to~\eqref{Furstenberg-result-proof-3}, one has
\begin{align}\label{Furstenberg-result-proof-4}
\big(\lambda-(1-s)^2\partial_s\big)\,s\,{G}(s)
\;=\;
\tilde{h}(s)\,.
\end{align}
Actually, it turns out that  ${G}$ has a continuous extension in $C^2([0,1])$ (see Lemma~\ref{Furstenberg-result-lemma} below). Hence, its  integral $g:[0,1]\rightarrow\mathbb{R}$ given by
\begin{align}\label{Furstenberg-result-proof-5}
g(s)
\;=\;
\int_0^s\textnormal{d}x\,G(x)
\end{align}
lies in $C^3([0,1])$. Now,~\eqref{Furstenberg-result-proof-4} and~\eqref{Furstenberg-result-proof-5} imply
\begin{align*}
(\mathcal{L}_{\lambda}g)(s)
\;=\;
\big(\lambda-(1-s)^2\partial_s\big)\,s\,\partial_s\,g(s)
\;=\;
\big(\lambda-(1-s)^2\partial_s\big)\,s\,{G}(s)
\;=\;
\tilde{h}(s)
\;.
\end{align*}
Therefore the identity~\eqref{overall-oscillation-argument-statement-2} of Lemma~\ref{overall-oscillation-argument} yields
\begin{align*}
\mathcal{D}\,\epsilon^2\int_{\overline{\mathbb{D}}}\mu^{\epsilon,\delta}(\textnormal{d}z)\textnormal{ }\tilde{h}(|z|^2)
\;=\;
\mathcal{D}\,\epsilon^2\int_{\overline{\mathbb{D}}}\mu^{\epsilon,\delta}(\textnormal{d}z)\textnormal{ }(\mathcal{L}_{\lambda}g)(|z|^2)
\;=\;
\mathcal{O}(\epsilon^3,\epsilon\delta,\delta^2)\,.
\end{align*}
Together with~\eqref{Furstenberg-result-proof-2} one infers
\begin{align*}
\int_{\overline{\mathbb{D}}}\mu^{\epsilon,\delta}(\textnormal{d}z)\textnormal{ }{h}(|z|^2)
& 
\;=\;
\int_0^1\textnormal{d}s\,\varrho_{\lambda}(s)\,h(s)
\;+\;
\int_{\overline{\mathbb{D}}}\mu^{\epsilon,\delta}(\textnormal{d}z)\textnormal{ }\tilde{h}(|z|^2)
\\
&
\;=\;
\int_0^1\textnormal{d}s\,\varrho_{\lambda}(s)\,h(s)
\;+\;
\mathcal{O}(\epsilon,\epsilon^{-1}\delta,\epsilon^{-2}\delta^2)\,.
\end{align*}
As this is of interest only for $\epsilon^{-1}\delta<1$, this proves~\eqref{Furstenberg-result-density}.
\hfill $\Box$

\vspace{2mm}

%%%%%%%%%%%%%%%%%%%%%%%%%%%%%%
\begin{lemma}\label{Furstenberg-result-lemma}
The map $G:(0,1)\rightarrow\mathbb{R}$, defined by
\begin{align*}
G(s)
\;=\;
\frac{1}{s}\exp\left[\frac{\lambda}{1-s}\right]\int_0^s\textnormal{d}x\,\frac{\tilde{h}(x)}{(1-x)^2}\exp\left[\frac{-\lambda}{1-x}\right]\,,
\end{align*}
where $\tilde{h}$ is given by~\eqref{Furstenberg-result-proof-2},
has a continuous extension that lies in $C^2([0,1])$. 
\end{lemma}
%%%%%%%%%%%%%%%%%%%%%%%%%%%%%%

\noindent\textbf{Proof.}
It is useful to factorize $G$ as
\begin{align}
G(s)\;=\;\frac{G_1(s)}{G_2(s)}\,,
\end{align}
where
\begin{align*}
G_1(s)\;=\;\int_0^s\textnormal{d}x\,\frac{\tilde{h}(x)}{(1-x)^2}\,\exp\left[\frac{-\lambda}{1-x}\right]\qquad\textnormal{and}\qquad
G_2(s)\;=\;s\,\exp\left[\frac{-\lambda}{1-s}\right]\,.
\end{align*}
Clearly $G_1$ and $G_2$ satisfy
\begin{align}\label{Furstenberg-result-lemma-proof-1}
\lim\limits_{s\downarrow 0}G_1(s)
\;=\;
\lim\limits_{s\downarrow 0}G_2(s)
\;=\;
0
\qquad\textnormal{and}\qquad
\lim\limits_{s\uparrow 1}G_2(s)
\;=\;
0\,,
\end{align}
and~\eqref{Furstenberg-result-proof-2} implies that
\begin{align}\label{Furstenberg-result-lemma-proof-2}
\lim\limits_{s\uparrow 1}G_1(s)\;=\;\int^1_0\textnormal{d}x\,\frac{\tilde{h}(x)}{(1-x)^2}\exp\left[\frac{-\lambda}{1-x}\right]\;=\;\big[\lambda\,e^{\lambda}\big]^{-1}\int^1_0\textnormal{d}s\,\varrho_{\lambda}(s)\,\tilde{h}(s)\;=\;0\,,
\end{align}
since $\varrho_{\lambda}$ is normalized.
Moreover, the derivatives of $G_1$ and $G_2$ are given~by 
\begin{align}\label{Furstenberg-result-lemma-proof-3}
G_1^{\prime}(s)
\;=\;
\frac{\tilde{h}(s)}{(1-s)^2}\,\exp\left[\frac{-\lambda}{1-s}\right]\qquad\textnormal{and}\qquad
G_2^{\prime}(s)
\;=\;
\left[1-\frac{\lambda\,s}{(1-s)^2}\right]\exp\left[\frac{-\lambda}{1-s}\right]\,.
\end{align}
Using~\eqref{Furstenberg-result-lemma-proof-1},~\eqref{Furstenberg-result-lemma-proof-2} and~\eqref{Furstenberg-result-lemma-proof-3} allows to apply L'H\^ospital's rule  to show
\begin{align*}
\lim\limits_{s\downarrow 0}G(s)
& \;=\;
\lim\limits_{s\downarrow 0}G_1^{\prime}(s)G_2^{\prime}(s)^{-1}
\;=\;
\tilde{h}(0)
\;,
\\
\lim\limits_{s\uparrow 1}G(s)
& \;=\;
\lim\limits_{s\uparrow 1}G_1^{\prime}(s)G_2^{\prime}(s)^{-1}
\;=\;-\,\lambda^{-1}\,\tilde{h}(1)\,.
\end{align*}
In particular, the limits $\lim\limits_{s\downarrow 0}G(s)$ and $\lim\limits_{s\uparrow 1}G(s)$ exist so that $G$ has a continuous extension to~$[0,1]$.

\vspace{1mm}

Next let us compute the derivative of $G$ for $s\in(0,1)$ and factorize it as follows:
\begin{align}\label{Furstenberg-result-lemma-proof-4}
G^{\prime}(s)
\;=\;
\frac{\tilde{G}_1(s)}{\tilde{G}_2(s)}\,,
\end{align}
where
\begin{align*}
\tilde{G}_1(s)\;=\;
\tilde{h}(s)+\left[\lambda\,s-(1-s)^2\right]G(s)\qquad\textnormal{and}\qquad
\tilde{G}_2(s)\;=\;
s(1-s)^2\,.
\end{align*}
Moreover, 
\begin{align}
&\frac{\textnormal{d}}{\textnormal{d}s}\left(\tilde{G}_1(s)\,s\,\exp\left[\frac{-\lambda}{1-s}\right]\right)
\nonumber
\\
&
\;=\;
\left(\Big[\tilde{h}^{\prime}(s)+\big(\lambda+2[1-s]\big)\,G(s)\Big]\,s+\left[1-\frac{\lambda\,s}{(1-s)^2}\right]\,\big[\tilde{G}_1(s)-\tilde{G}_2(s)\,G^{\prime}(s)\big]\right)\exp\left[\frac{-\lambda}{1-s}\right]
\nonumber
\\
&
\;=\;
\Big[\tilde{h}^{\prime}(s)+\big(\lambda+2[1-s]\big)\,G(s)\Big]\,s\,\exp\left[\frac{-\lambda}{1-s}\right]\,,
\label{Furstenberg-result-lemma-proof-5}
\end{align}
where \eqref{Furstenberg-result-lemma-proof-4} was used in the second step, and
\begin{align}\label{Furstenberg-result-lemma-proof-6}
\begin{split}
\frac{\textnormal{d}}{\textnormal{d}s}\left(\tilde{G}_2(s)\,s\,\exp\left[\frac{-\lambda}{1-s}\right]\right)
\;=\;
\Big[2\,\big(1-3s+2s^2\big)-\lambda\,s\Big]\,s\,\exp\left[\frac{-\lambda}{1-s}\right]\,.
\end{split}
\end{align}
Now the conditions
\begin{align*}
\lim\limits_{s\downarrow 0}\tilde{G}_l(s)\,\,s\,\exp\left[\frac{-\lambda}{1-s}\right]
\;=\;
0
\qquad\textnormal{and}\qquad
\lim\limits_{s\uparrow 1}\tilde{G}_l(s)\,\,s\,\exp\left[\frac{-\lambda}{1-s}\right]
\;=\;
0\,,
\end{align*}
which hold for $l=1$ and $l=2$, allow to apply L'H\^ospital's rule again to infer
\begin{align*}
\lim\limits_{s\downarrow 0}G^{\prime}(s)
&
\;=\;
\lim\limits_{s\downarrow 0}\left[\frac{\textnormal{d}}{\textnormal{d}s}\left(\tilde{G}_1(s)\,s\,\exp\left[\frac{-\lambda}{1-s}\right]\right)\right]\left[\frac{\textnormal{d}}{\textnormal{d}s}\left(\tilde{G}_2(s)\,s\,\exp\left[\frac{-\lambda}{1-s}\right]\right)\right]^{-1}\\
&
\;=\;
\frac{1}{2}\Big[\tilde{h}^{\prime}(0)+(2+\lambda)\,\lim\limits_{s\downarrow 0}G(s)\Big]
\end{align*}
and
\begin{align*}
\lim\limits_{s\uparrow 1}G^{\prime}(s)
&
\;=\;
\lim\limits_{s\uparrow 1}\left[\frac{\textnormal{d}}{\textnormal{d}s}\left(\tilde{G}_1(s)\,s\,\exp\left[\frac{-\lambda}{1-s}\right]\right)\right]\left[\frac{\textnormal{d}}{\textnormal{d}s}\left(\tilde{G}_2(s)\,s\,\exp\left[\frac{-\lambda}{1-s}\right]\right)\right]^{-1}\\
&
\;=\;
-\frac{1}{2}\Big[\lambda^{-1}\,\tilde{h}^{\prime}(1)+\lim\limits_{s\uparrow 1}G(s)\Big]
\end{align*}
by using~\eqref{Furstenberg-result-lemma-proof-4},~\eqref{Furstenberg-result-lemma-proof-5} and~\eqref{Furstenberg-result-lemma-proof-6}. In particular, $\lim\limits_{s\downarrow 0}G^{\prime}(s)$ and $\lim\limits_{s\uparrow 1}G^{\prime}(s)$ exist so that $G^{\prime}$ has a continuous extension to $[0,1]$. This implies that the continuous extension of $G$ lies in $C^1([0,1])$.

\vspace{1mm}

Finally let us consider the second derivative of $G$. Due to \eqref{Furstenberg-result-lemma-proof-4} it is given by   
\begin{align}\label{Furstenberg-result-lemma-proof-7}
G^{\prime\prime}(s)
\;=\;
\frac{\hat{G}_1(s)}{\hat{G}_2(s)}\,,
\qquad
s\in(0,1)
\;,
\end{align}
where
\begin{align}\label{Furstenberg-result-lemma-proof-8}
\hat{G}_1(s)
\;=\;
{\tilde{G}_1^{\prime}(s)\,\tilde{G}_2(s)-\tilde{G}_1(s)\,\tilde{G}^{\prime}_2(s)}\qquad\textnormal{and}\qquad 
\hat{G}_2(s)
\;=\;
{\tilde{G}_2(s)^2}\,.
\end{align}
Moreover, 
\begin{align}
&\frac{\textnormal{d}}{\textnormal{d}s}\left(\tilde{G}_1^{\prime}(s)\,\tilde{G}_2(s)\,s\,\exp\left[\frac{-\lambda}{1-s}\right]\right)
\nonumber
\\
&
\;=\;
\left(\left[1-\frac{\lambda\,s}{(1-s)^2}\right]\,\Big[\tilde{G}_1^{\prime}(s)\,\tilde{G}_2(s)-{G}^{\prime\prime}(s)\,\tilde{G}_2(s)^2\Big]+\tilde{G}^{\prime}_1(s)\,\tilde{G}^{\prime}_2(s)\,s\right)\,\exp\left[\frac{-\lambda}{1-s}\right]
\nonumber
\\
&\quad\;+\;
\Big[\tilde{h}^{\prime\prime}(s)-2\,G(s)+2\,\big[\lambda+2(1-s)\big]\,G^{\prime}(s)\Big]\,\tilde{G}_2(s)\,s\,\exp\left[\frac{-\lambda}{1-s}\right]
\label{Furstenberg-result-lemma-proof-9}
\end{align}
and
\begin{align}
\frac{\textnormal{d}}{\textnormal{d}s}\left(\tilde{G}_1(s)\,\tilde{G}^{\prime}_2(s)\,s\,\exp\left[\frac{-\lambda}{1-s}\right]\right)
&
\;=\;
\left(\left[1-\frac{\lambda\,s}{(1-s)^2}\right]\,\tilde{G}_1(s)\,\tilde{G}_2^{\prime}(s)+\tilde{G}^{\prime}_1(s)\,\tilde{G}^{\prime}_2(s)\,s\right)\,\exp\left[\frac{-\lambda}{1-s}\right]
\nonumber
\\
&
\quad\;+\;
G^{\prime}(s)\,\tilde{G}_2^{\prime\prime}(s)\,\tilde{G}_2(s)\,s\,\exp\left[\frac{-\lambda}{1-s}\right]\,.
\label{Furstenberg-result-lemma-proof-10}
\end{align}
Due to~\eqref{Furstenberg-result-lemma-proof-7} and~\eqref{Furstenberg-result-lemma-proof-8}, the first summands on the right sides of~\eqref{Furstenberg-result-lemma-proof-9} and~\eqref{Furstenberg-result-lemma-proof-10} are equal. Thus,
\begin{align*}
&\frac{\textnormal{d}}{\textnormal{d}s}\left(\hat{G}_1(s)\,s\,\exp\left[\frac{-\lambda}{1-s}\right]\right)\\
&
\;=\;
\frac{\textnormal{d}}{\textnormal{d}s}\left(\left[\tilde{G}^{\prime}_1(s)\,\tilde{G}_2(s)-\tilde{G}_1(s)\,\tilde{G}^{\prime}_2(s)\right]\,s\,\exp\left[\frac{-\lambda}{1-s}\right]\right)
\\
&
\;=\;
\left[\tilde{h}^{\prime\prime}(s)-2\,G(s)+\Big[2\lambda+4\,(1-s)-\tilde{G}_2^{\prime\prime}(s)\Big]\,G^{\prime}(s)\right]\,\tilde{G}_2(s)\,s\,\exp\left[\frac{-\lambda}{1-s}\right]\,.
\end{align*}
Moreover, one has
\begin{align*}
%\label{Furstenberg-result-lemma-proof-11}
\frac{\textnormal{d}}{\textnormal{d}s}\left(\hat{G}_2(s)\,s\,\exp\left[\frac{-\lambda}{1-s}\right]\right)
\;=\;
\Big[3\,(1-s)^2+\lambda\,s-4\,s\,(1-s)\Big]\,\tilde{G}_2(s)\,s\,\exp\left[\frac{-\lambda}{1-s}\right]\,.
\end{align*}
Now the conditions
\begin{align*}
\lim\limits_{s\downarrow 0}\hat{G}_l(s) \,s\, \exp\left[\frac{-\lambda}{1-s}\right]
\;=\;
0
\qquad\textnormal{and}\qquad
\lim\limits_{s\uparrow 1}\hat{G}_l(s)\,s\,\exp\left[\frac{-\lambda}{1-s}\right]
\;=\;
0\,,
\end{align*}
which hold for $l=1$ and $l=2$, allow to apply L'H\^ospital's rule a third time to infer
\begin{align*}
%\label{Furstenberg-result-lemma-proof-12}
%\begin{split}
\lim\limits_{s\downarrow 0}G^{\prime\prime}(s)&
\;=\;
\lim\limits_{s\downarrow 0}\left[\frac{\textnormal{d}}{\textnormal{d}s}\left(\hat{G}_1(s)\,s\,\exp\left[\frac{-\lambda}{1-s}\right]\right)\right]\left[\frac{\textnormal{d}}{\textnormal{d}s}\left(\hat{G}_2(s)\,s\,\exp\left[\frac{-\lambda}{1-s}\right]\right)\right]^{-1}\\
&
\;=\;
\frac{1}{3}\Big[\tilde{h}^{\prime\prime}(0)-2\,\lim\limits_{s\downarrow 0}G(s)+2(\lambda+4)\,\lim\limits_{s\downarrow 0}G^{\prime}(s)\Big]
%\end{split}
\end{align*}
and
\begin{align*}
%\label{Furstenberg-result-lemma-proof-13}
%\begin{split}
\lim\limits_{s\uparrow 1}G^{\prime\prime}(s)&
\;=\;\lim\limits_{s\uparrow 1}\left[\frac{\textnormal{d}}{\textnormal{d}s}\left(\hat{G}_1(s)\,s\,\exp\left[\frac{-\lambda}{1-s}\right]\right)\right]\left[\frac{\textnormal{d}}{\textnormal{d}s}\left(\hat{G}_2(s)\,s\,\exp\left[\frac{-\lambda}{1-s}\right]\right)\right]^{-1}\\
&\;=\;\frac{1}{\lambda}\Big[\tilde{h}^{\prime\prime}(1)-2\,\lim\limits_{s\uparrow 1}G(s)+2(\lambda-1)\,\lim\limits_{s\uparrow 1}G^{\prime}(s)\Big]
\;.
%\end{split}
\end{align*}
In particular, $\lim\limits_{s\downarrow 0}G^{\prime\prime}(s)$ and $\lim\limits_{s\uparrow 1}G^{\prime\prime}(s)$ exist so that $G^{\prime\prime}$ has a continous extension to $[0,1]$. This implies that the continuous extension of $G^{\prime}$ lies in $C^1([0,1])$ and, all in all, that the contiuous extension of $G$ to $[0,1]$ lies indeed in $C^2([0,1])$.
\hfill $\square$

%%%%%%%%%%%%%%%%%%%%%%%%%%%%%%%
\section{The Lyapunov exponent}\label{Lyapunov}

The Lyapunov exponent $\gamma^{\epsilon,\delta}$ can be expressed by the Furstenberg formula
\begin{align}\label{Lyapunov-Furstenberg-relation}
\gamma^{\epsilon,\delta}\;=\;\int_{\mathbb{S}^1_{\mathbb{C}}}\nu^{\epsilon,\delta}(\textnormal{d}x)\textnormal{ }\mathbb{E}\log\left\|T^{\epsilon,\delta}_{\sigma} x\right\|\,,
\end{align}
where $\nu^{\epsilon,\delta}$ is some invariant probability measure on $\mathbb{S}^{1}_{\mathbb{C}}$ corresponding to $\mu^{\epsilon,\delta}$ (and satisfying $\pi_*(\nu^{\epsilon,\delta})=\mu^{\epsilon,\delta}$).
For the proof of Theorem~\ref{main-result}, one has to express the term $\log\left\|T^{\epsilon,\delta}_{\sigma} x\right\|$ appearing in~\eqref{Lyapunov-Furstenberg-relation} in terms of the stereographic projection of $x$. This is carried out in Lemma~\ref{Lyapunov-1}.

%%%%%%%%%%%%%%%%%%%%%%%%%%%%%%%
\begin{lemma}\label{Lyapunov-1}
Let $x\in\mathbb{S}_{\mathbb{C}}^1$ and $z=\pi(x)$. Then, 
\begin{align}\label{Lyapunov-1-statement}
\begin{split}
\log\left\|T^{\epsilon,\delta}_{\sigma} x\right\|&\;=\;2\,\epsilon\,\frac{\Re\mathfrak{e}(\beta_{\sigma}z)}{1+|z|^2}+\epsilon^2\,|\beta_{\sigma}|^2\,\frac{1+|z|^4}{(1+|z|^2)^2}+\delta\, q_{3,\sigma}\,\frac{1-|z|^2}{1+|z|^2}\\
&\qquad+2\,\epsilon^2\,\left[\frac{\Re\mathfrak{e}([\beta^{\prime}_{\sigma}+\imath p_{3,\sigma}\beta_{\sigma}]z)}{1+|z|^2}-\frac{\Re\mathfrak{e}(\beta_{\sigma}^2z^2)}{(1+|z|^2)^2}\right]\;+\;\mathcal{O}(\epsilon^3,\epsilon\delta,\delta^2)\,.
\end{split}
\end{align}
\end{lemma}
%%%%%%%%%%%%%%%%%%%%%%%%%%%%%%%

\noindent\textbf{Proof.} Let us begin by computing
\begin{align}\label{Lyapunov-1-1}
%\begin{split}
& \left\|T^{\epsilon,\delta}_{\sigma} x\right\|^2\;=\;\left\|\exp\left[\epsilon {P}_{\sigma}+\epsilon^2{P}^{\prime}_{\sigma}+\imath\delta {Q}_{\sigma}\;+\;\mathcal{O}(\epsilon^3,\epsilon\delta,\delta^2)\right]x\right\|^2
\nonumber
\\
&\;\;\;
\;=\;\left\langle x,\left[\mathbf{1}+\epsilon ({P}_{\sigma})_s+\epsilon^2({P}^{\prime}_{\sigma})_s+\delta (\imath{Q}_{\sigma})_s+\frac{1}{2}\epsilon^2({P}^2_{\sigma})_s+\epsilon^2|{P}_{\sigma}|^2\right]x\right\rangle\;+\;\mathcal{O}(\epsilon^3,\epsilon\delta,\delta^2)\,,
%\end{split}
\end{align}
where the notation $(A)_s=A+A^*$ and $|A|^2=A^*A$ was used. Next, one verifies the formulae
\begin{align*}
%\label{Lyapunov-1-2}
%\begin{split}
({P}_{\sigma})_s&\;=\;2\left[p_{1,\sigma}B_1+ p_{2,\sigma}B_2\right]\,,\qquad ({P}_{\sigma}^{\prime})_s\;=\;2\left[p_{1,\sigma}^{\prime}B_1+ p_{2,\sigma}^{\prime}B_2\right]\,,\qquad (\imath{Q}_{\sigma})_s\;=\;2\imath q_{3,\sigma}B_3\,,\\
({P}_{\sigma}^2)_s&\;=\;2(|\beta_{\sigma}|^2-p_{3,\sigma}^2)\mathbf{1}\,,\quad\qquad\qquad |{P}_{\sigma}|^2\;=\;(|\beta_{\sigma}|^2+p_{3,\sigma}^2)\mathbf{1}-2p_{3,\sigma}\left[p_{1,\sigma}B_2-p_{2,\sigma}B_1\right]\quad
%\end{split}
\end{align*}
and 
\begin{align*}
%\label{Lyapunov-1-3}
\langle v,B_1 v\rangle\;=\;\frac{2\Re\mathfrak{e}(z)}{1+|z|^2}\,,\qquad\langle v,B_2 v\rangle\;=\;\frac{2\,\Im\mathfrak{m}(z)}{1+|z|^2}\,,\qquad\langle v,B_3 v\rangle\;=\;\imath\frac{|z|^2-1}{1+|z|^2}\,.
\end{align*}
Combining them yields
\begin{align*}
%\label{Lyapunov-1-4}
%\begin{split}
\langle x,({P}_{\sigma})_s x\rangle\;=\;\frac{4\Re\mathfrak{e}(\beta_{\sigma}z)}{1+|z|^2}\,,\qquad
\langle x,({P}_{\sigma}^{\prime})_s x\rangle\;=\;\frac{4\Re\mathfrak{e}(\beta^{\prime}_{\sigma}z)}{1+|z|^2}\,,\qquad
\langle x,(\imath{Q}_{\sigma})_s x\rangle\;=\;2q_{3,\sigma}\frac{1-|z|^2}{1+|z|^2}\,,\\
\langle x,({P}^2_{\sigma})_s x\rangle\;=\;2(|\beta_{\sigma}|^2-p_{3,\sigma}^2)\,,\qquad\qquad\langle x,|{P}_{\sigma}|^2 x\rangle\;=\;|\beta_{\sigma}|^2+p_{3,\sigma}^2-\frac{4p_{3,\sigma}\,\Im\mathfrak{m}(\beta_{\sigma}z)}{1+|z|^2}\,.
%\end{split}
\end{align*}
Using these identities one can now rewrite \eqref{Lyapunov-1-1} as
\begin{align*}
%\label{Lyapunov-1-5}
%\begin{split}
\left\|T^{\epsilon,\delta}_{\sigma} x\right\|^2\,=\,1+4\epsilon\frac{\Re\mathfrak{e}(\beta_{\sigma}z)}{1+|z|^2}+2\delta q_{3,\sigma}\frac{1-|z|^2}{1+|z|^2}+2\epsilon^2\left[|\beta_{\sigma}|^2+2\frac{\Re\mathfrak{e}([\beta^{\prime}_{\sigma}+\imath p_{3,\sigma}\beta_{\sigma}]z)}{1+|z|^2}\right]+\mathcal{O}(\epsilon^3,\epsilon\delta,\delta^2)\,.
%\end{split}
\end{align*}
In view of $\log(1+a)=a-\frac{a^2}{2}\;+\;\mathcal{O}(a^3)$, this implies
\begin{align*}
%\label{Lyapunov-1-6}
%\begin{split}
\log\left[\left\|T^{\epsilon,\delta}_{\sigma} x\right\|^2\right]&\;=\;4\epsilon\,\frac{\Re\mathfrak{e}(\beta_{\sigma}z)}{1+|z|^2}\;-\;8\epsilon^2\,\frac{\Re\mathfrak{e}(\beta_{\sigma}z)^2}{(1+|z|^2)^2}\;+\;2\delta q_{3,\sigma}\frac{1-|z|^2}{1+|z|^2}\\
&\qquad +\;2\epsilon^2\left[|\beta_{\sigma}|^2+2\,\frac{\Re\mathfrak{e}([\beta^{\prime}_{\sigma}+\imath p_{3,\sigma}\beta_{\sigma}]z)}{1+|z|^2}\right]\;+\;\mathcal{O}(\epsilon^3,\epsilon\delta,\delta^2)\,.
%\end{split}
\end{align*}
This implies \eqref{Lyapunov-1-statement} because $\Re\mathfrak{e}(\beta_{\sigma}z)^2=\frac{1}{2}\Re\mathfrak{e}(\beta_{\sigma}^2z^2)+\frac{1}{2}|\beta_{\sigma}|^2|z|^2$ and $1-\frac{2|z|^2}{(1+|z|^2)^2}=\frac{1+|z|^4}{(1+|z|^2)^2}$.
\hfill $\Box$

\vspace{2mm}

\noindent {\bf Proof} of Theorem~\ref{main-result}. Thanks to Lemma~\ref{Lyapunov-1}, equation~\eqref{Lyapunov-Furstenberg-relation} can be rewritten as
\begin{align}\label{Lyapunov-2}
\begin{split}
\gamma^{\epsilon,\delta}&\;=\;2\,\epsilon\,\Re\mathfrak{e}\left(\mathbb{E}(\beta_{\sigma})\int_{\overline{\mathbb{D}}}\mu^{\epsilon,\delta}(\textnormal{d}z)\textnormal{ }\frac{z}{1+|z|^2}\right)\;+\;\epsilon^2\textnormal{ }\mathbb{E}(|\beta_{\sigma}|^2)\int_{\overline{\mathbb{D}}}\mu^{\epsilon,\delta}(\textnormal{d}z)\textnormal{ }\frac{1+|z|^4}{(1+|z|^2)^2}\\
&\qquad+\;\delta\textnormal{ }\mathcal{C}\int_{\overline{\mathbb{D}}}\mu^{\epsilon,\delta}(\textnormal{d}z)\textnormal{ }\frac{1-|z|^2}{1+|z|^2}\;+\;2\,\epsilon^2\textnormal{ }\Re\mathfrak{e}\left(\mathbb{E}(\beta^{\prime}_{\sigma}+\imath p_{3,\sigma}\beta_{\sigma})\int_{\overline{\mathbb{D}}}\mu^{\epsilon,\delta}(\textnormal{d}z)\textnormal{ }\frac{z}{1+|z|^2}\right)\\
&\qquad-\;2\,\epsilon^2\textnormal{ }\Re\mathfrak{e}\left(\mathbb{E}(\beta_{\sigma}^2)\int_{\overline{\mathbb{D}}}\mu^{\epsilon,\delta}(\textnormal{d}z)\textnormal{ }\frac{z^2}{1+|z|^2}\right)\;+\;\mathcal{O}(\epsilon^3,\epsilon\delta,\delta^2)\,.
\end{split}
\end{align}
According to Lemma~\ref{oscillation-argument}, the fourth and the fifth summand of the r.h.s.  of~\eqref{Lyapunov-2} are of the order $\mathcal{O}(\epsilon^3,\epsilon^2\delta)$. Together with equation~\eqref{oscillation-argument-first-order-corollary-statement}, which is applicable to the first summand of the r.h.s.  of~\eqref{Lyapunov-2}, this implies
\begin{align*}
%\label{Lyapunov-4}
%\begin{split}
\gamma^{\epsilon,\delta}&\;=\;\mathcal{D}\,\epsilon^2\,\int_{\overline{\mathbb{D}}}\mu^{\epsilon,\delta}(\textnormal{d}z)\textnormal{ }\frac{2(1+|z|^4)}{(1+|z|^2)^2}\;+\;\mathcal{C}\;\delta\int_{\overline{\mathbb{D}}}\mu^{\epsilon,\delta}(\textnormal{d}z)\textnormal{ }\frac{1-|z|^2}{1+|z|^2}\;+\;\mathcal{O}(\epsilon^3,\epsilon\delta,\delta^2)\\
&\;=\;\mathcal{D}\,\epsilon^2\,\int_{\overline{\mathbb{D}}}\mu^{\epsilon,\delta}(\textnormal{d}z)\textnormal{ }\left[1+\left[\frac{1-|z|^2}{1+|z|^2}\right]^2\right]\;+\;\mathcal{C}\;\delta\int_{\overline{\mathbb{D}}}\mu^{\epsilon,\delta}(\textnormal{d}z)\textnormal{ }\left[1-\frac{2|z|^2}{1+|z|^2}\right]\;+\;\mathcal{O}(\epsilon^3,\epsilon\delta,\delta^2)\,.
%\end{split}
\end{align*}
Due to~\eqref{Lyapunov-auxiliary-equation-statement}, this implies~\eqref{main-result-statement}. 
\hfill $\Box$

%%%%%%%%%%%%%%%%%%%%%%%%%%%%%%%
\section{The support of the Furstenberg measure}
\label{sec-FuerstenbergExample}

The statement~\eqref{Furstenberg-result-density} of Theorem~\ref{Furstenberg-result} approximates the radial distribution of $\mu^{\epsilon,\delta}$ as long as $\delta=o(\epsilon)$. Since the approximate radial density $\varrho_{\lambda}$ given by~\eqref{Furstenberg-radial-density} is supported on $[0,1]$, it is natural to presume that $\mu^{\epsilon,\delta}$ is supported by the whole closed unit disc $\overline{\mathbb{D}}$ in that case. Of course, statement~\eqref{Furstenberg-result-density} does {\it not} imply that presumption. In the complementary case $\epsilon=o(\delta)$, however, the support of $\mu^{\epsilon,\delta}$ can be proven to be a strict subset of the unit disc under some supplementary assumption. 

\begin{proposi}\label{center-support}
Suppose that $q_{3,\sigma}>0$ holds for all $\sigma\in\Sigma$. Then, one has
\begin{align}\label{center-support-statement}
\textnormal{supp}(\mu^{\epsilon,\delta})
\;\subset\;
\Big\{z\in\overline{\mathbb{D}}\,:\, |z|^2\leq \esssup\limits_{\sigma\in\Sigma}|\xi_{\sigma}|q_{3,\sigma}^{-1}\;+\;\mathcal{O}(\epsilon\delta^{-1},\delta)\Big\}\,.
\end{align}
\end{proposi}

\noindent\textbf{Proof.} One may assume that
$$
\zeta\;=\;\esssup\limits_{\sigma\in\Sigma}\big[q_{\sigma}-|\xi_{\sigma}|\big]\;>\;0\,,
$$
as~\eqref{center-support-statement} is trivial otherwise. For $n\in\mathbb{N}_0$, let us compute by using~\eqref{Baker} and~\eqref{exponentials} 
\begin{align*}
|z_{n+1}|^2-|z_n|^2\;=\;2\delta\,\Im \mathfrak{m}(\xi_{n+1}z_n)[1+|z_n|^2]-4\delta q_{3,n+1} |z_n|^2\;+\;\mathcal{O}(\delta^2,\epsilon)\,,
\end{align*}
which implies
\begin{align*}
%\label{center-support-1}
%\begin{split}
|z_{n+1}|^2&\;=\;|z_n|^2\big(1-2\delta\big[q_{3,n+1}-\,\Im\mathfrak{m}(\xi_{n+1}z_n)\big]\big)+2\delta\left[\Im\mathfrak{m}(\xi_{n+1}z_n)-q_{3,n+1}|z_n|^2\right]\;+\;\mathcal{O}(\delta^2,\epsilon)\\
&\;\leq\; |z_n|^2\big(1-2\delta\zeta\big)+2\delta\left[|\xi_{n+1}|-q_{3,n+1}|z_n|^2\right]\;+\;\mathcal{O}(\delta^2,\epsilon)\,.
%\end{split}
\end{align*}
This shows
\begin{align}\label{center-support-2}
|z_{n+1}|^2\;\leq\; |z_n|^2\big(1-2\delta\zeta\big)
\end{align}
whenever
\begin{align*}
2\delta\left[|\xi_{n+1}|-q_{3,n+1}|z_n|^2\right]\;+\;\mathcal{O}(\delta^2,\epsilon)\;\leq\; 0\,,
\end{align*}
which is equivalent to
\begin{align}\label{center-support-3}
|z_n|^2\;\geq\; |\xi_{n+1}|q^{-1}_{3,n+1}\;+\;\mathcal{O}(\delta,\epsilon\delta^{-1})\,.
\end{align}
In conclusion, if $z_n$ is not contained in the r.h.s.  of~\eqref{center-support-statement}, then it obeys~\eqref{center-support-3} and thus~\eqref{center-support-2},~\textit{i.e.}, the modulus is properly decreased by a uniform factor. Hence, the dynamics $(z_n)_{n\in\mathbb{N}}$ runs deterministically into the set on the r.h.s.  of~\eqref{center-support-statement} for all starting points $z_0\in\mathbb{D}$. This implies~\eqref{center-support-statement}.~\hfill $\Box$

\vspace{2mm}

In general, however, the relation of $\epsilon$ and $\delta$ does {\it not} shrink the support of $\mu^{\epsilon,\delta}$ in any manner.
To illustrate this, an elementary example is given in Proposition~\ref{full-support}, in which the Furstenberg measure is supported by the whole (closed) unit disc, regardless of the relation of $\epsilon>0$ and $\delta>0$. The assumptions enforce $\mathcal{C}>0$, but both $\mathcal{D}=0$ and $\mathcal{D}>0$ is possible.

\begin{proposi}\label{full-support}
Let $\epsilon,\delta>0$ and $\mathsf{p},\mathsf{q}>0$ and $\mathsf{k}\in\mathbb{R}\setminus\pi\mathbb{Q}$ and suppose that
\begin{align*}
T^{\epsilon,\delta}_{\sigma}\;=\;R_{\eta_{\sigma}}\exp\left[\epsilon P_{\sigma}+\imath\delta Q_{\sigma}\right]\,,\qquad\qquad\textnormal{where}\qquad P_{\sigma}\;=\;p_{1,\sigma}B_1\,,\qquad Q_{\sigma}\;=\;q_{3,\sigma}B_3\,,
\end{align*}
where $\{(\mathsf{k},0,0),(0,\mathsf{p},0),(0,0,\mathsf{q})\}\subset\textnormal{supp}((\eta_{\sigma},p_{1,\sigma},q_{3,\sigma}))$.  Then, the $(T^{\epsilon,\delta}_{\sigma}\cdot)$-invariant probability measure is uniquely given by the Furstenberg measure $\mu^{\epsilon,\delta}$ and $\mu^{\epsilon,\delta}$ satisfies $\textnormal{supp}(\mu^{\epsilon,\delta})=\overline{\mathbb{D}}$.
\end{proposi}

\noindent\textbf{Proof.} Due to the assumption $\{(\mathsf{k},0,0),(0,\mathsf{p},0),(0,0,\mathsf{q})\}\subset\textnormal{supp}((\eta_{\sigma},p_{1,\sigma},q_{3,\sigma}))$, the matrices
\begin{align*}
R_{\mathsf{k}}\;=\;\begin{pmatrix}e^{\imath\mathsf{k}}&0\\0&e^{-\imath \mathsf{k}}\end{pmatrix}\,,\quad \exp\left[\epsilon \mathsf{p}B_1\right]\;=\;\begin{pmatrix}\cosh(\epsilon\mathsf{p})&\sinh(\epsilon\mathsf{p})\\\sinh(\epsilon\mathsf{p})&\cosh(\epsilon\mathsf{p})\end{pmatrix}\,, \quad\exp\left[\imath\delta\mathsf{q}B_3\right]\;=\;\begin{pmatrix}e^{-\delta\mathsf{q}}&0\\0&e^{\delta\mathsf{q}}\end{pmatrix}\,
\end{align*}
lie in the support of $T^{\epsilon,\delta}_{\sigma}$.\\\\
\textit{\underline{Step 1.} The $(T^{\epsilon,\delta}_{\sigma}\cdot)$-invariant probability measure is unique.}\\
Since $\mathsf{k}\in\mathbb{R}\setminus\pi\mathbb{Q}$, one has $e^{\imath\mathsf{k}m_1}\neq e^{\imath\mathsf{k}m_2}$ whenever $m_1,m_2\in\mathbb{N}$ and $m_1\neq m_2$. Therefore, the set
$\left((R_{\mathsf{k}})^{m}\cdot z\right)_{m\in\mathbb{N}}=z\left(e^{2\imath\mathsf{f}m}\right)_{m\in\mathbb{N}}$ is infinite if $z\in{\mathbb{C}}\setminus\{0\}$. But $\exp\left[\epsilon \mathsf{p}B_1\right]\cdot 0=\tanh(\epsilon\mathsf{p})$ and $\exp\left[\epsilon \mathsf{p}B_1\right]\cdot \infty=\coth(\epsilon\mathsf{p})$ lie in $\mathbb{C}\setminus\{0\}$. Hence, no finite subset of $\overline{\mathbb{C}}$ is left invariant under the M{\"o}bius action of $\textnormal{supp}(T^{\epsilon,\delta}_{\sigma})$,~\textit{i.e.}, $T^{\epsilon,\delta}_{\sigma}$ fulfils condition (viii). Moreover, one has $\|\exp\left[\imath\delta \mathsf{q}B_3\right]^n\|=e^{\delta\mathsf{q}n}\rightarrow\infty$ as $n\rightarrow\infty$. Therefore, the semigroup generated by $\textnormal{supp}(T^{\epsilon,\delta}_{\sigma})$ is not relatively compact,~\textit{i.e.}, $T^{\epsilon,\delta}_{\sigma}$ fulfills condition (ix). All in all, (viii) and (ix) imply that the $(T^{\epsilon,\delta}_{\sigma}\cdot$)-invariant probability measure is uniquely given by the Furstenberg measure $\mu^{\epsilon,\delta}$ (see~\cite{BQ}).~\hfill$\diamond$

\vspace{.1cm}

\noindent\textit{\underline{Step 2.} The support of $\mu^{\epsilon,\delta}$ is a subset of $\overline{\mathbb{D}}$.}\\
Given some Borel probability measure $\varrho_0$ on $\overline{\mathbb{C}}$ as initial distribution, each weak limit point of
\begin{align}\label{generate-invariant-measure}
(\xi_N)_{N\in\mathbb{N}}\,,\qquad \textnormal{where}\qquad\xi_N\;=\;\frac{1}{N}\sum\limits_{n=1}^N\varrho_n\,,\qquad\qquad\textnormal{with}\qquad \varrho_n\;=\;\left(\left(T^{\epsilon,\delta}_{\sigma}\cdot\right)_*\right)^{(n)}(\varrho_0)\,,
\end{align}
is again a Borel probability measure on $\overline{\mathbb{C}}$ and is, moreover, invariant under the M{\"o}bius action of $T^{\epsilon,\delta}_{\sigma}$ (see~\cite{BL}, Part A, Chapter I, Lemma 3.5). Here,
$\left(\left(T^{\epsilon,\delta}_{\sigma}\cdot\right)_*\right)^{(n)}$ is the $n$-th iterate of the pushforward of the M{\"o}bius action of $T^{\epsilon,\delta}_{\sigma}$. Due to the compactness of $\overline{\mathbb{C}}$,~\eqref{generate-invariant-measure} has a weakly convergent subsequence and, therefore, each initial distribution $\varrho_0$ produces at least one $(T_{\sigma}^{\epsilon,\delta}\cdot)$-invariant Borel probability measure in that manner. In fact, it was shown in \textit{Step 1} that $\mu^{\epsilon,\delta}$ is the unique $(T^{\epsilon,\delta}_{\sigma}\cdot)$-invariant probability measure and, therefore, $\mu^{\epsilon,\delta}$ is the only weak limit point of $(\xi_N)_{N\in\mathbb{N}}$, regardless of the choice of the initial distribution $\rho_0$. Now, if one chooses the initial distribution $\rho_0$ to be supported by a subset of the unit disc $\mathbb{D}$, then, all $\xi_N$ are supported by a subset of $\mathbb{D}$, since the M{\"o}bius action of $T^{\epsilon,\delta}_{\sigma}$ leaves $\mathbb{D}$ invariant. In particular, the weak limit point $\mu^{\epsilon,\delta}$ of $(\xi_{N})_{N\in\mathbb{N}}$ is supported by a subset of the closed unit disc $\overline{\mathbb{D}}$. \hfill $\diamond$

\vspace{.1cm}

\noindent\textit{\underline{Step 3.} For all $\zeta>0$ and all $z,z^{\prime}\in\overline{\mathbb{D}}$, there exists some $\vartheta>0$ and $N\in\mathbb{N}$ such that one~has}
\begin{align}\label{full-support-step-3-statement}
\mathbb{P}\left(\{(T_N\cdots T_1)\cdot \tilde{z}\in B_{\zeta}(z^{\prime})\}\right)\;>\;0\qquad\forall\textnormal{ }\tilde{z}\in B_{\vartheta}(z)\,,
\end{align}
\textit{where $(T_n)_{n=1}^N$ are independent copies of $T^{\epsilon,\delta}_{\sigma}$ and $B_{\vartheta}(z)$ and $B_{\zeta}(z^{\prime})$ are balls of radius $\vartheta$ and $\zeta$ around $z$ and $z^{\prime}$, respectively.}

\vspace{.1cm}

First, every angle shift can be approximated arbitrarily well while the radius is preserved:

\vspace{.1cm}

\noindent \textit{\dashuline{Step 3a:} For all $\kappa>0$, $r\in [0,1]$ and $\Delta\varphi\in [0,2\pi)$, there exists a number $N_{\Delta\varphi}^{\kappa}\in\mathbb{N}$ such that}
\begin{align}\label{arbitrary-angle-change}
\left|{R}_{\mathsf{k}}^{N^{\kappa}_{\Delta\varphi}}\cdot (re^{\varphi})-re^{\varphi+\Delta\varphi}\right|\;<\;\kappa\,.
\end{align}
Indeed, since $\mathsf{k}\not\in\pi\mathbb{Q}$, the sequence $((2n\mathsf{k})\hspace{-1.75mm}\mod (2\pi))_{n\in\mathbb{N}}$ lies dense in $[0,2\pi)$, which implies~\eqref{arbitrary-angle-change}.

\vspace{.1cm}

Second, within $\mathbb{D}$, arbitrary radius growth is possible at the expense of some angle change: 

\vspace{.1cm}

\noindent \textit{\dashuline{Step 3b.} For all $r\in [0,1]$ and $C\in 1+\sinh(\epsilon\mathsf{p})^2[0,1]$, there exists an angle $\varphi\in [0,2\pi)$ such~that}
\begin{align}\label{arbitrary-radius-growth-1}
1-\big|e^{\epsilon \mathsf{p}B_1}\cdot (re^{\imath\varphi})\big|^2\;=\;[1-r^2]C^{-1}\,.
\end{align}
To prove~\eqref{arbitrary-radius-growth-1}, let us observe that all $r\in [0,1]$ and $\varphi\in [0,2\pi)$ satisfy
\begin{align}\label{arbitrary-radius-growth-2}
\frac{1-r^2}{1-\big|e^{\epsilon \mathsf{p}B_1}\cdot (re^{\imath\varphi})\big|^2}\;=\;1+r\cos(\varphi)\sinh(2\epsilon \mathsf{p})+[1+r^2]\sinh(\epsilon \mathsf{p})^2\;=:\;h(\cos(\varphi))
\end{align}
and one has
$$
h(\pm 1)\;=\;1\pm r\sinh(2\epsilon \mathsf{p})+[1+r^2]\sinh(\epsilon \mathsf{p})^2\,,
$$
which implies 
\begin{align}\label{arbitrary-radius-growth-3}
h(1)\;\geq\; 1+\sinh(\epsilon\mathsf{p})^2\;\geq\; 1\;\geq\; h(-1)\,.
\end{align}
In view of~\eqref{arbitrary-radius-growth-3} and the continuity of $h$, there is some $\varphi\in[0,2\pi)$ obeying $h(\cos\varphi)=C$. With this choice,~\eqref{arbitrary-radius-growth-2} implies that~\eqref{arbitrary-radius-growth-1} is indeed satisfied.

\vspace{.1cm}

Third, from any point in $\overline{\mathbb{D}}$, the origin can be approached arbitrarily closely:

\vspace{.1cm}

\noindent \textit{\dashuline{Step 3c.} For all $\kappa>0$ and $z\in \overline{\mathbb{D}}$, there exists a number $N\in\mathbb{N}$ such~that}
\begin{align}\label{convergence-to-origin}
\left|\exp\left[\imath\delta\mathsf{q} B_3\right]^N\cdot z\right|\;<\;\kappa\,.
\end{align}
Inequality~\eqref{convergence-to-origin} follows from $\big|\exp\left[\imath\delta\mathsf{q} B_3\right]^N\cdot z\big|=e^{-\delta\mathsf{q} N}|z|$ for arbitrarily large $N$.

\vspace{.1cm}

Now, let $\zeta>0$ and $z,z^{\prime}\in\mathbb{D}$. In view of \textit{Step 3a}, \textit{Step 3b}, \textit{Step 3c} and due to the continuity of the M{\"o}bius action, there exists a finite sequence $(\mathsf{T}_n)_{n=1}^N\subset \{R_{\mathsf{k}},\exp\left[\epsilon\mathsf{p}B_1\right],\exp\left[\imath\delta\mathsf{q}B_3\right]\}$ and a positive number $\vartheta>0$ for which all $\tilde{z}\in B_{\vartheta}(z)$ satisfy
\begin{align}\label{full-support-step-3-almost-the-statement}
\mathsf{T}_N\cdots\mathsf{T}_1\cdot \tilde{z}
\;\in\; 
B_{\frac{\zeta}{2}}(z^{\prime})\,.
\end{align}
Since the matrices $R_{\mathsf{k}}$, $\exp\left[\epsilon\mathsf{p}B_1\right]$ and $\exp\left[\imath\delta\mathsf{q}B_3\right]$ lie in the support of $T^{\epsilon,\delta}_{\sigma}$, the inclusion~\eqref{full-support-step-3-almost-the-statement} allows to infer~\eqref{full-support-step-3-statement}, again by taking the continuity of the M{\"o}bius action into account. \hfill $\diamond$

\vspace{.1cm}

\noindent\textit{\underline{Step 4.} The support of $\mu^{\epsilon,\delta}$ is a superset of $\overline{\mathbb{D}}$.}\\
Let $z^{\prime}\in\mathbb{D}$ and $\zeta>0$. By \textit{Step 2}, the (non-empty) support of $\mu^{\epsilon,\delta}$ is a subset of $\overline{\mathbb{D}}$. Therefore, one can pick some $z\in\textnormal{supp}(\mu^{\epsilon,\delta})$, for which the statement of \textit{Step 3} implies the existence of some $\vartheta>0$ and some $N\in\mathbb{N}$ that satisfy~\eqref{full-support-step-3-statement}. Now, since $z\in \textnormal{supp}(\mu^{\epsilon,\delta})$, one has $\mu^{\epsilon,\delta}(B_{\vartheta}(z))>0$. Combined with~\eqref{full-support-step-3-statement} and the invariance property of $\mu^{\epsilon,\delta}$, this implies
\begin{align}\label{positive-measure-of-ball}
\mu^{\epsilon,\delta}(B_{\zeta}(z^{\prime}))
&
\;=\;\int_{\overline{\mathbb{C}}}\textnormal{d}\mu^{\epsilon,\delta}(\tilde{z})\textnormal{ }\mathbb{P}\left(T_N\cdots T_1\cdot \tilde{z}\in B_{\zeta}(z^{\prime})\right)
\nonumber
\\
&
\;\geq\;
\int_{B_{\vartheta}(z)}\hspace{-1mm}\textnormal{d}\mu^{\epsilon,\delta}(\tilde{z})\textnormal{ }\mathbb{P}\left(T_N\cdots T_1\cdot \tilde{z}\in B_{\zeta}(z^{\prime})\right)\;>\;0\,.
\end{align}
Since $\zeta>0$ was arbitrary,~\eqref{positive-measure-of-ball} implies that $z^{\prime}$ lies in the support of $\mu^{\epsilon,\delta}$. Now, since $z^{\prime}\in\mathbb{D}$ was also arbitrary, the (closed) support of $\mu^{\epsilon,\delta}$ is a superset of the closure $\overline{\mathbb{D}}$ of $\mathbb{D}$. \hfill $\diamond$

\vspace{.1cm}

The statements of \textit{Step 1}, \textit{Step 2} and \textit{Step 4} imply the claim.
\hfill $\Box$

%%%%%%%%%%%%%%%%%%%%%%%%%%%%%%%%%%%%%%%%%%
\section{Complex energies for random Jacobi matrices}
\label{sec-Anderson}

A random Jacobi matrix is a family $(H_\omega)_{\omega\in\Omega}$ of Jacobi operators on $\ell^2(\ZM)$ indexed by a compact dynamical system $(\Omega,\tau,\ZM,\PP)$ specified by a compact set $\Omega$ equipped with a $\ZM$ action $\tau$ and a $\tau$-invariant and ergodic probability measure $\PP$ on $\Omega$, which satisfies the covariance relation
$$
U_n^*\,H_\omega U_n\;=\;
H_{\tau_n\omega}
\;.
$$
Here $U_n$ is the left shift on $\ell^2(\ZM)$ by $n$. A Jacobi operator is a selfadjoint tridiagonal operator, namely it is of the form
\begin{equation}
\label{eq-polymerHam}
(H_\omega\psi)(n)
\;=\;
-\,t_{\omega}(n+1) \psi(n+1)
\,+\,v_{\omega}(n)\psi(n)\,-\,t_{\omega}(n)\psi(n-1)
\mbox{ , }
\qquad
\psi\in\ell^2(\ZZ)
\;,
\end{equation}
with sequences $(t_\omega(n))_{n\in\ZM}$ and $(v_\omega(n))_{n\in\ZM}$ of compactly supported, positive and real random numbers, respectively, called the hopping and potential values. Here, we will focus on particular kinds of random Jacobi matrices, namely so-called random polymer models \cite{JSS,DSB}. In these models, $H_\omega$ is built from independently drawn blocks of random length $K$. Each such block is called a polymer and is given by the data $\sigma=(K,\baret_{\sigma}(1),\ldots,\baret_{\sigma}(K),\barev_{\sigma}(1),\ldots,\barev_{\sigma}(K))$ containing the length, as well as the hopping and potential values of the polymer. Hence, $\sigma\in\Sigma$ with $\Sigma\subset\bigcup\limits_{K=1}^L\{K\}\times \mathbb{R}_+^K\times \mathbb{R}^K$, which is supposed to be compact and equipped with a probability measure $\PM_\Sigma$. How to construct the dynamical system $\Omega$ as a Palm measure is explained in detail in \cite{JSS,DSB}, but this is not relevant for the following. The best known example is the Anderson model in which $K=1$ and $t_\omega(n)=1$ and only the potential values are random and given by an i.i.d. sequence $(v_\omega(n))_{n\in\ZM}$ of compactly supported, real-valued random variables.

\vspace{.2cm}

Solutions of the Schr\"odinger equation $H_\omega \psi=E\psi$ for $E\in\CM$ are usually \cite{BL,CL} studied using transfer matrices
\begin{align*}
S_{v-E,t}
\;=\;
\frac{1}{t}\,
\left(\begin{array}{cc} v-E & {-}t^2 \\ 1 & 0 \end{array} \right)
\;.
\end{align*}
In case of real energies $E\in\mathbb{R}$, the matrices $S_{v-E,t}$ lie in $\textnormal{SL}(2,\mathbb{R})=\left\{A\in\mathbb{R}^{2\times 2}: \det(A)=1\right\}$.  A basis of the Lie algebra $\textnormal{sl}(2,\RM)=\{A\in\RM^{2\times 2}:\Tr(A)=0\}$ of $\textnormal{SL}(2,\mathbb{R})$ is given by $\left\{B_1,\imath B_2,\imath B_3\right\}$.
For polymer models it is then natural to consider the polymer transfer matrices $S^E_\sigma$ defined by
\begin{equation}
\label{eq-transfer}
S^E_\sigma
\;=\;
\prod\limits_{k=1}^K
S_{\barev_{{\sigma}}(k)-E,\baret_{{\sigma}}(k)} 
\;.
\end{equation}
%

%%%%%%%%%%%%%%%%%%%%%%%%%%%%%%%%%%%%%%%%%%%%%%
\begin{defini}[\cite{JSS}]
\label{def-critical}
An energy $E_c\in\RR$ is called a {\rm critical energy} for the random family
$(H_\omega)_{\omega\in\Omega}$ of polymer Hamiltonians if
the polymer transfer matrices  $S_\sigma^{E_c}$ 
commute for all $\sigma,\sigma'\in\Sigma$:
\begin{equation}
\label{eq-critical}
[S^{E_c}_\sigma,S^{E_c}_{\sigma'}]\;=\;0
\;.
\end{equation}
The critical energy is called {\rm elliptic} if for all $\sigma$ one has either $|\mbox{\rm Tr}(S_\sigma^{E_c})|<2$ or  $S_\sigma^{E_c}=\pm{\bf 1}$. 
%The critical energy is called hyperbolic if for all $\sigma$ one has either $|\mbox{\rm Tr}(S_\sigma^{E_c})|>2$ or  $S_\sigma^{E_c}=\pm{\bf 1}$. 
\end{defini}
%%%%%%%%%%%%%%%%%%%%%%%%%%%%%%%%%%%%%%%%%%%%%%

The definition of an elliptic critical energy implies that there is a basis change $M^{\prime}\in\textnormal{SL}(2,\mathbb{R})$ that transforms all polymer transfer matrices simultaneously into rotations:
\begin{equation}
\label{eq-tric-prelim}
M^{\prime}S^{E_c}_{\sigma}(M^{\prime})^{-1}
\;=\;
\begin{pmatrix}
\cos(\eta_{\sigma}) & -\sin(\eta_{\sigma}) \\ \sin(\eta_{\sigma}) & \cos(\eta_{\sigma})
\end{pmatrix}
\;.
\end{equation}

For further use let us next introduce the notations
\begin{align}\label{Cayley}
J
\;=\;
\begin{pmatrix}
1 & 0 \\ 0 & -1 
\end{pmatrix}
\;,
\qquad
I
\;=\;
\begin{pmatrix}
0 & -1  \\ 1 & 0 
\end{pmatrix}
\;,
\qquad
C
\;=\;
{\sqrt{\frac{-\imath}{2}}}
\begin{pmatrix}
1 & -\imath \\ 1 & \imath
\end{pmatrix}
\;.
\end{align}
The matrix $C$ is also referred to as the Cayley transform. It satisfies $\imath I=C^*JC$ so that $C\,\textnormal{SL}(2,\mathbb{R})\,C^* =\textnormal{SU}(1,1)$. Here it yields as basis change $M=CM^{\prime}$ that transforms all polymer transfer matrices simultaneously into diagonal matrices:
\begin{equation}
\label{eq-tric}
MS^{E_c}_{\sigma}M^{-1}
\;=\;
R_{\eta_\sigma}
\;.
\end{equation}
Now, for energies $E=E_c+\epsilon-\imath\delta$ in the vicinity of $E_c$, let us compute for all $k\in\{1,\dots,K\}$
\begin{align}\label{Anderson-SU_leq}
\begin{split}
[M & S_{\barev_{{\sigma}}(k)-(E_c+\epsilon-\imath\delta),\baret_{{\sigma}}(k)}M^{-1}]^*J[MS_{\barev_{{\sigma}}(k)-(E_c+\epsilon-\imath\delta),\baret_{{\sigma}}(k)}M^{-1}]\\
&\;=\;(M^{-1})^*[S_{\barev_{{\sigma}}(k)-(E_c+\epsilon-\imath\delta),\baret_{{\sigma}}(k)}(M^{\prime})^*C^*JCM^{\prime}S_{\barev_{{\sigma}}(k)-(E_c+\epsilon-\imath\delta),\baret_{{\sigma}}(k)}]M^{-1}\\
&\;=\;\imath (M^{-1})^*[S_{\barev_{{\sigma}}(k)-(E_c+\epsilon-\imath\delta),\baret_{{\sigma}}(k)}(M^{\prime})^* IM^{\prime}S_{\barev_{{\sigma}}(k)-(E_c+\epsilon-\imath\delta),\baret_{{\sigma}}(k)}]M^{-1}\\
&\;=\;\imath (M^{-1})^*[S_{\barev_{{\sigma}}(k)-(E_c+\epsilon-\imath\delta),\baret_{{\sigma}}(k)}IS_{\barev_{{\sigma}}(k)-(E_c+\epsilon-\imath\delta),\baret_{{\sigma}}(k)}]M^{-1}\\
&\;=\;(M^{-1})^*[\imath I-2\delta\hspace{1mm} \baret_{{\sigma}}(k)^{-2}\hspace{1mm}\textnormal{diag}(1,0)]M^{-1}\\
&\;=\; \imath (C^{-1})^*({(M^{\prime})}^{-1})^*I{(M^{\prime})}^{-1}C^{-1}-2\delta\hspace{1mm} \baret_{{\sigma}}(k)^{-2}\hspace{1mm}(M^{-1})^*\textnormal{diag}(1,0)M^{-1}\\
&\;=\; \imath (C^{-1})^*IC^{-1}-2\delta\hspace{1mm} \baret_{{\sigma}}(k)^{-2}\hspace{1mm}(M^{-1})^*\textnormal{diag}(1,0)M^{-1}\\
&\;=\; J-2\delta\hspace{1mm} \baret_{{\sigma}}(k)^{-2}\hspace{1mm}(M^{-1})^*\textnormal{diag}(1,0)M^{-1}\,,
\end{split}
\end{align}
where the equation ${(M^{\prime})}^*I{M^{\prime}}=I$ was used. The identity~\eqref{Anderson-SU_leq} implies that the matrices
$MS_{\barev_{{\sigma}}(k)-(E_c+\epsilon-\imath\delta),\baret_{{\sigma}}(k)}M^{-1}$ are $\textnormal{SU}_{\leq}(1,1)$-valued if $\delta\geq 0$ and are even $\textnormal{SU}(1,1)$-valued if $\delta=0$. Since $\textnormal{SU}_{\leq}(1,1)$ and $\textnormal{SU}(1,1)$ are semi-groups, one has 
$$
T^{\epsilon,\delta}_{\sigma}\;=\;MS^{E_c+\epsilon-\imath\delta}_{\sigma}M^{-1}\;\in\;\textnormal{SU}_{\leq}(1,1)\qquad \textnormal{if}\quad \delta\geq 0
$$
and
$$
MS^{E_c+\epsilon}_{\sigma}M^{-1}\;\in\;\textnormal{SU}(1,1)
$$
and 
$$
MS^{E_c-\imath(\imath\delta)}_{\sigma}M^{-1}\;\in\;\textnormal{SU}(1,1)\,.
$$
All in all, if $E_c$ is an elliptic critical energy of a random polymer model, then the random~matrices
\begin{equation}
\label{eq-TChoice1}
T^{\epsilon,\delta}_\sigma
\;=\;
MS^{E_c+\epsilon-\imath\delta}_{\sigma}M^{-1}
\end{equation}
satisfy the assumptions (i)-(vii) stated in the introduction, where the (negative) imaginary part of the energy plays the role of the parameter $\delta$.

\vspace{.2cm}

For a system stemming from a random Jacobi matrix, the Lyapunov exponent can be obtained from the density of states $\Nn$ via the so-called Thouless formula (see~\cite{CL}, p. 376):
\begin{align}\label{Thouless}
\gamma^{\epsilon,\delta}
\;=\;
\int\mathcal{N}(\textnormal{d}\lambda)\textnormal{ }\log(|E-\lambda|)\;+\;\textnormal{const.}\,,\qquad E=E_c+\epsilon-\imath\delta\,.
\end{align}
Due to~\eqref{Thouless}, the Lyapunov exponent increases if the energy diverges from the real line. Indeed,
\begin{align}
\label{Thouless-consequence}
\gamma^{\epsilon,\delta}\,-\,\gamma^{\epsilon,0}
\;=\;
\frac{1}{2}\int\mathcal{N}(\textnormal{d}\lambda)\textnormal{ }\log\left(1+\frac{\delta^2}{(E_c+\epsilon-\lambda)^2}\right)\;>\;0\,,\qquad E\in\mathbb{R}\,,\quad\delta>0\,.
\end{align}
Theorem~\ref{main-result} makes a more precise statement on the l.h.s.  of~\eqref{Thouless-consequence}.

\vspace{2mm}

Explicit examples of random polymer models with $K\geq 2$ that have such an elliptic critical energy are given in \cite{JSS}. For the Anderson model where $K=1$ and $t_\omega(n)=1$ for all $n$, all energies in the interval $(-2,2)$ are critical in the sense of Definition~\ref{def-critical}. One can thus also work with the family \eqref{eq-TChoice1}. There is, however, a more interesting choice when the potentials are small and of the form $\hat{v}_\sigma(1)=\epsilon  w_\sigma$, where $w_\sigma$ is a compactly supported, real-valued random variable, which is independent from $\epsilon$ and $\delta$. Then for any fixed $E\in(-2,2)$ also
\begin{equation}
\label{eq-TChoice2}
T^{\epsilon,\delta}_\sigma
\;=\;
M 
\begin{pmatrix} \epsilon w_\sigma-E+\imath\delta & -1\\ 1 & 0 \end{pmatrix}
M^{-1}
\end{equation}
satisfies the assumptions (i)-(v). Let us also give the explicit form of $M$ in this case. First~set
$$
M^{\prime}
\;=\;
\frac{1}{\sqrt{\sin(k)}}\begin{pmatrix}
\sin(k)&0\\-\cos(k)&1
\end{pmatrix}\;,
\qquad 
k
\;=\;
\arccos\left(-\frac{E}{2}\right)\in (0,\pi)
$$
and deduce
\begin{align}\label{Anderson-basis-change}
M^{\prime} 
\begin{pmatrix} \epsilon w_\sigma-E+\imath\delta & -1\\ 1 & 0 \end{pmatrix}
(M^{\prime})^{-1}
\;=\;
\begin{pmatrix}
\cos(k) & -\sin(k) \\ \sin(k) & \cos(k)
\end{pmatrix}
\exp\left[-\frac{\imath\delta+\epsilon w_\sigma}{\sin(k)}\begin{pmatrix}
0&0\\1&0
\end{pmatrix}\right]
\;.
\end{align}
A further conjugation by the Cayley transform given by~\eqref{Cayley} yields
\begin{align}\label{Anderson-explicit}
T^{\epsilon,\delta}_{\sigma}\;=\;M\begin{pmatrix} \epsilon w_\sigma-E+\imath\delta & -1\\ 1 & 0 \end{pmatrix}M^{-1}\;=\;\begin{pmatrix}e^{-\imath k}&0\\0&e^{\imath k}\end{pmatrix}\exp\left[\frac{\imath\delta+\epsilon w_\sigma}{2\sin(k)}(B_2+B_3)\right]\,,
\end{align}
where $M=CM^{\prime}$. Then, one has
\begin{align}\label{Anderson-explicit-coefficients}
\eta_{\sigma}\;=\;-k\,,\qquad P_{\sigma}\;=\;\frac{w_{\sigma}}{2\sin(k)} (B_2+B_3)\,,\qquad P^{\prime}_{\sigma}\;=\;0\,,\qquad Q_{\sigma}\;=\;\frac{1}{2\sin(k)} (B_2+B_3)
\end{align}
without any terms of higher order in the exponent of~\eqref{Anderson-explicit}. Therefore,~\eqref{Anderson-explicit} also satisfies the assumptions (vi) and (vii) stated in the introduction.

\vspace{.2cm}

The numerics in Figures~1-4 were done with the single-site Anderson model \eqref{eq-TChoice2} with the choice $E=-2\cos(2)$ and the random variable $w_{\sigma}$ being distributed uniformly on $[-1,1]$. In that case, one readily computes $\mathcal{C}=[2\,\sin(2)]^{-1}$ and $\mathcal{D}=[24\,\sin(2)^2]^{-1}$.
In Figures~\ref{fig-drop-of-v} and~\ref{fig-drop-of-v-and-C=0}, the last matrix in~\eqref{Anderson-explicit-coefficients} was additionally multiplied with an independent random factor $d_{\sigma}$ uniformly drawn from $[-1,3]$ and $[-1,1]$, respectively. Then $Q_{\sigma}=(B_2+B_3)d_{\sigma}[2\sin(k)]^{-1}$ and assumption~(v) is violated.  The choices of $d_{\sigma}$ yield $\mathcal{C}=[2\,\sin(2)]^{-1}$ and $\mathcal{C}=0$, respectively.

\vspace{.2cm}

It is next shown in Proposition~\ref{strenghten-positivity-of-q_3} that the positivity of the coefficients $q_{3,\sigma}$ as given in Remark~\ref{non-negativity-of-q_3} can be strengthened for matrices $T^{\epsilon,\delta}_{\sigma}$ that arise from random polymer models. We encourage the reader to have a look at Proposition~\ref{center-support} again and compare with Proposition~\ref{strenghten-positivity-of-q_3}.

%%%%%%%%%%%%%%%%%%%%%%%%%%%%%%%
\begin{proposi}\label{strenghten-positivity-of-q_3}
If $T^{\epsilon,\delta}_{\sigma}$ arises from a random polymer model, then all $\sigma\in\Sigma$ satisfy $q_{3,\sigma}\geq |\xi_{\sigma}|$.
\end{proposi}
%%%%%%%%%%%%%%%%%%%%%%%%%%%%%%%

\noindent\textbf{Proof.} The statement follows by mimicking the proof of Proposition 3 in~\cite{DSB}. For this, let us observe that one has $M^*JM=\imath I$ due to $M\in C[\textnormal{SL}(2,\mathbb{R})]$. This allows to compute
\begin{align*}
%\label{positive-matrix}
%\begin{split}
&-(T^{0,0}_{\sigma})^*J\partial_{\delta}T^{0,\delta}_{\sigma}\Big|_{\delta=0}\\
&\quad\;=\;-\imath (S^{E_c}_{\sigma}M^{-1})^*I(\partial_{\delta}S^{E_c+\imath\delta}_{\sigma})M^{-1}\Big|_{\delta=0}\\
&\quad\;=\;(S^{E_c}_{\sigma}M^{-1})^*I\mbox{\small $\sum\limits_{k=1}^K\left[\prod\limits_{m=k+1}^KS_{\hat{v}_{\sigma}(m)-E_c-\imath\delta,\hat{t}_{\sigma}(m)}\right]$}\mbox{\footnotesize $\begin{pmatrix}\hat{t}_{\sigma(k)}^{-1}&0\\0&0\end{pmatrix}$}\mbox{\small $\left[\prod\limits_{m=1}^{k-1}S_{\hat{v}_{\sigma}(m)-E_c-\imath\delta,\hat{t}_{\sigma}(m)}\right]$}M^{-1}\Bigg|_{\delta=0}\\
&\quad\;=\;(M^{-1})^*\mbox{\footnotesize $\sum\limits_{k=1}^K$}\mbox{\footnotesize $\left[\prod\limits_{m=1}^KS_{\hat{v}_{\sigma}(m)-E_c,\hat{t}_{\sigma}(m)}\right]^*$}I\mbox{\footnotesize $\left[\prod\limits_{m=k+1}^KS_{\hat{v}_{\sigma}(m)-E_c,\hat{t}_{\sigma}(m)}\right]$}\mbox{\footnotesize $\begin{pmatrix}\hat{t}_{\sigma(k)}^{-1}&0\\0&0\end{pmatrix}\left[\prod\limits_{m=1}^{k-1}S_{\hat{v}_{\sigma}(m)-E_c,\hat{t}_{\sigma}(m)}\right]$}M^{-1}\\
&\quad\;=\;(M^{-1})^*\mbox{\small $\sum\limits_{k=1}^K$}\mbox{\small $\left[\prod\limits_{m=1}^{k-1}S_{\hat{v}_{\sigma}(m)-E_c,\hat{t}_{\sigma}(m)}\right]^*$}S_{\hat{v}_{\sigma}(m)-E_c,\hat{t}_{\sigma}(m)}\textnormal{ }I\textnormal{ }\mbox{\footnotesize $\begin{pmatrix}\hat{t}_{\sigma(k)}^{-1}&0\\0&0\end{pmatrix}$}\mbox{\small $\left[\prod\limits_{m=1}^{k-1}S_{\hat{v}_{\sigma}(m)-E_c,\hat{t}_{\sigma}(m)}\right]$}M^{-1}\\
&\quad\;=\;(M^{-1})^*\mbox{\small $\sum\limits_{k=1}^K$}\mbox{\small $\left[\prod\limits_{m=1}^{k-1}S_{\hat{v}_{\sigma}(m)-E_c,\hat{t}_{\sigma}(m)}\right]^*$}\mbox{\footnotesize $\begin{pmatrix}\hat{t}_{\sigma(k)}^{-2}&0\\0&0\end{pmatrix}$}\mbox{\small $\left[\prod\limits_{m=1}^{k-1}S_{\hat{v}_{\sigma}(m)-E_c,\hat{t}_{\sigma}(m)}\right]$}M^{-1}\,.
%\end{split}
\end{align*}
Clearly, the r.h.s. of this equation is non-negative, and thus also the l.h.s.. Thus, its determinant
\begin{align*}
\begin{split}
\det\mbox{\footnotesize $\left(-(T^{0,0}_{\sigma})^*J\partial_{\delta}T^{0,\delta}_{\sigma}\Big|_{\delta=0}\right)$}
&
\;=\;\det\left(-\imath R_{\eta_{\sigma}}^*JR_{\eta_{\sigma}}Q_{\sigma}\right)
\\
&
\;=\;\det\left(-\imath JQ_{\sigma}\right)
\\
&
\;=\;\det\mbox{\footnotesize $\begin{pmatrix}q_{3,\sigma}&-\imath\overline{\xi_{\sigma}}\\\imath\xi_{\sigma}&q_{3,\sigma}\end{pmatrix}$}\\
&
\;=\;q_{3,\sigma}^2-|\xi_{\sigma}|^2
\end{split}
\end{align*}
and its trace $\Tr\mbox{\footnotesize $\left(-(T^{0,0}_{\sigma})^*J\partial_{\delta}T^{0,\delta}_{\sigma}\Big|_{\delta=0}\right)$}=2q_{3,\sigma}$ are non-negative, which implies $q_{3,\sigma}\geq |\xi_{\sigma}|$.
\hfill $\Box$

\vspace{.2cm}

Moreover, if $\epsilon$ is non-zero and the potential $w_{\sigma}$ is non-trivial, then~\eqref{Anderson-explicit} fulfills also conditions (viii) and (ix) so that the $(T^{\epsilon,\delta}_{\sigma}\cdot)$-invariant probability measure is unique. In fact, to insure condition (ix), it is also sufficient that the imaginary part of the energy is non-zero. All this is carried out in the following Propositions~\ref{condition-viii} and~\ref{condition-ix}.

\begin{proposi}\label{condition-viii}
Suppose that $\epsilon\neq 0$ and $\textnormal{card}(\textnormal{supp}(w_{\sigma}))>1$. Then,~\eqref{Anderson-explicit} fulfils condition~\textnormal{(viii)}.
\end{proposi}

\noindent\textbf{Proof.} By assumption, there exist $\sigma_1,\sigma_2\in\Sigma$ such that $w_{\sigma_1},w_{\sigma_2}\in\textnormal{supp}(w_{\sigma})$ satisfy $w_{\sigma_1}\neq w_{\sigma_2}$. \\
\noindent\textit{\underline{Case 1:} Both $T_{\sigma_1}^{\epsilon,\delta}$ and $T_{\sigma_2}^{\epsilon,\delta}$ are diagonalizable.}\\
In that case, there exist some $U_{\sigma_1}\in\textnormal{SL}(2,\mathbb{C})$ and $U_{\sigma_2}\in\textnormal{SL}(2,\mathbb{C})$ for which one has
\begin{align*}
\tilde{T}^{\epsilon,\delta}_{\sigma_l}\;:=\;U_{\sigma_l}^{-1}T^{\epsilon,\delta}_{\sigma_l}U_{\sigma_l}\;=\;\begin{pmatrix}
r_{\sigma_l}e^{\imath\varphi_{\sigma_l}}&0\\0&r_{\sigma_l}^{-1}e^{-\imath\varphi_{\sigma_l}}
\end{pmatrix}\,,\qquad r_{\sigma_l}> 0\,,\quad \varphi_{\sigma_l}\in [0,2\pi)\,,\qquad l\in\{1,2\}\,.
\end{align*}
Now, if $(\varphi_l,r_l)\not\in\pi\mathbb{Q}\times\{1\}$, the singeltons $\{0\}$ and $\{\infty\}$ specify all finite orbits of $\tilde{T}^{\epsilon,\delta}_{\sigma_l}\cdot$. Otherwise, one has for each couple $(\varrho,\vartheta)\in (0,\infty)\times [0,2\pi)$ a further finite orbit $\varrho\exp [\imath(\vartheta+\varphi_l\mathbb{N})]$. By symmetry, a finite subset of $\overline{\mathbb{C}}$ is an orbit of $\tilde{T}^{\epsilon,\delta}_{\sigma_l}\cdot$ if and only if it is an orbit of $(\tilde{T}^{\epsilon,\delta}_{\sigma_l})^{-1}\cdot$. Accordingly, a finite  subset of $\overline{\mathbb{C}}$ is an orbit of $T^{\epsilon,\delta}_{\sigma_l}\cdot$ if and only if it is an orbit of $(T^{\epsilon,\delta}_{\sigma_l})^{-1}\cdot$. Moreover, any finite $(({T}^{\epsilon,\delta}_{\sigma_l})^{\pm 1}\cdot)$-invariant set is a union of finitely many finite $(({T}^{\epsilon,\delta}_{\sigma_l})^{\pm 1}\cdot)$-invariant orbits. Hence, if $F$ were a finite subset $F$ of $\overline{\mathbb{C}}$ invariant both under ${T}^{\epsilon,\delta}_{\sigma_1}\cdot$ and ${T}^{\epsilon,\delta}_{\sigma_2}\cdot$, then, it would also be invariant under $({T}^{\epsilon,\delta}_{\sigma_1})^{-1}\cdot$ and $({T}^{\epsilon,\delta}_{\sigma_2})^{-1}\cdot$ and, in particular, under the M{\"o}bius action~of
$$
(T^{\epsilon,\delta}_{\sigma_2})^{-1}T^{\epsilon,\delta}_{\sigma_1}\;=\;\begin{pmatrix}1-\imath\frac{\xi}{2}&-\imath\frac{\xi}{2}\\\imath\frac{\xi}{2}&1+\imath\frac{\xi}{2}\end{pmatrix}\,,\qquad\qquad\textnormal{where}\qquad \xi\;=\;\epsilon [w_{\sigma_2}-w_{\sigma_1}]\sin(k)^{-1}\neq 0\,.
$$
Therefore, $C^{-1}\cdot F$ (would also be finite and) would be invariant under the M{\"o}bius action~of
\begin{align}\label{condition-viii-1}
C^{-1}(T^{\epsilon,\delta}_{\sigma_2})^{-1}T^{\epsilon,\delta}_{\sigma_1}C\;=\;\begin{pmatrix}1 & 0 \\ \xi &1\end{pmatrix}\,,
\end{align}
which clearly satisfies
\begin{align*}
%\label{condition-viii-2}
\lim\limits_{N\rightarrow\infty}\left[C^{-1}(T^{\epsilon,\delta}_{\sigma_2})^{-1}T^{\epsilon,\delta}_{\sigma_1}C\right]^N\cdot z\;=\;0\quad\forall\textnormal{ }z\in\overline{\mathbb{C}}\quad\mbox{and}\quad 
\Big(
C^{-1}(T^{\epsilon,\delta}_{\sigma_2})^{-1}T^{\epsilon,\delta}_{\sigma_1}C\cdot z\;=\;0\textnormal{ }\Longleftrightarrow\textnormal{ } z\;=\;0\Big)\,.
\end{align*}
These properties, the finiteness of $C^{-1}\cdot F$ and its invariance under the M{\"o}bius action of~\eqref{condition-viii-1} would imply $C^{-1}\cdot F=\{0\}$,~\textit{i.e.}, $F=\{C\cdot 0\}=\{-1\}$. But, $T^{\epsilon,\delta}_{\sigma}\cdot (-1)=-e^{2\imath k}\neq -1$, as $k\in (0,\pi)$. \hfill $\diamond$ 

\vspace{2mm}

\noindent\textit{\underline{Case 2:} For some $l\in\{1,2\}$, the matrix $T_{\sigma_j}^{\epsilon,\delta}$ is {\it not} diagonalizable.}\\
First of all, let us remark that this case is of minor relevance, since it can only occur if $\delta=0$ and $|\epsilon|$ is sufficiently large. One can assume without loss of generality that $T_{\sigma_1}^{\epsilon,0}$ is {\it not} diagonalizable. Then, $T^{\epsilon,0}_{\sigma_1}$ has either $+1$ or $-1$ as its only eigenvalue, namely with geometric multiplicity $1$. Moreover, the conjugation of $T^{\epsilon,0}_{\sigma_1}$ with some $U\in\textnormal{SL}(2,\mathbb{C})$ yields the Jordan form of $T_{\sigma_l}^{\epsilon,\delta}$,
$$
U^{-1}T^{\epsilon,0}_{\sigma_1}U\;=\;\begin{pmatrix}
\pm 1& 1\\0&\pm 1
\end{pmatrix}
$$
which clearly satisfies
\begin{align*}
\lim\limits_{N\rightarrow\infty}(U{T}^{\epsilon,0}_{\sigma_1}U^{-1})^N\cdot z\;=\;\infty\quad\forall\textnormal{ }z\in\overline{\mathbb{C}}\qquad\mbox{and}\qquad 
\Big(
U{T}^{\epsilon,0}_{\sigma_1}U^{-1}\cdot z\;=\;\infty\quad \Longleftrightarrow\quad z\;=\;\infty
\Big)\,.
\end{align*}
Thus, the only finite $(T^{\epsilon,0}_{\sigma_1}\cdot)$-invariant subset of $\overline{\mathbb{C}}$ is given by $\{U^{-1}\cdot\infty\}$. Now, if $\{U^{-1}\cdot\infty\}$ were also $(T^{\epsilon,0}_{\sigma_2}\cdot)$-invariant, then one would have $T_{\sigma_2}^{\epsilon,0}\cdot(U^{-1}\cdot\infty)=U^{-1}\cdot\infty$, which is equivalent to
\begin{align}\label{condition-viii-3}
U(T_{\sigma_2}^{\epsilon,0})^{-1}U^{-1}\cdot \infty\;=\;\infty
\end{align}
Combining~\eqref{condition-viii-3} with $UT^{\epsilon,0}_{\sigma_1}U^{-1}\cdot\infty=\infty$ would yield $U(T_{\sigma_2}^{\epsilon,0})^{-1}T_{\sigma_1}^{\epsilon,0}U^{-1}\cdot\infty=\infty$ and, due to~\eqref{condition-viii-1},
$$
UC\begin{pmatrix}1&0\\\xi &1\end{pmatrix}(UC)^{-1}\cdot \infty\;=\;\infty
$$
or, equivalently,
\begin{align}\label{condition-viii-4}
\begin{pmatrix}1&0\\\xi &1\end{pmatrix}\cdot ((UC)^{-1}\cdot \infty)\;=\;(UC)^{-1}\cdot \infty\,.
\end{align}
Since $0$ is the only fixed point of the M{\"o}bius action of~\eqref{condition-viii-1}, the equation~\eqref{condition-viii-4} would imply $(UC)^{-1}\cdot\infty=0$, which is equivalent to $U^{-1}\cdot\infty =C\cdot 0$. It would follow that $C\cdot 0=-1$ is a fixed point of $T_{\sigma_1}^{\epsilon,0}\cdot$ and $T_{\sigma_2}^{\epsilon,0}\cdot$. But, $T^{\epsilon,\delta}_{\sigma}\cdot (-1)=-e^{2\imath k}\neq -1$, as $k\in (0,\pi)$. \hfill $\diamond$  $\Box$

\begin{proposi}\label{condition-ix}
If $\delta\neq 0$ or $\epsilon\neq 0$ and $\textnormal{card}(\textnormal{supp}(w_{\sigma}))>1$, then~\eqref{Anderson-explicit} fulfils condition \textnormal{(ix)}.
\end{proposi}

\noindent\textbf{Proof.} It suffices to show that the semigroup $\Ss_\PM$ generated by $\textnormal{supp}(T^{\epsilon,\delta}_{\sigma})$ contains a non-elliptic matrix, since all non-elliptic matrices $\mathsf{T}\in\textnormal{SL}(2,\mathbb{C})$ satisfy $\|\mathsf{T}^N\|\rightarrow\infty $ as $N\rightarrow\infty$. If one has either $\delta\neq 0$ or $|\epsilon w-E|>2$ for some $w\in\textnormal{supp}(w_{\sigma})$, the associated matrix $\mathsf{T}$ given by \eqref{eq-TChoice2} is hyperbolic due to $\textnormal{tr}(\mathsf{T})\in\mathbb{C}\setminus [-2,2]$.
If $\delta=0$ and $|\epsilon w-E|=2$ for some $w\in\textnormal{supp}(w_{\sigma})$, then some $\mathsf{T}$ in $\textnormal{supp}(T^{\epsilon,\delta}_{\sigma})$ satisfies $\textnormal{tr}(\mathsf{T})\in\{\pm 2\}$ and thus it is parabolic. It remains to consider the case $\delta=0$ and $\epsilon\neq 0$ under the assumption $|\epsilon w-E|<2$ for all $w\in\textnormal{supp}(w_{\sigma})$.  While this is the standard case \cite{BL,CL}, let us nevertheless sketch an argument. By hypothesis (and an energy shift by $-\epsilon w_{\sigma_1}$), one can assume that $w_{\sigma_1}=0$ and $w_{\sigma_2}\neq 0$. Then
$$
T^{\epsilon,0}_{\sigma_1}=\begin{pmatrix}e^{-\imath k}&0\\0&e^{\imath k}\end{pmatrix}\,,\quad T^{\epsilon,0}_{\sigma_2}=\begin{pmatrix}e^{-\imath k}&0\\0&e^{\imath k}\end{pmatrix}\exp\left[\frac{\imath\delta+\epsilon w_{\sigma_2}}{2\sin (k)}(B_2+B_3)\right]=T^{\epsilon,0}_{\sigma_1}\begin{pmatrix}1+\imath\kappa&\imath\kappa\\-\imath\kappa&1-\imath\kappa\end{pmatrix}\,,
$$
where $\kappa=\epsilon w_{\sigma_2}\left[2\sin(k)\right]^{-1}\neq 0$. Now for any $N\in\NM$ and with $\vartheta=Nk$, one has
\begin{align*}
\mathsf{T}_\vartheta
\;=\;
(T^{\epsilon,0}_{\sigma_1})^{N-1}T^{\epsilon,0}_{\sigma_2}
\;=\;
\begin{pmatrix}e^{-\imath \vartheta} & 0 \\ 0 &e^{\imath \vartheta}\end{pmatrix}
\begin{pmatrix}1+\imath\kappa&\imath\kappa\\-\imath\kappa&1-\imath\kappa\end{pmatrix}
\;\in\;\Ss_\PM
\,.
\end{align*}
The second factor is parabolic. For rational $2\pi k$, one can achieve $\vartheta=0$. For irrational $2\pi k$, one can obtain arbitrarily small $\vartheta$ with desired sign. As $\vartheta\mapsto \mathsf{T}_\vartheta$ undergoes a Krein collision at $0$, one always finds a hyperbolic element in $\Ss_\PM$ in this manner. 
\hfill $\square$

\vspace{.3cm}

\noindent {\bf Acknowledgements:} We thank two anonymous referees for constructive comments. F.~D. received funding from the {\it Studienstiftung des deutschen Volkes}. This work was also partly supported by the~DFG grant SCHU 1358/6-2.

%%%%%%%%%%%%%%%%%%%%%%%%%%%%%%%%%%%%%%%%%%%%%%%%%%%

\end{document}